\documentclass[aps,prc,twocolumn,groupedaddress,showpacs]{revtex4-1}
\usepackage{color,soul}
\usepackage{graphicx}

\newcommand {\la} {\langle}\newcommand {\ra} {\rangle}
\newcommand {\beq} {\begin{eqnarray}}
\newcommand {\eeq} {\end{eqnarray}}
\newcommand {\eeqn} [1] {\label{#1} \end{eqnarray}}
\newcommand {\eol} {\nonumber \\}
\newcommand {\ve} [1] {\mbox{\boldmath $#1$}}
\newcommand {\dem} {\mbox{$\frac{1}{2}$}}
\newcommand {\third} {\mbox{$\frac{1}{3}$}}

\begin{document}

\title{Three-nucleon force contribution to the deuteron channel in $(d,p)$ reactions}
\author{N. K. Timofeyuk, M. J. Dinmore and J.S. Al-Khalili }
\affiliation{Department of Physics, Faculty of Engineering and Physical Sciences, University of Surrey
Guildford, Surrey GU2 7XH, United Kingdom }
\date{June 2020}

\begin{abstract}
The contribution of a three-nucleon (3N) force, acting between the neutron and proton in the incoming deuteron with a target nucleon, to the  deuteron-target potential  in the entrance channel of the $A(d,p)B$ reaction has been calculated within the adiabatic distorted wave approximation (ADWA). Four different 3N  interaction sets from local chiral effective field theory ($\chi$EFT) at next-to-next-to-leading order (N2LO) were used. Strong sensitivity of the adiabatic deuteron-target potential to the choice of the 3N force format has been found, which originates from the enhanced sensitivity to the short-range physics of nucleon-nucleon (NN) and 3N interactions in the ADWA.
Such a sensitivity is reduced when a Watanabe folding model is used  to generate $d$-$A$ potential instead of ADWA. The impact of the 3N force contribution on  $(d,p)$ cross sections depends on assumptions made about the  $p$-$A$ and $n$-$A$ optical potentials used to calculate the distorted $d$-$A$ potential in the entrance channel. It is different for local and nonlocal optical  potentials and depends on whether the induced three-body force arising due to neglect of target excitations is included or not.
\end{abstract}

\maketitle

\section{Introduction}

The three-nucleon (3N) force is  important for understanding the structure and dynamics of atomic nuclei and nuclear matter from first principles and  it is routinely used in many {\it ab-initio} structure calculations \cite{Bar13,Lyn16,Tew16,Pia20}. It is also included in calculations describing reactions involving nuclei that can be modelled as few-body systems \cite{Del15,Lyn16,Viv18}. However, this force is not considered in analyses of experimental data involving direct reactions with complex nuclei. One particular class of such reactions, deuteron stripping $(d,p)$ and pick-up $(p,d)$, is an important experimental tool for testing the shell-model picture of atomic nuclei, which is often used for indirect determination of nucleon capture reaction rates at astrophysical energies \cite{Bar16}. 

The $(d,p)$ and $(p,d)$ reactions are
usually described either within the Distorted Wave Born Approximation (DWBA) \cite{Austern} or the Adiabatic Distorted Wave Approximation (ADWA) \cite{JT}. Both use phenomenological optical potentials where  the contribution from the 3N force is present implicitly. However, there are two other ways for  the 3N force to  manifest itself in $(d,p)$ and $(p,d)$ reactions. One of them is through an additional term in the DWBA or ADWA $T$-matrix element  containing the $n$-$p$-$i$ interaction $W_{npi}$, where $i$ belongs to the target, in the transition operator \cite{Tim18}. Its importance has been assessed  in \cite{Tim18} using  a simplified hypercentral contact interaction model only. The second way for the 3N force to exercise its influence is through the interaction of a nucleon in the target $A$ with both the neutron and proton in the incoming deuteron. Such a contribution can be important in the ADWA, which accounts for deuteron breakup and uses neutron-target  and proton-target optical potentials. A bare 3N force would create a new $n$-$p$-$A$ three-body force in addition to the $n-A$ and $p-A$ optical potentials. It would complement the  effective induced $n$-$p$-$A$ three-body force that arises due to neglect of  target excitation in the three-body description of deuteron stripping and pickup reactions \cite{Din19}. The $n$-$p$-$A$ force arising from bare 3N interactions has not been considered before. Including 3N force has been listed in \cite{Tim20} among  outstanding tasks aimed at the further development of deuteron stripping and pick-up reaction theory.

In this work we will present the first calculations of the 3N contribution to the deuteron adiabatic potential for the $A(d,p)B$ reactions as defined in Sec. II. We will start in Sec. III with an investigation of the general sensitivity of the ADWA potential to the strength and range of this force using a simple hypercentral model, similar to that used in \cite{Tim18}, showing the importance of the consistency between the NN and 3N models. We then proceed with a local chiral EFT at N2LO model with parameters taken from  \cite{Lyn16}
that reproduce well the properties of light nuclei and nuclear matter. We evaluate the contributions from the contact, one-pion-exchange plus contact and two-pion-exchange parts of these interaction in Sec. IV,  V, and VI, respectively.  We demonstrate influence of the choice of the 3N force format and reveal the role of the deuteron $d$-state. The ADWA calculations will then be presented in Sec. VII for $^{40}$Ca and $^{27}$Al targets using a range of nucleon-target optical potentials. In Sec. VIII we calculate  the 3N force contribution in the Watanabe folding $d-A$ model where the coupling to deuteron breakup states is missing, but 3N effects could be expected to be more similar to those one might expect beyond the ADWA. We will show that the 3N contribution has a different effect on the Watanabe $d-A$ folding potential and the corresponding $(d,p)$ cross sections. The obtained results are discussed in Sec. IX where  conclusions will also be given and the need for specific future developments described. The Appendix provides expressions for a few functions needed to calculate the 2$\pi$-exchange contribution.

\section{Adiabatic model with 3N folding potential}

\begin{figure}[t]
\includegraphics[width=0.45\textwidth]{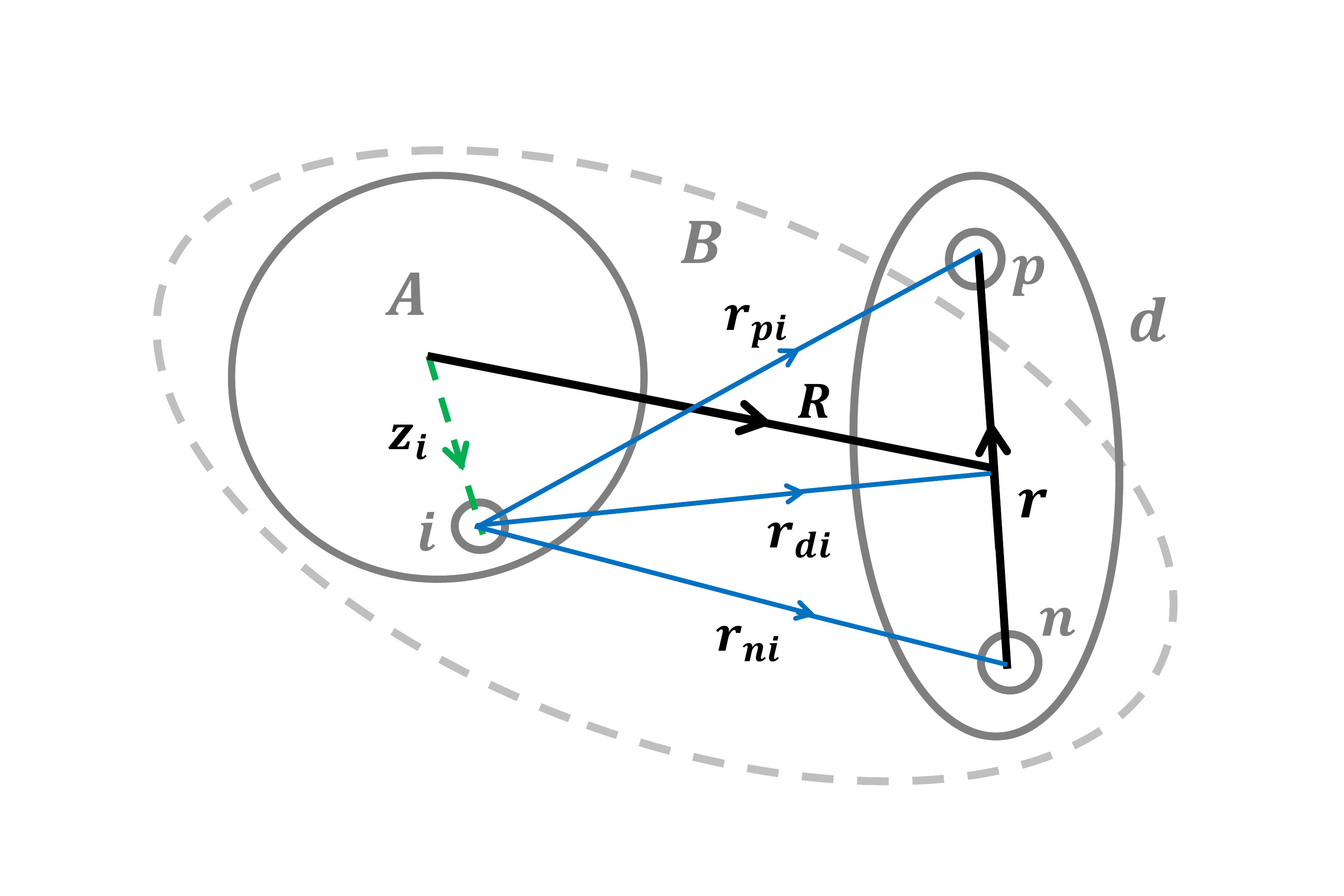}
\caption{
Coordinates associated with 3N force in the $A(d,p)B$ reaction:  radius vectors $\ve{r}_{ni}$, $\ve{r}_{pi}$ and $\ve{r}_{di}$ connecting a target nucleon $i$ with the neutron, proton  and the deuteron center-of-mass, respectively; $\ve{r} = \ve{r}_{ni} - \ve{r}_{pi}$;   coordinate $\ve{z}_i$ of nucleon $i$ with respect to the target center-of-mass  and   the radius-vector $\ve{R}$ connecting the target and deuteron centers of mass.
}
\label{fig:coordinates}
\end{figure} 

The ADWA assumes that the wave function of the $d$--$A$ system is described by a $n+p+A$ three-body model and that it can be represented by the first term in the  Weinberg expansion  \cite{JT}, which has been shown to dominate the $(d,p)$ cross sections due to the short-range $n$-$p$ potential in the $(d,p)$ $T$-matrix \cite{Pan13}. The ADWA approximates the first Weinberg component by a product of the adiabatic distorted-wave function $\chi_{ad}(\ve{R})$ and a deuteron wave function $\phi_0$. 
The distorted wave $\chi_{ad}(\ve{R})$ is found from the solution of the two-body Schr\"odinger equation
\beq
(T + \la \phi_1 \mid U_{nA}  + U_{pA} \mid \phi_0\ra  - E_d) \chi_{ad}(\ve{R}) = 0,
\eeqn{adSE}
where $T$ is the kinetic energy operator associated with the coordinate $\ve{R}$ of the relative $d$--$A$ motion, $E_d$ is the centre-of-mass deuteron energy and $U_{nA}$ and $U_{pA}$ are the nucleon-target optical potentials that depend on coordinates $\ve{R} + \ve{r}/2$ and $\ve{R} - \ve{r}/2$, respectively, where $r$ is the $n$-$p$ separation, shown in Fig. \ref{fig:coordinates}. Also in Eq.(\ref{adSE}), $\phi_0$ is the deuteron wave function and
\beq
\phi_1 = \frac{V_{np} \phi_0} {\la \phi_0  \mid V_{np} \mid \phi_0\ra}.
\eeqn{phi1}
The ADWA, as well as other direct reaction theories, was  derived on the assumption that the  nuclear Hamiltonian contains NN interactions only. In the presence the 3N force one can expect that the adiabatic equation (\ref{adSE}) should include an additional term
\beq
U_{3N}(\ve{R}) =
\la \phi_1 \phi_A \mid \sum_{i \in A} W_{npi} \mid \phi_A \phi_0 \ra,
\eeqn{3Nfold}
where $W_{npi}$ is the 3N interaction between $n$, $p$ and a nucleon $i$ belonging to the target $A$, $\phi_A$ is the many-body wave function describing $A$ and the integration in the matrix element (\ref{3Nfold}) is carried out over all internal coordinates of $A$ and over $\ve{r}$. We will start with considering the matrix element
\beq
W^{\rm eff}_{di}(\ve{r}_{di}) =
\la \phi_1  \mid   W_{npi} \mid  \phi_0 \ra,
\eeqn{2Neff}
where integration is done over  $\ve{r}$. It has a meaning of  an effective two-body potential between nucleon $i$ and the centre-of-mass of deuteron. The 3N  contribution $U_{3N}$ to the adiabatic potential is then described by the well-known folding potential
\beq
U_{3N}(\ve{R}) = \int d\ve{s} \rho_A(\ve{s}) W^{\rm eff}_{di}(\ve{R}-\ve{s}),
\eeqn{3Nfold2}
where $\rho_A$ is the target density for which we will use some available phenomenological representations. The adiabatic distorted wave $\chi_{ad}(\ve{R})$ is now found from 
\beq
(T + \la \phi_1 \mid U_{nA}  + U_{pA} \mid \phi_0\ra + U_{3N}(\ve{R})   - E_d) \chi_{ad}(\ve{R}) = 0. \eol
\eeqn{adSE3N}

\section{Adiabatic $d-A$ potential with a hypercentral 3N interaction}

To have a general idea about the sensitivity of the $d-A$ potential  $U_{3N}(\ve{R})$ to the 3N force we  first consider a hypercentral 3N interaction
\beq
W(\rho_{ijk}) = W_0 e^{-\frac{\rho_{ijk}^2}{\rho_0^2}},
\eeqn{hc3N}
where $\rho_{ijk}^2 = (r_{ij}^2 + r_{ik}^2 + r_{jk}^2)/3$, with $r_{ij}$ being the distance between nucleons $i$ and $j$, and $\rho_0$ is the range of the 3N force. We can also consider another format of this force, such as used in \cite{Tim18},
\beq
W^{(\tau)}(\rho_{ijk}) = \third (\ve{\tau}_i \cdot \ve{\tau}_j + \ve{\tau}_i \cdot \ve{\tau}_k + \ve{\tau}_k \cdot \ve{\tau}_j )W_0^{(\tau)} e^{-\frac{\rho_{ijk}^2}{\rho_0^2}}, \eol 
\eeqn{hc3Ntau}
where $\ve{\tau}_i$ is the isospin Pauli matrix. The two different representations of the hypercentral force are reminiscent of different possible formats of the 3N contact interactions in $\chi$EFT at N2LO that we will use in the sections below.
We notice that the matrix element in Eq.(\ref{2Neff}) that determines $W^{\rm eff}_{di}$ uses $W_{npi}$ together with the functions $\phi_0$ and $\phi_1$, which correspond to zero deuteron isospin.
In this case the isospin $T_{npi}$ of the $i-d$ system is equal to $\frac{1}{2}$ so that
\beq
(\ve{\tau}_n \cdot \ve{\tau}_p + \ve{\tau}_n \cdot \ve{\tau}_i + \ve{\tau}_p \cdot \ve{\tau}_i ) \chi_{d}\chi_{i}\qquad\qquad\qquad\eol
 \!\!\! = 2(\ve{T}_{npi}^2 -\ve{t}^2_n - \ve{t}^2_p - \ve{t}^2_i)\chi_{d}\chi_{i} = -3\, \chi_{d}\chi_{i},
\eeqn{tautau}
where $\chi_d$ and $\chi_i$ are the isospin functions of the deuteron and nucleon $i$, respectively, and $\ve{T}_{npi}$ and $\ve{t}_i$ are the isospin operators of the three-body $n$-$p$-$i$ system and nucleon $i$, respectively. Therefore, using the  formats (\ref{hc3N}) and (\ref{hc3Ntau}) of the 3N force with $W_0^{(\tau)}= -W_0  $ will give identical results.

The  hyperradius  $\rho_{npi}$ can be expressed  via normalized Jacobi coordinates $\ve{x}_1 = \ve{r}/\sqrt{2}$ and $\ve{x}_2 = \sqrt{\frac{2}{3}} \ve{r}_{di}$ as $\rho_{npi}^2 = x_1^2 + x_2^2$ (see Fig. 1 for coordinate definitions). Then using (\ref{hc3N}) or (\ref{hc3Ntau}) in Eq. (\ref{2Neff}) we obtain
\beq
W_{di}^{\rm eff} (\ve{r}_{di}) = W_0 e^{-\frac{2}{3} \frac{r_{di}^2}{\rho_0^2}}
\int d\ve{r}  \phi_1^*(\ve{r}) e^{-\frac{1}{2} \frac{r^2}{\rho_0^2}} \phi_0(\ve{r}). \,\,\,\,\,\,\,\,\,\,\,\,\,
\eeqn{W3Nhc}

Let is consider first a zero-range 3N force with a strength fixed by the volume integral $I_3 = 3^{\frac{3}{2}}W_0 \pi^3 \rho_0^6$, consistent with the definition given by Eq. (45) of Ref. \cite{Tim18}.
Such a force is analogous to the unregularized contact interaction of the $\chi$EFT at N2LO \cite{Lyn16}. In the $\rho_0 \rightarrow 0$ limit we obtain
\beq
W_{di}^{\rm eff} (\ve{r}_{di}) = I_3 \left(\frac{2}{3}\right)^{\frac{3}{2}} \delta \left(\sqrt{\frac{2}{3}}\ve{r}_{di}\right)\phi_1(0) \phi_0(0).
\eeqn{W3NZR}
This leads to the folding 3N force
\beq 
U_{3N}(\ve{R}) = I_3 \phi_1(0) \phi_0(0) \rho_A(\ve{R}),
\eeqn{U3NZR}
which has exactly the same shape as the target density profile with the absolute values determined by the volume integral $I_3$ of the 3N interaction and the values of the deuteron wave function $\phi_0$ and the vertex function $\phi_1$ at zero $n$-$p$ separations. The deuteron wave function $\phi_0(0)$ is very sensitive to high $n$-$p$ momentum content in the deuteron and is strongly dependent on the NN model used, which is illustrated in Table I of Ref. \cite{Tim18}. The same is valid for $\phi_1(0)$. This value can be obtained from $\phi_0(\ve{r})$ by using the Schr\"odinger equation $(T-\epsilon_d)\phi_0 = -\phi_1 \la \phi_0| V_{np}| \phi_0 \ra$, where $\epsilon_d$ is the deuteron binding energy. Taking $\phi_0$ from  the N2LO $\chi$EFT   \cite{Lyn16} we deduce $\phi_1(\ve{r})$  and obtain  
$\phi_1(0) = -1.17$ fm$^{-3/2}$. We also use $\phi_0(0) =0.0796$ fm$^{-3/2}$ from the same NN model and  assume $I_3$ = 2186 MeV$\cdot$fm$^6$, which is consistent with the set II of the N2LO $\chi$EFT considered below. Then taking an average density $\rho_A(0) \approx 0.16 $ fm$^{-3}$  we obtain  $U_{3N}(0) \approx -$33 MeV. 
The origin of this attraction is the negative value of $\phi_1$ at $r=0$. However, beyond zero-range approximation, with increasing range  $\rho_0$, the integral in (\ref{W3Nhc}) would span a larger fraction of both $\phi_0$ and $\phi_1$ which can affect the depth of the folding potential $U_{3N}(R)$.

\begin{figure}[t]
\includegraphics[width=0.45\textwidth]{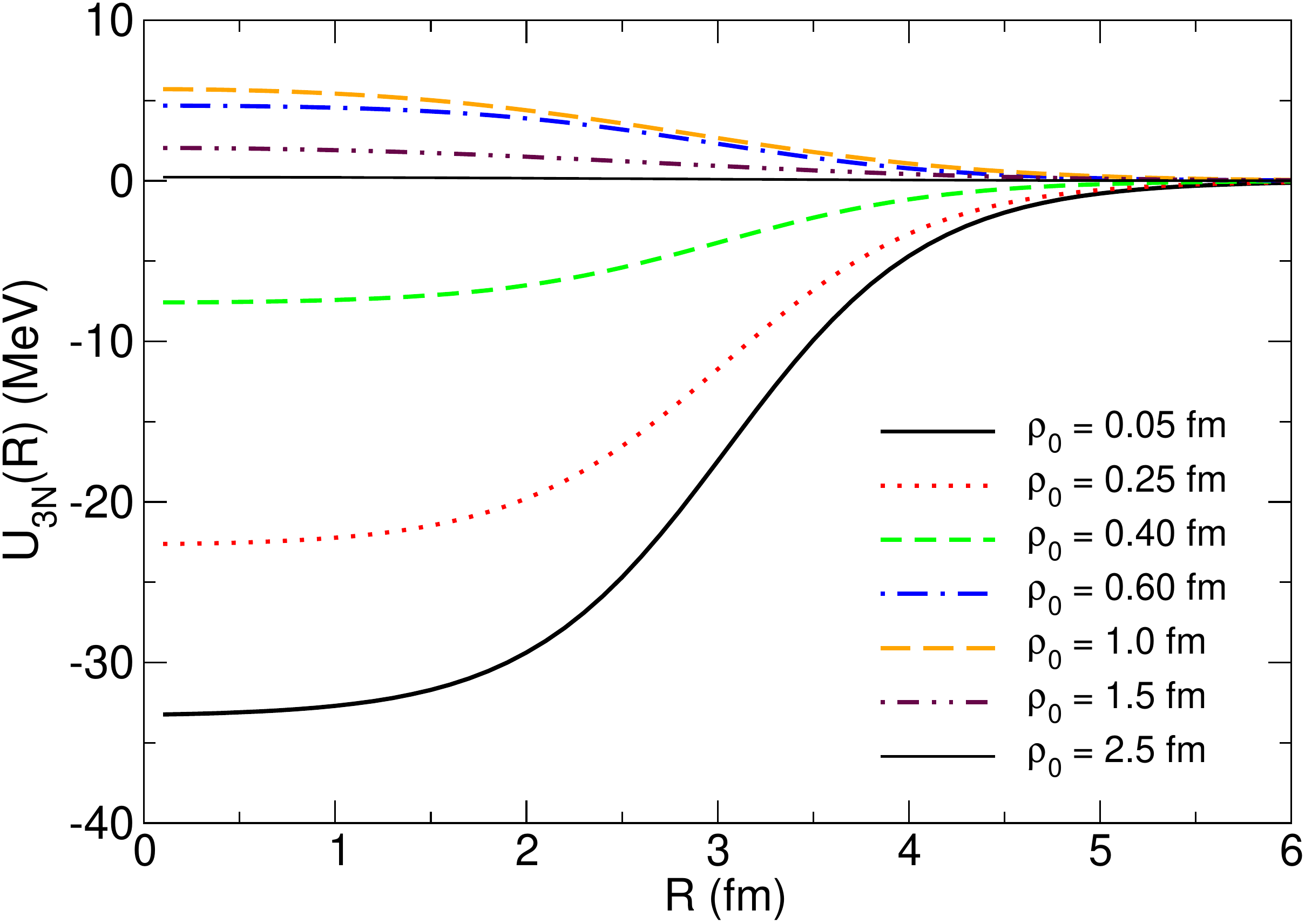}
\caption{
The adiabatic potential $U_{3N}(R)$ calculated for $d +^{26}$Al system using hypercentral 3N potential of a fixed volume integral and several values of the hyperradius $\rho_0$.
}
\label{fig:dA-hc}
\end{figure}

In Fig. \ref{fig:dA-hc} we show the $U_{3N}$ potential calculated for the $d+^{26}$Al system. The $^{26}$Al matter density was assumed to be the same as the two-parameter Fermi charge density of $^{27}$Al obtained in \cite{deJ74} from the analysis of electron scattering.
The same strength $I_3 = 2186$ MeV$\cdot$fm$^{6}$ was used while the range $\rho_0$ was varied. One can see that the choice of $\rho_0$ has a dramatic effect on $U_{3N}(R)$. First, its depth decreases from 33 to 0 MeV when
$\rho_0$ increases from 0 to  $\sim$0.5 fm. 
Then $U_{3N}$ changes sign, becoming repulsive and increasing to $\sim 5.7$ MeV at $ 0.5 \lesssim \rho_0 \lesssim 0.8$ fm after which it decreases again. 

\begin{table}[b]
\caption{The regulator $R_{3N}$ of the two-nucleon formfactor (in fm) and the low-energy constants $c_E$ and $c_D$ for a chosen format of the contact 3N forces taken from \cite{Lyn16}.
}
\centering
\begin{tabular} {p{1.5 cm } p {1.5 cm} p{ 1.2 cm} p{ 1.2 cm} p{ 1.2 cm}  }  
set & format &  $R_{3N}$ & $c_E$ & $c_D$ \\
\hline
\hline
I   & $E1$        & 1.0 & 0.62  & 0.5 \\
II  & $E\tau$     & 1.0 & $-$0.63 & 0.0 \\
III & $E\tau$     & 1.2 &    0.09 & 3.5 \\
IV  & $E_{\cal P}$ & 1.0 & 0.59  & 0.0 \\
\hline\hline
\end{tabular}
\label{tab:sf}
\end{table}

This simple exercise demonstrates the important role of the range of the 3N interaction and shows the necessity of using a 3N force model compatible with the NN interaction model.
Below, we will proceed with using both of them from the  N2LO $\chi$EFT of \cite{Lyn16},  where they have been  consistently fitted to describe some properties of light nuclei, neutron-$\alpha$ scattering and neutron matter.
 This 3N force has  contact, one-pion-contact (1$\pi$-c)  and double-pion (2$\pi$) exchange parts. The strength of the first two of these is governed by low-energy constants $c_E$ and $c_D$, respectively. Several best sets of $(c_E,c_D)$ are available in Ref. \cite{Lyn16}, corresponding to different relative contributions from these two components, and they  are shown in Table I. The low-energy constants defining the 2$\pi$-exchange  contribution are fixed from pion-nucleon or NN scattering  \cite{Epe05}.

\section{Adiabatic $d-A$ potential with contact  3N potential }

We will use for $W_{npi}$ the contact $\chi$EFT interaction at N2LO from \cite{Lyn16} where it was represented by three different formats,  
\beq
V_{E1 }& =& \frac{ c_E}{ \Lambda_{\kappa}F^4_{\pi} } \sum_{cyc}  
 \delta_{R_{3N}}(\ve{r}_{ij}) \delta_{R_{3N}}(\ve{r}_{kj}),
  \\
  V_{E\tau}&=& \frac{ c_E}{ \Lambda_{\kappa}F^4_{\pi} } 
\sum_{cyc} (\ve{\tau}_i\cdot \ve{\tau_k} )\,
 \delta_{R_{3N}}(\ve{r}_{ij}) \delta_{R_{3N}}(\ve{r}_{kj}), 
 \\
 V_{E{\cal P} }& =& \frac{ c_E}{ \Lambda_{\kappa}F^4_{\pi} } \sum_{cyc}  {\cal P}
 \delta_{R_{3N}}(\ve{r}_{ij}) \delta_{R_{3N}}(\ve{r}_{kj}),
\eeqn{3NFcontact}
where $cyc$ denotes all possible cyclic permutations in the $n$-$p$-$i$ system, $\Lambda_{\kappa} = 700$ MeV, $F_{\pi} = 92.4$ MeV and the operator ${\cal P}$ will be discussed later. The different formats are a consequence of the need for regularization of the $\delta$-functions present in the $\chi$EFT \cite{Lyn16,Tew16}. The regularized $\delta$-functions of range $R_{3N}$
used here are given by 
\beq
\delta_{R_{3N}}(\ve{r}) = \frac{1}{\pi\Gamma(3/4) R^3_{3N}} e^{-(r/R_{3N})^4}.
\eeqn{deltadef}

\begin{figure}[t]
\includegraphics[width=0.45\textwidth]{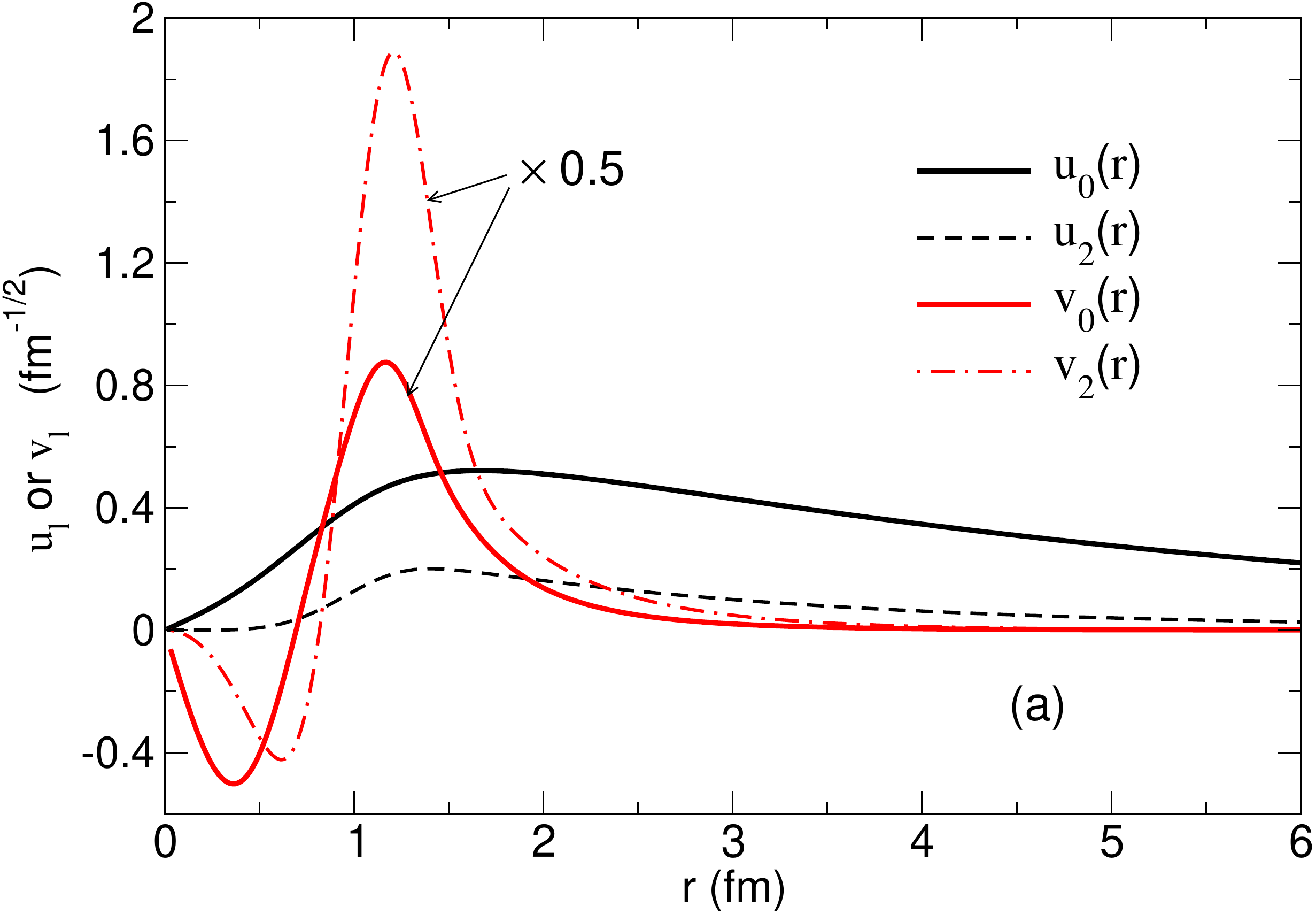}
\includegraphics[width=0.45\textwidth]{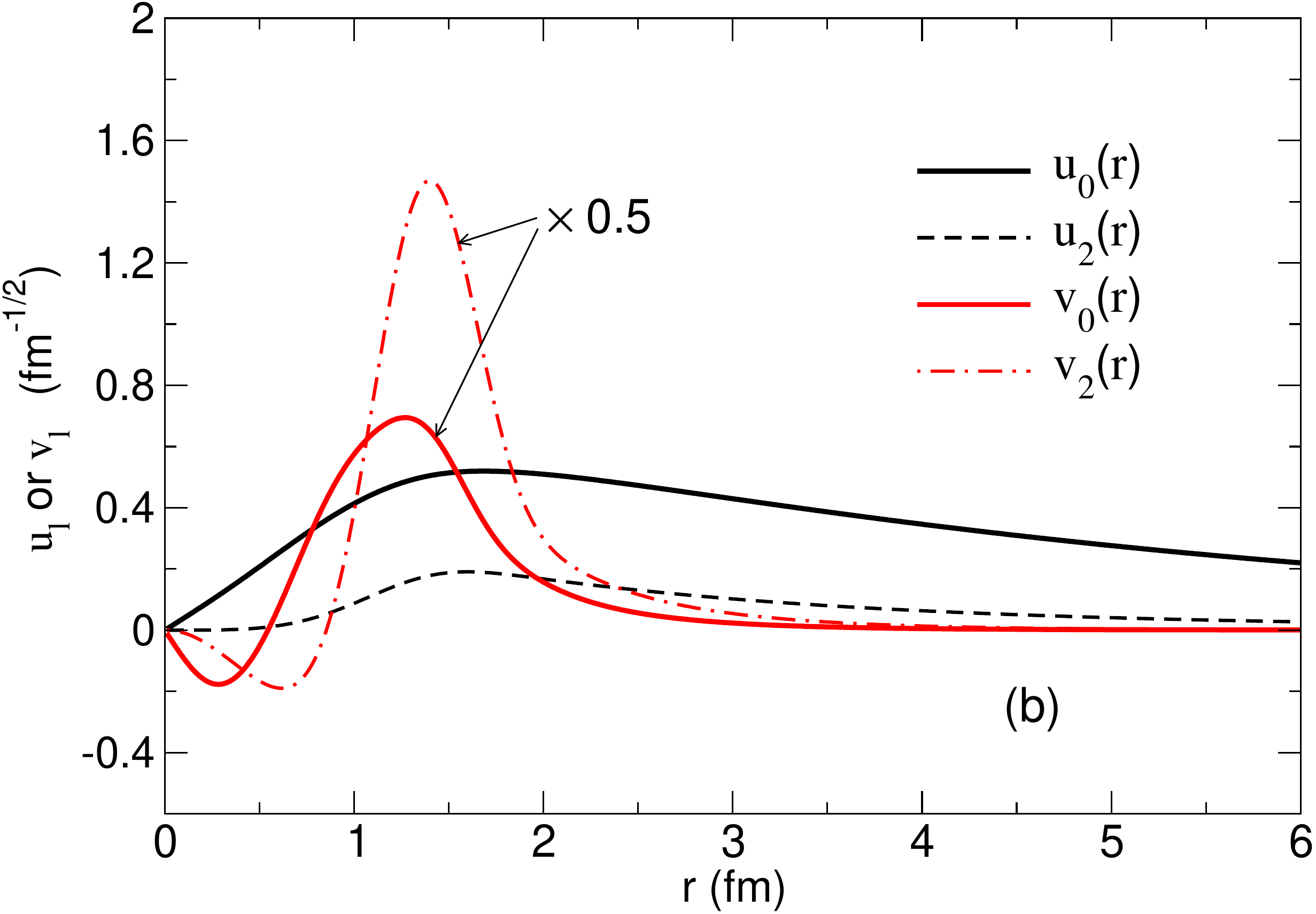}
\caption{
Deuteron $s$-wave and $d$-wave functions $u_0$ and $u_2$ and the corresponding components $v_0$ and $v_2$  of $\phi_1$ in $\chi$EFT at N2LO   shown for two different regulators: $R_{3N} =$ 1.0 fm ($a$) and $R_{3N} = 1.2$ fm ($b$).
}
\label{fig:dwfs}
\end{figure}

Let us first start with the $V_{E1}$ force. It does not contain any isospin operators. All one needs is to use in Eqs. (\ref{3Nfold}) and (\ref{2Neff}) the following quantities:
\beq
 D(r_{di},r) \equiv \delta_{R_{3N}}(\ve{r}_{ni}) \delta_{R_{3N}}(\ve{r}_{pi})
 = \frac{e^{-\frac{16r_{di}^4+ 24 r^2 r_{di}^2 + r^4}{8R_{3N}^4}}}{ \pi^2\Gamma^2\left(\frac{3}{4}\right) R^6_{3N} }  \,\,\,\,\,\,\,\,\,\,\,\,\,  \eeqn{fnfp}
and 
\beq
\left[ \delta_{R_{3N}}(\ve{r}_{ni}) +\delta_{R_{3N}}(\ve{r}_{pi})\right]
 &=&
 8\pi  
 \sum_{\lambda={\rm even}, \mu}   \delta_{\lambda}\left(r_{di},\frac{r}{2}\right) 
 \eol &\times &
 Y^*_{\lambda \mu} (\hat{r}_{di}) Y_{\lambda \mu}(\hat{r}) ,
 \eeqn{deltas+partial}
 where
 \beq
\delta_{\lambda}\left(r_{di},\frac{r}{2}\right) = \frac{1}{2} \int_{-1}^1 d\mu \, P_{\lambda}(\mu) \delta_{R_{3N}} \left( \left|\ve{r}_{di} - \frac{\ve{r}}{2} \right| \right). \,\,\,\,\,\,\,\,
\eeqn{partdel}
Here, the integration is performed over the cosine of the angle between vectors $\ve{r}$ and $\ve{r}_{di}$.
Then, the corresponding effective $d$-$i$ force is  
\beq
   W^{\rm eff}_{E1}(\ve{r}_{di})   \equiv   \la \phi_1 | V_{E1} |\phi_0 \ra   \,\,\,\,\,\,\,\,\,\,\,\,\,\,\,\, \,\,\,\,\,\,\,\,\,\,\,\,\,\,\,\,
   \,\,\,\,\,\,\,\,\,\,\,\,\,\,\,\,\,\,\,\,\,\,\,\,\,\,\,\,\,\,\,\,
   \,\,\,\,\,\,\,\,\,\,\,\,\,\,\,\,
\eol = \sqrt{4\pi}
\sum_{\lambda \mu } (\lambda \mu  J_d M_{d} | J'_d M'_{d})  W^{\rm eff }_{E1,\lambda}(r_{di}) 
Y^*_{\lambda \mu} (\hat{r}_{di}), \,\,\,\,\,\,\,\,\,\,\,\,\,\,\,\,
 \eeqn{WeffE1lam}  
where  $\lambda$ takes only even values,  $J_d(J'_d)$ and $M_d(M'_d)$ are the ket (bra) deuteron angular momentum and its projection, respectively, coupled by the Clebsch-Gordan coefficient  (we have to note that ADWA only deals  with $J'_d = J_d$), while the radial part of $W^{\rm eff }_{E1}(\ve{r}_{di})$ is
\beq  
 W^{\rm eff }_{E1,\lambda}(r_{di})  = 
 \frac{ c_E  }{ \Lambda_{\kappa}F^4_{\pi} }  \,\hat{\lambda}\hat{J}_d \sum_{ll'} \hat{l} (l 0 \lambda 0 | l'0) \qquad\qquad\qquad
\eol  
\!\!\!\! \times W(\lambda l J'_d 1 ; l' J_d) \int_0^{\infty} dr  \,  v_{l'}(r)u_l(r)   \,\,\,\,\,\,\,\,\,\,\,\,\,\,
\eol \times
\left[ D(r_{di},r) \delta_{\lambda,0} + 2 \delta_{R_{3N}}(r)\delta_{\lambda}\left(r_{di}, \frac{r}{2}\right) \right],
\,\,\,\,\,\,\,\,\,
\eeqn{WeffE1}
where $W$ is the Racah coefficient. Also, the $u_l(r)$ and $v_l(r)$ are the radial parts of the deuteron wave function $r\phi_0(\ve{r})$ and vertex function $r\phi_1(\ve{r})$, respectively, in the partial wave $l$. They are shown in Fig. \ref{fig:dwfs} for two regulators $R_{3N}=$1.0 fm and 1.2 fm. 
For realistic deuteron wave function, with non-zero contribution from the $d$-state, the effective $d$-$i$ interaction also has  a quadrupole  component.

The second force, $V_{E\tau}$,   includes isospin operators. Let us consider the first of them: $\ve{\tau}_i \cdot \ve{\tau}_n$. To evaluate $W^{\rm eff}_{di}$, we need the following matrix element:
\beq
\la {T'_d M'_d}(n,p),  {\dem\tau'}(i) | \ve{\tau}_i \cdot \ve{\tau}_n |  {T_d M_d}(n,p), {\dem\tau}(i) \ra, 
\eeqn{MEtau}
where $|T_dM_d(n,p)\ra$ represents the deuteron isospin state vector 
with isospin $T_d$ and its projection $M_d$, and $|\dem\tau(i) \ra$ represents the isospin state of nucleon $i$ with projection $\tau$.  The ADWA involves the deuteron isospin $T_d = 0$  both in the entrance and exit channels  and $\tau'(i) = \tau(i)$. It is easy to show that in this case the contribution from $\ve{\tau}_i \cdot \ve{\tau}_n$ vanishes. For the same reason there will be no contribution from  $\ve{\tau}_i \cdot \ve{\tau}_p$, while the contribution from $\ve{\tau}_n \cdot \ve{\tau}_p$ is calculated in a way similar to Eq. (\ref{tautau}),
\beq
(\ve{\tau}_n \cdot\ve{\tau} _p ) \,\chi_d \chi_i = 2 (T^2_{np} - t^2_n - t^2_p) \chi_d \chi_i
= -3 \, \chi_d \chi_i. \,\,\,\,\,\,\,\,\,\,
\eeqn{tautau2}
Therefore, the $W^{\rm eff}_{E\tau}$ interaction has a simpler structure than $W^{\rm eff}_{E1}$ being 
\beq
W^{\rm eff}_{E\tau} (r_{di})  = -
 \frac{ 3c_E  }{ \Lambda_{\kappa}F^4_{\pi} } \sum_l
\int_0^{\infty} dr  \,  v_{l}(r)u_l(r)   D(r_{di},r). \,\,\,\,\,\,\,\,\,\,\,\,\,\,
\eeqn{WeffEtau}

The third force, $V_{E{\cal P}}$, contains the operator\\ ${\cal P} \!=\! \frac{1}{36} \left(3 - \sum\limits_{i<j} \ve{\sigma}_i\cdot \ve{\sigma}_j \right)\left(3 - \sum\limits_{i<j} \ve{\tau}_i\cdot \ve{\tau}_j \right)$ acting on $(S=\frac{1}{2}, T=\frac{1}{2})$ state of the $n$-$p$-$i$ system only. If we take Eq. (\ref{tautau}) into account, then for $T_d=0$ this operator reduces to ${\cal P} = \frac{1}{6} \left(3 - \sum\limits_{i<j} \ve{\sigma}_i\cdot \ve{\sigma}_j \right)$ which is needed in the context of the matrix element
\beq
\la S'_d M'_d (n,p), \dem \sigma'_i(i) \mid {\cal P} \mid  S_d M_d (n,p), \dem \sigma_i(i) \ra
\eol
 = \sum_{M_3} (S'_d M'_d \dem \sigma'_i | \dem M_3) (S_d M_d \dem \sigma_i | \dem M_3),
 \eeqn{calP}
 where $|S_dM_d(n,p)\ra$ represents the deuteron spin state with  spin $S_d$ and projection $M_d$ and $|\dem\sigma(i) \ra$ is the spin state of nucleon $i$ with projection $\sigma$.
In general, such a matrix element leads to an effective $d$-$i$ force that depend on spin projections $\sigma_i$ of nucleon $i$, which would require the corresponding  densities of the target $A$. Nondiagonal on spin projection parts of such  densities can be nonzero and, moreover, they can explicitly depend on $\sigma_i$. However, in a specific case when the expectation value of the operator $S_A^2 = \left(\sum\limits_{i=1}^A \vec{s}_i \right)^2$ is zero, which is a good approximation for  double-closed shell nuclei, we have $\sigma'_i = \sigma_i$ and 
\beq
\sum_{\sigma_i} \la S_d M'_d , \dem \sigma_i \mid {\cal P} \mid  S_d M_d , \dem \sigma_i \ra = \frac{2}{3} \delta_{S'_d S_d}\delta_{M'_d M_d}, \,\,\,\,\,\,\,\,\,\,
 \eeqn{calPmagic}
 assuming that density distribution of spin projections $\dem$ and $-\dem$ are exactly the same. Therefore, we can conclude that the effective $d$-$i$ interaction $W^{\rm eff}_{E{\cal P}}$ associated with the force $V_{E{\cal P}}$ is given by Eqs. (\ref{WeffE1}) and (\ref{WeffE1lam}),  the same as in the case of $V_{E1}$ but multiplied by 2/3. We stress that this conclusion is valid only for a target that has  spin-zero component.

\begin{figure}[t]
\includegraphics[width=0.45\textwidth]{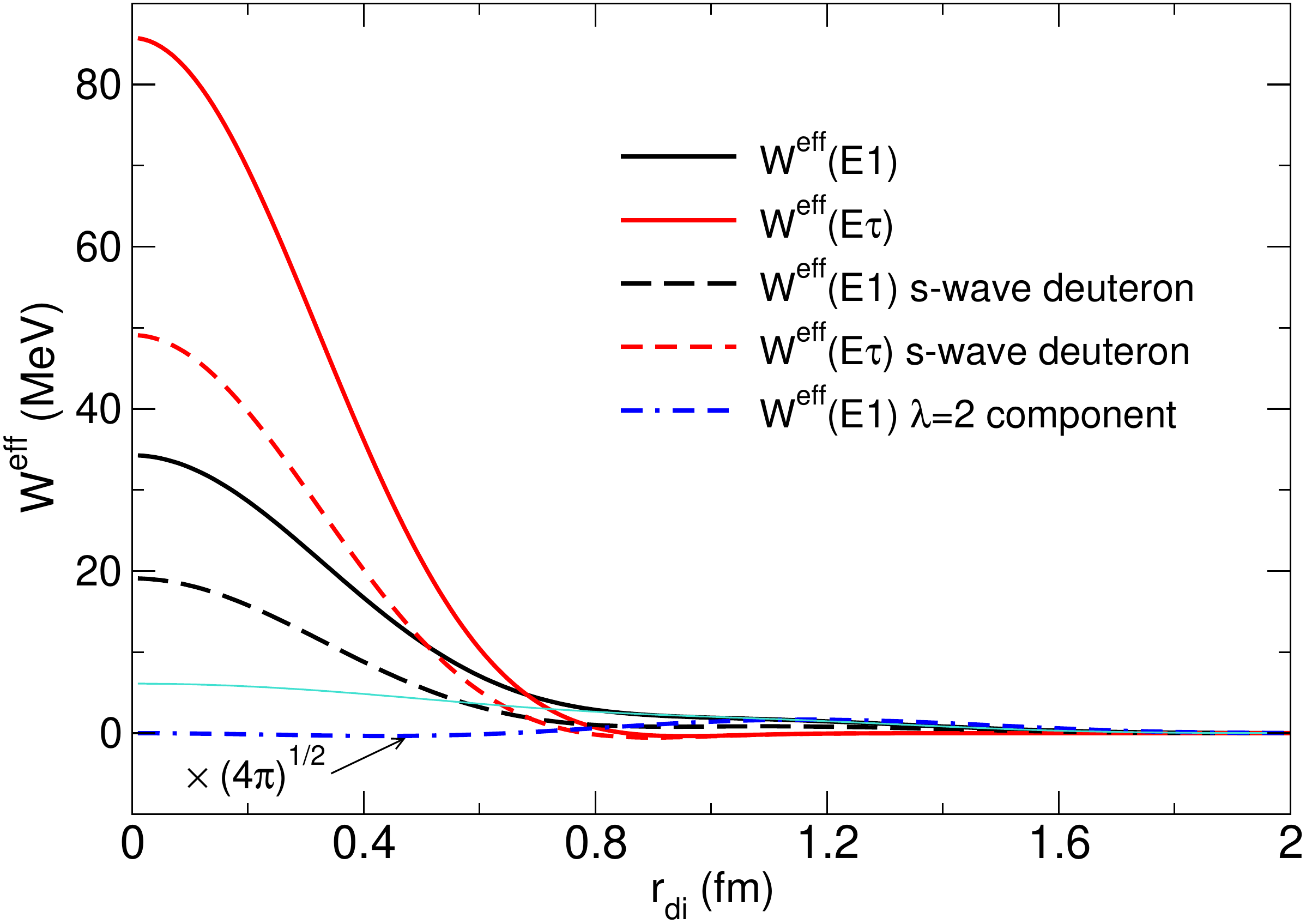}
\caption{
Effective $d$-$i$ potential $W^{\rm eff}_{di}$ obtained with the contact $\chi$EFT 3N force at N2LO in $E1$ and $E\tau$ formats using parameters from sets I and II from Table I, respectively. The calculations without deuteron $d$-wave are shown by dashed lines. A thin blue line shows a contribution from the interaction between $i$ and $n$($p$) via $p(n)$ missing in $E\tau$. The dash-dot line shows quadrupole part of $W^{\rm eff}_{E1}$, multiplied by $\sqrt{4\pi}$.
}
\label{fig:weffad}
\end{figure} 
 
 Figure \ref{fig:weffad}  compares  $ W^{\rm eff}_{E1}$ and $W^{\rm eff}_{E\tau}$  corresponding to set I and II  from Table I, respectively. Both sets have the same regulator $R_{3N} = 1.0$ fm and very similar strengths given by low-energy constants $c_E$. However,    $ W^{\rm eff}_{E1}$  is significantly larger than $W^{\rm eff}_{E\tau}$ at $r_{di} \approx 0$. Although both of them have a common part containing $D(r_{di},r) \delta_{\lambda,0}$, this part is three times stronger for $W^{\rm eff}_{E\tau}$ due to the contribution from $\ve{\tau}_n \cdot \ve{\tau}_p$ that benefits from a factor of $-3$. However,
  $W^{\rm eff}_{E\tau}$ receives the contribution from the $n$ and $p$ interacting via the target nucleon $i$ only, while $W^{\rm eff}_{E1}$ also includes interaction of $n$($p$) with $i$ via $p$($n$). The latter are of a significantly  longer range, shown on Fig. \ref{fig:weffad} by a thin line. As a result, the volume integrals of the $\lambda=0$ components in $W^{\rm eff}_{E\tau}$ and $W^{\rm eff}_{E1}$ are similar, being 31.1 and 37.4 MeV$\cdot$fm$^{3}$, respectively.
 Figure \ref{fig:weffad}  also presents results obtained with the deuteron $s$-state only. One can see that the $d$-state contribution is very important, which is a consequence of a large magnitude of  $v_2(r)$. The $d$-state gives rise to a
  quadrupole $\lambda=2$ component in $W^{\rm eff}_{E1}$ but it is very small. No quadrupole component is present in $W^{\rm eff}_{E\tau}$.

\begin{figure}[b]
\includegraphics[width=0.45\textwidth]{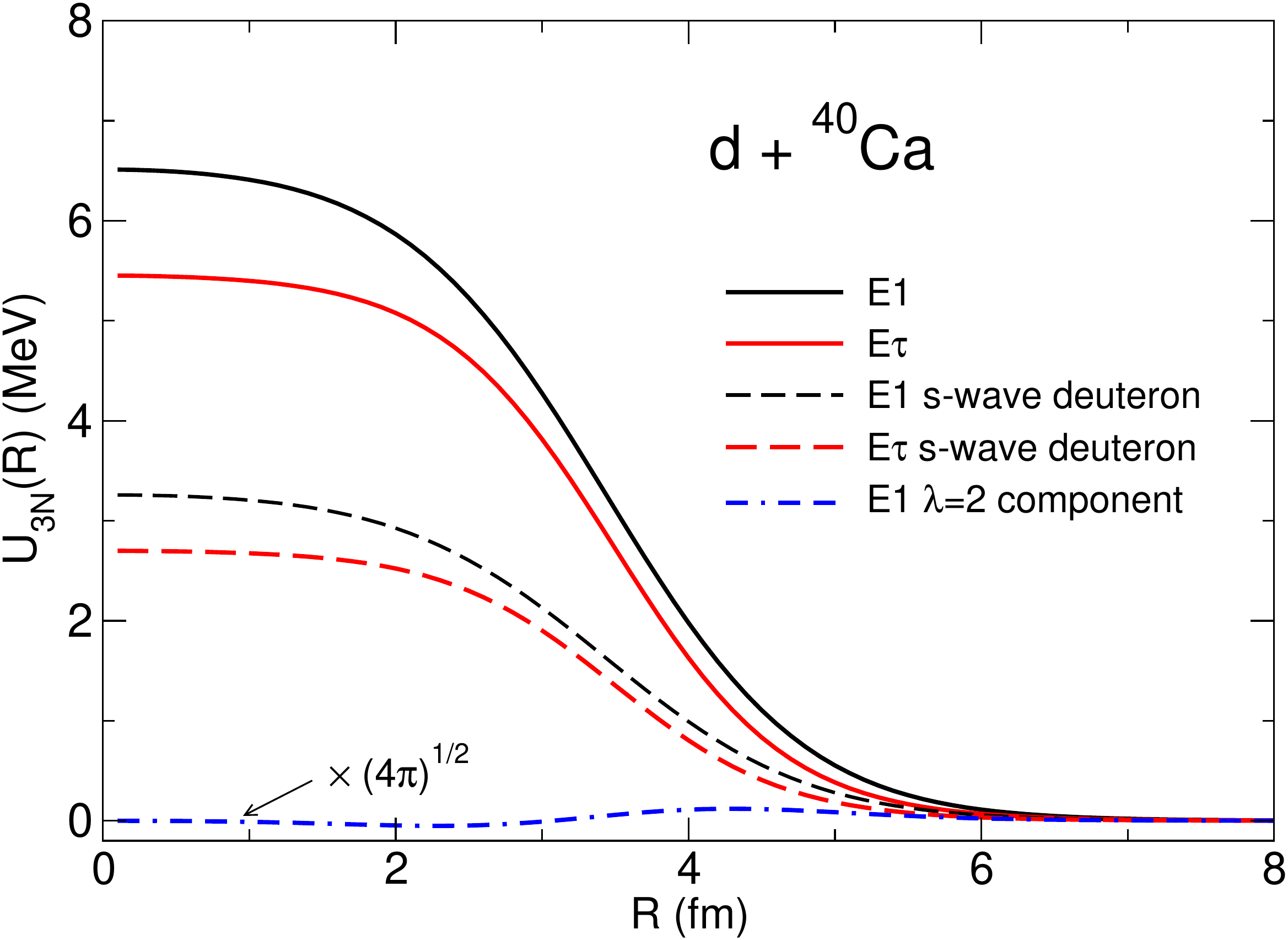}
\caption{
The adiabatic potential $U_{3N}(R)$ calculated for $d +^{40}$Ca system for contact $E1$ and $E\tau$ interactions. The calculations without the contribution from the deuteron $d$-wave are shown by dashed lines. The $\lambda=2$ contribution for $E1$ is also shown, multiplied by $\sqrt{4\pi}$. 
}
\label{fig:ca40-ad}
\end{figure}
 
For  any $W^{\rm eff}_{di}$ interaction the folding $d$-$A$ potential can be represented by a partial wave expansion
 \beq
 U_{3N}(\ve{R}) = 
 \sqrt{4\pi} \sum_{\lambda \mu}
 (\lambda \mu J_dM_d | J'_d M'_d)U_{3N}^{(\lambda)} (R)Y^*_{\lambda \mu} (\hat{R}), \,\,\,\,\,\,\,\,\,\,
\eeqn{folding}
where the radial part is
 \beq
 U_{3N}^{(\lambda)} (R) = 4\pi\int_0^{\infty}ds \,s^2  W^{\rm eff}_{di,\lambda}(s) \rho_{\lambda}(R,s), \eeqn{folding_l}
 with $\rho_{\lambda}\left(R,s\right)$  given by  Eq. (\ref{partdel}) in which $\delta$ is replaced by density $\rho$.  Fig. \ref{fig:ca40-ad} shows  $U_{3N}(R)$ for the $d-^{40}$Ca system, calculated using sets I and II of the contact 3N force, given by the $E1$ and $E\tau$ formats, respectively. The parameterisation of the $^{40}$Ca density has been taken from electron scattering studies using the two-parameter Fermi model of \cite{deJ74}. One can see that the  $d$-$^{40}$Ca potentials originating from the contact 3N force differ by about 20$\%$, which is explained by a similar difference in the corresponding  volume integrals of $W^{\rm eff}_{di}$. However, for  $E1$ the $U_{3N}$ potential is wider. Once again, we see an important contribution (about 50$\%$) comes from the deuteron $d$-state, which is demonstrated in Fig. \ref{fig:ca40-ad} by plotting the calculations retaining the deuteron $s$-wave state only. 
 We did not show the contribution from  the 3N force in the $E{\cal P}$ format (set IV) but  it should be the same as the $E1$ contribution  scaled down by a factor of 2/3 and corrected for the corresponding value of the low-energy constant $c_E$. The $E1$ format also give rise to
the quadrupole part of the $d$-$^{40}$Ca potential but this component is small.

\section{Contribution from  one-pion-contact exchange 3N force}

Following \cite{Lyn16} we will consider the 1$\pi$-c force in the $n$-$p$-$i$ system defined as
\beq
V_D &=& \frac{g_A c_D m_{\pi}^2}{96 \pi \Lambda_{\kappa}F^4_{\pi} }
\sum_{cyc} \ve{\tau}_i\cdot \ve{\tau}_k \, \left[ \delta_{R_{3N}}(\ve{r}_{ij}) +\delta_{R_{3N}}(\ve{r}_{kj}) \right]
\eol &\times&
\left[ X_{ik} (\ve{r}_{ik})- \frac{4\pi}{m^2_{\pi}}\ve{\sigma}_i\cdot \ve{\sigma}_k \delta_{R_{3N}}(\ve{r}_{ik})
\right]\ ,
\eeqn{3NF2pi}
where $g_A = 1.267$, the pion mass $m_{\pi} = 138.03$ MeV/$c^2$,
\beq
X_{ik} (\ve{r}) = [S_{ik} (\ve{r}) T(r) +  \ve{\sigma}_i\cdot \ve{\sigma}_k ]Y(r)
\eeqn{Xdef}
is the coordinate-space pion propagator that contains the tensor operator
\beq
S_{ik} (\ve{r}) = 3(\ve{\sigma}_i \cdot \hat{\ve{r}}) (\ve{\sigma}_k \cdot \hat{\ve{r}} ) -  \ve{\sigma}_i\cdot \ve{\sigma}_k,
\eeqn{Sdef}
and the tensor Yukawa functions, $T$ and $Y$, are defined as $T(r) = 1 + 3/(m_{\pi} r) + 3/(m_{\pi}r)^2$ and $Y(r) = (1 - e^{-(r/R_{3N})^4})e^{-m_{\pi} r}/r$, where the latter contains the long-range regulator. As in the case of the contact interaction, the cyclic sum runs over all cyclic permutations of the $n$-$p$-$i$ system. We should note that one more format of the 1$\pi$-c force has  been considered in \cite{Lyn16} but the best fit of the data returned a value of $c_D=0$ for that force.

\begin{figure}[b]
\includegraphics[width=0.45\textwidth]{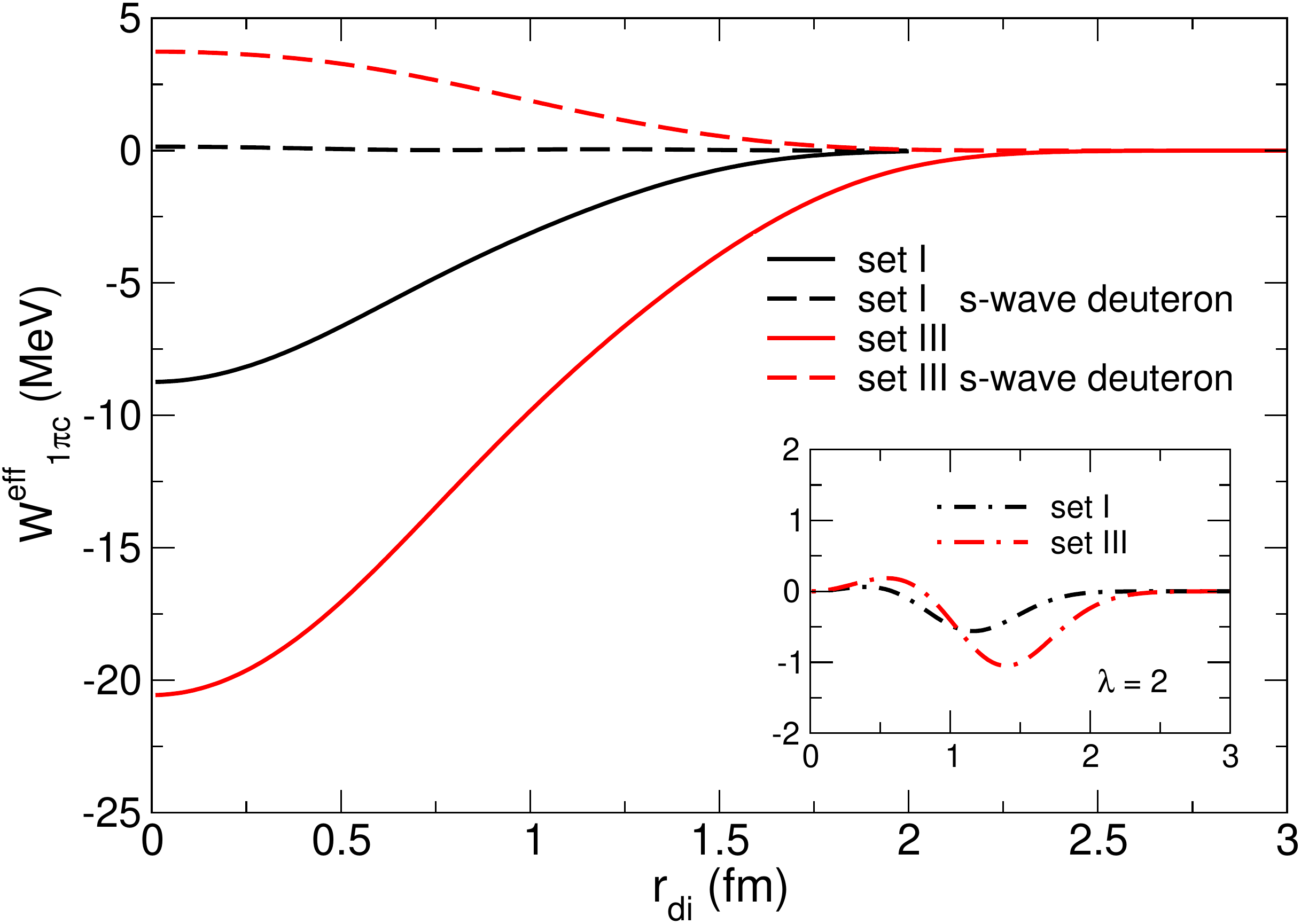}
\caption{
The $\lambda=0$ component of the adiabatic effective $d$-$i$ potential $W^{\rm eff}_{1\pi c}$   arising from the 1$\pi$-c force, obtained with sets I and III. Calculations with and without the deuteron $d$-state are shown by solid and dashed lines, respectively. The inset represent the quadrupole part of this force multiplied by $\sqrt{4\pi}$.
}
\label{fig:Weff-1pic}
\end{figure}

We will need to act with the operator $V_D$ on the product of isospin function $\chi_i$ of a nucleon $i$ in $A$ and the deuteron isospin function $\chi_d$ and, as in the case of the contact interaction, to evaluate the matrix element
 (\ref{MEtau}). The contributions from $\ve{\tau}_n\cdot \ve{\tau}_i$ and $\ve{\tau}_p\cdot \ve{\tau}_i$ to this matrix element in the ADWA are zero. Therefore, this selects in Eq. (\ref{3NF2pi}) the $\ve{\tau}_n\cdot \ve{\tau}_p$ and $\ve{\sigma}_n\cdot \ve{\sigma}_p$ operators only. The latter, acting on the deuteron spin-function $\chi_1$ with the deuteron spin $S_d = 1$, gives
\beq
 (\ve{\sigma}_n\cdot \ve{\sigma}_p ) \chi_1 = 2(\hat{S}_d^2 - \hat{s}_1^2 - \hat{s}_2^2) \chi_1 =   \chi_1.
 \eeqn{chichi}
The corresponding effective interaction between nucleon $i$ and the deuteron is then
\beq
W^{\rm eff}_{1\pi c}(\ve{r}_{di}) &=& -
\frac{g_A c_D m_{\pi}^2}{32\pi \Lambda_{\kappa}F^4_{\pi} } \la \phi_1 (\ve{r})  |
    [ X_{np} (\ve{r}) 
     - \frac{4\pi}{m^2_{\pi}}   \delta_{R_{3N}}(\ve{r})  ]
   \eol &\times&
\left[ \delta_{R_{3N}}(\ve{r}_{ni}) +\delta_{R_{3N}}(\ve{r}_{pi}) \right] | \phi_0(\ve{r}) \ra.
\eeqn{MEdef}
To evaluate the matrix element we will need
\beq
\la \left[ Y_{l'}(\hat{r}) \otimes \chi_1\right]^{J'_d}_{M'_d} |(\ve{\sigma}_n \cdot\ve{\hat{r}}) (\ve{\sigma}_p \cdot\ve{\hat{r}} )Y_{\lambda \mu} (\hat{\ve{r}})| \left[ Y_l(\hat{r}) \otimes \chi_1\right]^{J_d}_{M_d}\ra
\eol
= (4\pi)^{-1/2}(\lambda \mu J_dM_d | J'_d M'_d) A_{\lambda ll'},
\,\,\,\,\,\,\,\,\,\,\,\,\,\,\,\,\,\,\,\,\,\,\,\,\,\,\,\,
\eeqn{MEXY}
where
\beq
A_{\lambda ll'}
 = 
18 \hat{\lambda}\hat{l} \hat{J}_d \sum_{\lambda'  \lambda''}  
 (1 0 1 0 | \lambda' 0 ) (\lambda' 0 l 0 | \lambda'' 0 ) (\lambda'' 0 \lambda 0 |l' 0 ) \,\,\,\,\,\,\,\,\,\,\,\,\,\,
 \eol
 \times \hat{\lambda}'  \hat{\lambda}''
   W( \lambda'' \lambda' J_d 1; l 1) W(\lambda \lambda'' J'_d 1; l'J_d)
\left\{ \begin{array}{lll} {\frac{1}{2}} &{ \frac{1}{2} }&{1}\\
{\frac{1}{2} }&{\frac{1}{2}}&{ 1 }
\\{ 1}&{1}&{\lambda'}\end{array}\right\}. \,\,\,\,\,\,\,\,\,\,
 \eeqn{coefA}
 We will also need
 \beq
\la \left[ Y_{l'}(\hat{r}) \otimes \chi_1\right]^{J'_d}_{M'_d} |(\ve{\sigma}_n \cdot\ve{\sigma}_p  )Y_{\lambda \mu} (\hat{\ve{r}})| \left[ Y_l(\hat{r}) \otimes \chi_1\right]^{J_d}_{M_d}\ra
\eol
= (4\pi)^{-1/2}(\lambda \mu J_dM_d | J'_d M'_d) B_{\lambda ll'}
\,\,\,\,\,\,\,\,\,\,\,\,\,\,\,\,\,\,\,\,\,\,\,\,\,\,\,\,
\eeqn{MEss}
with
\beq
B_{\lambda ll'} =    \hat{\lambda} \hat{l}\hat{J}_d(l 0 \lambda 0 | l' 0) W(\lambda lJ'_d1; l'J_d).
\eeqn{coefB}
Assuming the same partial wave decomposition of $W^{\rm eff}_{1\pi c}$ as given by Eq. (\ref{WeffE1lam}) for  $W^{\rm eff}_{E1}$ and using (\ref{MEXY}) and (\ref{MEss}) in Eq. (\ref{MEdef}) we obtain
\beq
W^{\rm eff}_{1\pi c,\lambda}(r_{di}) = &-& 
\frac{    g_A c_D m_{\pi}^2}{16\pi  \Lambda_{\kappa}F^4_{\pi} } \sum_{ll'}
\int_0^{\infty} dr   \,   v_{l'}(r)u_l(r) 
\eol
&\times&
\delta_{\lambda}\left(r_{di}, \frac{r}{2}\right)f_{\lambda ll'}(r),  
\eeqn{Weff1pilam}
where
\beq
f_{\lambda ll'}(r)& =
&     3 A_{\lambda ll'} T(r)Y(r) 
     \eol
     &-& B_{\lambda ll'}\left((T(r)-1)Y(r) +   \frac{4\pi}{m_{\pi}^2}  \delta_{R_{3N}}(r) \right).  
  \eol
\eeqn{flll}

Figure \ref{fig:Weff-1pic} shows the $W^{\rm eff}_{1\pi c}$ potential for two 3N interactions, given by set I and set III from Table I. The contribution from the deuteron $s$-state only is also shown.  With the $s$-state only, the matrix element associated with the tensor operator $S_{np}$ is zero so that $W^{\rm eff}_{1\pi c}$ depends only on $(\ve{\sigma}_n \cdot \ve{\sigma}_p )\left( Y(r)- \frac{4\pi}{m^2_{\pi}} \delta_{3N}(\ve{r})\right) $. 
This contribution is positive and it does not exceed 5 MeV at zero $d$-$i$ separation. Thus, the main contribution to $W^{\rm eff}_{1\pi c}$ comes from the $d$-state and is determined by the $1\pi$-c  part associated with the tensor operator $S_{np}$. The
$\lambda=0$ monopole part of $W^{\rm eff}_{1\pi}$ dominates.
Contrary to the case of the contact interaction, the $W^{\rm eff}_{1\pi c}$ is attractive.

\section{Contribution from two-pion-exchange 3N force.}

The 2$\pi$-exchange force $V_C$ has  contributions from three terms, $V_{C,c1}$, $V_{C,c3}$ and  $V_{C,c4}$  but the latter will not contribute to the $d$-$A$ potential because of isospin arguments similar to those explained in the sections above.
The potentials  $V_{C,c1}$  and $V_{C,c3}$  are given in Ref. \cite{Tew16} by
 \beq
 V_{C,{c_1}} &=& \frac{c_1m_{\pi}^4 g_A^2}{2f^4_{\pi}(4\pi)^2} \sum_{cyc} (\ve{\tau}_i \cdot \ve{\tau}_k) (\ve{\sigma}_i \cdot \hat{\ve{r}}_{ij}) 
 (\ve{\sigma}_k \cdot \hat{\ve{r}}_{kj}) 
 \eol
 &\times&
 U(r_{ij})Y(r_{ij})U(r_{kj})Y(r_{kj}) \,\,\,\,\,\,\,\,\,\,
 \eeqn{VCc1}
 and
   \beq
  V_{C,c_3} &=& \frac{c_3 g_A^2}{36f^4_{\pi}} \sum_{ijk} (\tau_i \cdot \tau_k) \left[ \frac{m_{\pi}^4}{(4\pi)^2} X_{ij}(\ve{r}_{ij}) X_{kj}(\ve{r}_{kj}) \right.
  \eol &-& 
   \frac{m_{\pi}^2}{4\pi} X_{ik}(\ve{r}_{ij}) \delta_{R_{3N}}(\ve{r}_{kj})
 - \frac{m_{\pi}^2}{4\pi} X_{ik}(\ve{r}_{kj}) \delta_{R_{3N}}(\ve{r}_{ij})
  \eol  
 \eol
  &+&  \left.
\sigma_i \cdot \sigma_k \,\delta_{R_{3N}}(\ve{r}_{ij}) \delta_{R_{3N}}(\ve{r}_{kj}) \right],
  \eeqn{force_c3}
  where  $c_1 = -0.81$ MeV$^{-4}$ and $c_3 = -3.4$ MeV$^{-4}$ are the low-energy constants \cite{Epe05},   $U(r) = 1+1/(m_\pi r)$ and $\delta_{R_{3N}}$ and $X$ are defined in the sections above. As in the previous sections, only $(\tau_n \cdot \tau_p)$ will contribute to $V_{C,c1}$ and $V_{C,c3}$ for isospin reasons. 
 To obtain expressions of the corresponding contributions to   $W^{\rm eff}_{di}$  we use
 \beq
 (\ve{\sigma}_n \cdot \hat{\ve{r}}_{ni}) 
 (\ve{\sigma}_p \cdot \hat{\ve{r}}_{pi}) = \frac{1}{r_{ni}r_{pi}} \sum_{\lambda=0,2;\mu}(-)^{\lambda+\mu} [\sigma_n \times \sigma_p]^{\lambda}_{-\mu} 
 \eol \times
 \left[ [\ve{r}_{di} \times \ve{r}_{di}]^{\lambda}_{\mu}  + \frac{1-(-)^{\lambda}}{2}  [\ve{r}_{di} \times \ve{r}]^{\lambda}_{\mu}- \frac{1}{4}  [\ve{r} \times \ve{r}]^{\lambda}_{\mu} \right] \,\,\,\,\,\,\,\,\,\,\,\,
 \eeqn{sr1sr2}
and the fact that an arbitrary function $H^{(i)}$ that depends on $r_{ni}$ and $r_{pi}$ only  can be expanded as  
 \beq
H^{(i)}(r_{ni},r_{pi})
 = 4\pi \sum_{\lambda' \mu'} {\cal H}_{\lambda'}^{(i)}\left(r_{di},\frac{r}{2}\right) Y^*_{\lambda' \mu' }(\hat{\ve{r}}_{di})
 Y_{\lambda' \mu'}(\hat{\ve{r}}), \eol
 \eeqn{f1f2}
 where
 \beq
{\cal H}^{(i)}_{\lambda}\left(r_{di},\frac{r}{2}\right) =
 \frac{1}{2} \int_{-1}^1 d\mu P_{\lambda}(\mu) H^{(i)}(x_-,x_+) ,    
 \eeqn{flambda} 
 with $x_{\pm} = \sqrt{r^2_{di} \pm r_{di}r \mu +  \frac{r^2}{4}}$. Then, performing standard Racah algebra we obtain:
 \beq
W^{\rm eff}_{2\pi,\lambda}(r_{di}) &=&  \sum_{ll'}
\int_0^{\infty} dr   \,   v_{l'}(r)u_l(r) 
\eol &\times&
\left[ {\cal F}_{\lambda ll'}^{(c1)} \left(r_{di},\frac{r}{2} \right)+ {\cal F}_{\lambda ll'}^{(c3)} \left(r_{di},\frac{r}{2} \right)\right], \,\,\,\,\,\,\,\,\,\,\,\,\,
\eeqn{Weff2pilam}
where  
 \beq
  {\cal F}_{\lambda ll'}^{(c1)} = \sum_{\lambda'} {\cal C}^{(1)}_{ll'\lambda\lambda'}
  \left[ r_{di}^2{\cal H}^{(0)}_{\lambda'}-
 \left( \frac{r}{2}\right)^2{\cal H}^{(0)}_{\lambda}\right], \,\,\,\,\,\,\,\,\,\,\,\,\,\,
 \,\,\,\,\,\,\,\,\,\,\,\,\,\,
 \,\,\,\,\,\,\,\,\,\,\,\,\,\,
  \eeqn{Fc1}
   \beq
  {\cal F}_{\lambda ll'}^{(c3)}  &=&   \sum_{\lambda'} \left[
  {\cal C}^{(1)}_{ll'\lambda\lambda'}
  \left( r_{di}^2{\cal H}^{(1)}_{\lambda'} + r^2 {\cal H}^{(2)}_{\lambda}\right) - r r_{di} {\cal C}^{(2)}_{ll'\lambda\lambda'}
{\cal H}^{(3)}_{\lambda'}\right]
  \eol &+& \hat{\lambda} \hat{l} \hat{J}_d (l 0 \lambda 0 | l' 0) W(\lambda l J'_d 1; l' J_d) {\cal H}^{(4)}_{\lambda}.
  \eeqn{Fc3}
 Here ${\cal F}^{(c1,c3)}_{\lambda ll'}$ as well as all ${\cal H}^{(i)}_{\lambda}$ are functions of  $\left(r_{di},\frac{r}{2} \right)$.  The ${\cal H}^{(i)}_{\lambda}$  are obtained from Eq. (\ref{flambda}) using functions $H^{(i)}(x_-,x_+)$ whose expressions can be found in the Appendix. The other quantities in Eqs. (\ref{Fc1}) and (\ref{Fc3}) are the coefficients ${\cal C}^{(1,2)}_{ll'\lambda \lambda'}$ given by
  \beq
  {\cal C}^{(1)}_{ll'\lambda \lambda'} &=&
  18 \hat{l} \hat{J}_d \hat{{\lambda}'}^2 ( l 0 \lambda' 0 | l' 0)  \sum_{\lambda''=0,2 }   \hat{\lambda}''(1 0 1 0 | \lambda'' 0)  
 \eol &\times&
 (\lambda'' 0 \lambda' 0 | \lambda 0)\left\{ \begin{array}{lll} {\dem} & {\dem} &{1}\\{ \dem}&{\dem}&{1} \\
{1 }&{1}&{ \lambda'' }
\end{array}\right\}
 \left\{ \begin{array}{lll} {l} & {l'} &{\lambda'}\\{ 1}&{1}&{\lambda''} \\
{J_d }&{J'_d}&{ \lambda }
\end{array}\right\}
  \eol
  \eeqn{Cc1}
  and
\beq
  {\cal C}^{(2)}_{ll'\lambda \lambda'} &=&
  18 \hat{l} \hat{J}_d \hat{{\lambda}'}^2   \sum_{L,\lambda''=0,2 }  \hat{L}  \hat{\lambda}''^2  ( l 0 L 0 | l' 0)
 \eol&\times&
 ( 1 0 \lambda' 0 | \lambda 0)
( 1 0 \lambda' 0 | L 0) W(\lambda 1 L 1 ; \lambda' \lambda'')
\eol &\times&
  \left\{ \begin{array}{lll} {\dem} & {\dem} &{1}\\{ \dem}&{\dem}&{1} \\
{1 }&{1}&{ \lambda'' }
\end{array}\right\}
 \left\{ \begin{array}{lll} {l} & {l'} &{L}\\{ 1}&{1}&{\lambda''} \\
{J_d }&{J'_d}&{ \lambda }
\end{array}\right\}.
  \eeqn{Cc2}

\begin{figure}[t]
\includegraphics[width=0.45\textwidth]{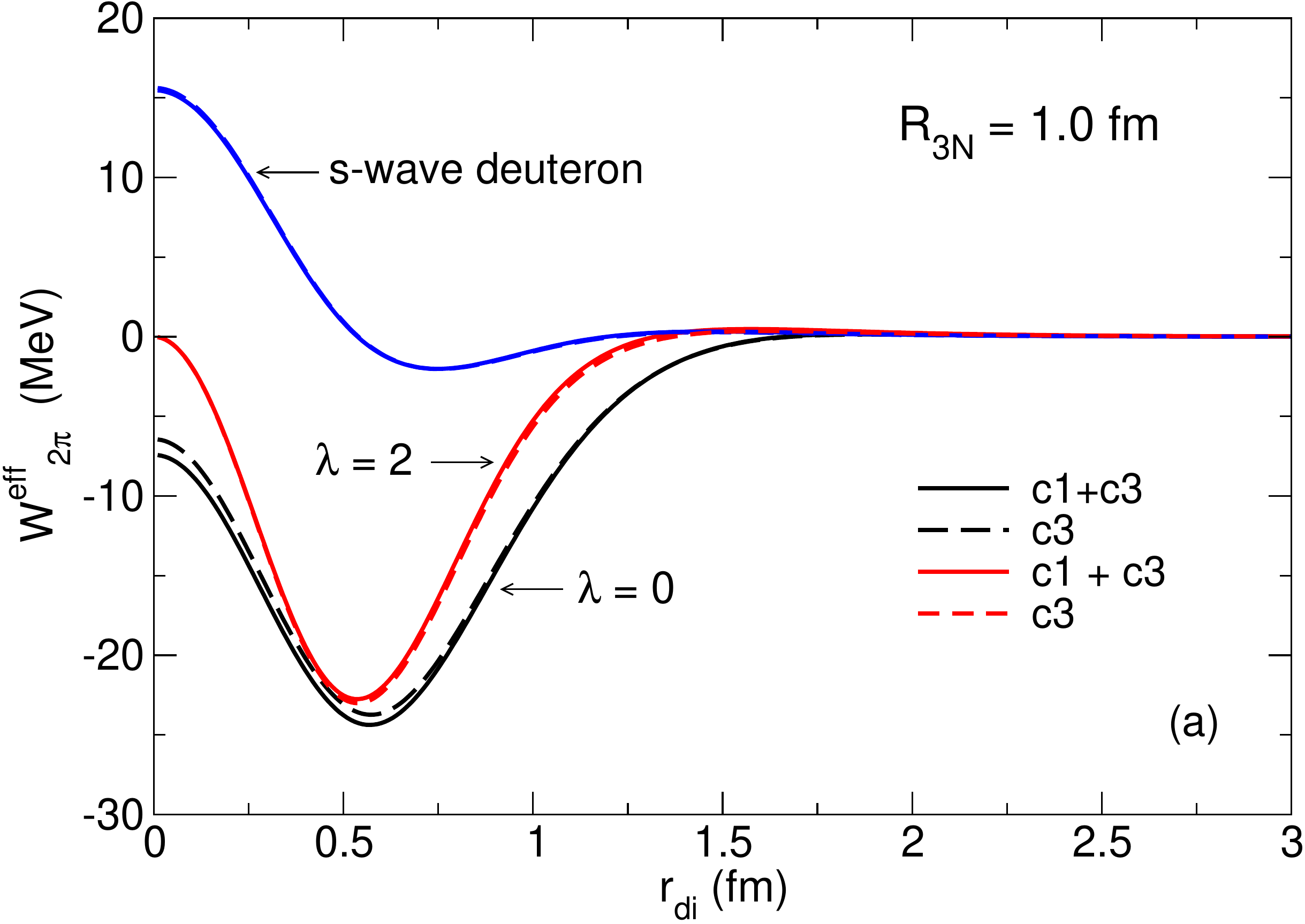}
\includegraphics[width=0.45\textwidth]{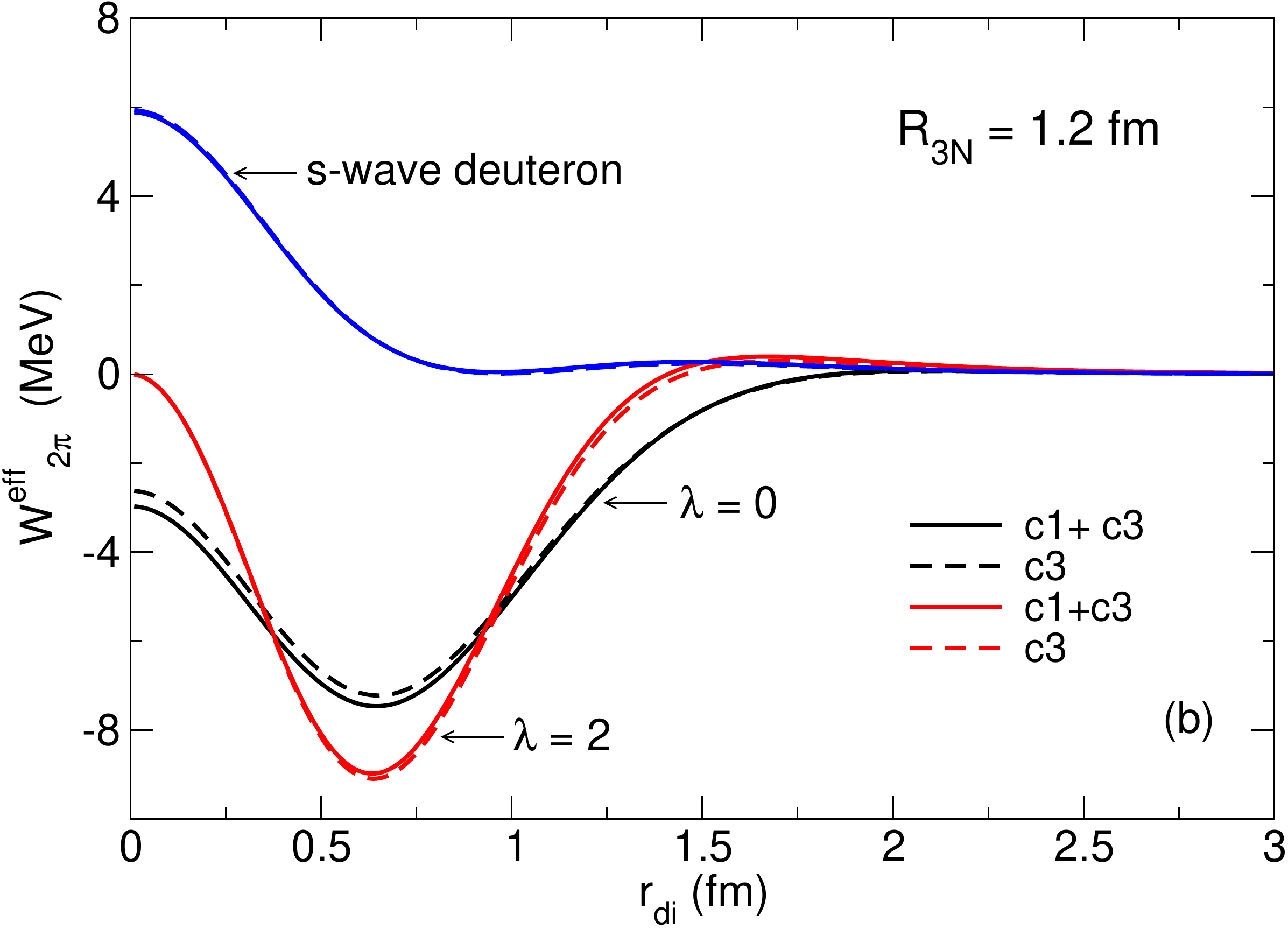}
\caption{
The monopole and quadrupole adiabatic effective $d$-$i$ potentials $W^{\rm eff}_{2\pi,\lambda}$ calculated with $V_{C,c3}$ force only (dashed lines) and with full $V_C$ potential (solid lines)  for two different regulators:  ($a$) 1.0 fm and  ($b$) 1.2 fm.   
}
\label{fig:Weff-2pi}
\end{figure}

Numerical calculations have shown that the contribution from the $V_{C,c1}$ force is negligible for both regulators used (see Fig. \ref{fig:Weff-2pi}). The contribution from $V_{C,c3}$ strongly depends on $R_{3N}$ and it is dominated by contribution from the deuteron $d$-state. Without this state the $W^{\rm eff}_{2\pi}$ is repulsive, while adding it makes the $d$-$i$ potential attractive. It should be noted that this potential has a large quadrupole part which, unlike in the case of the potential 1$\pi$-c, is plotted in  Fig. \ref{fig:Weff-2pi} without multiplication by a factor of $\sqrt{4\pi}$. However, for the spherical targets considered here its contribution to the folding $d$-$A$ potential is very small.

\begin{figure}[b]
\includegraphics[width=0.45\textwidth]{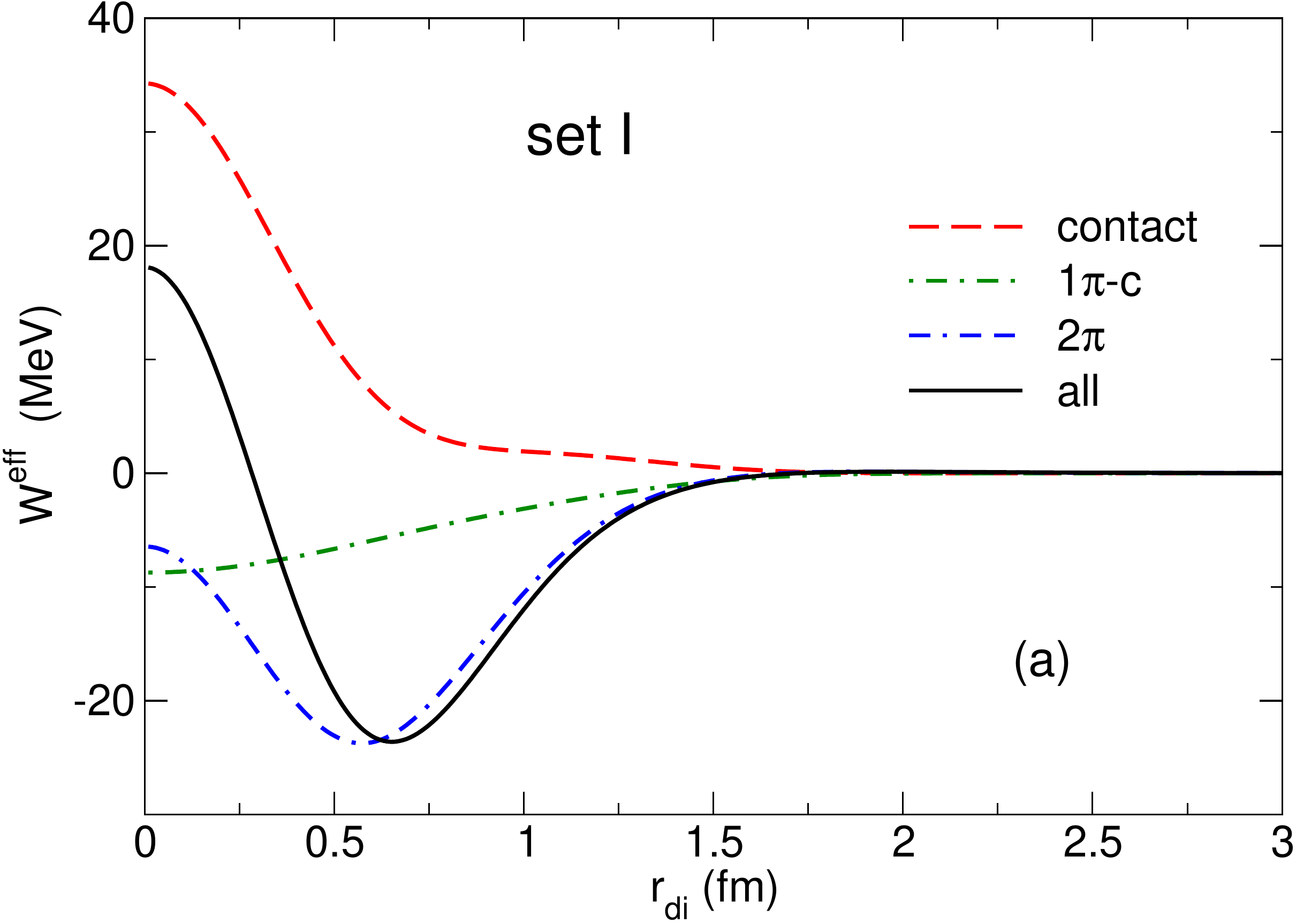}
\includegraphics[width=0.45\textwidth]{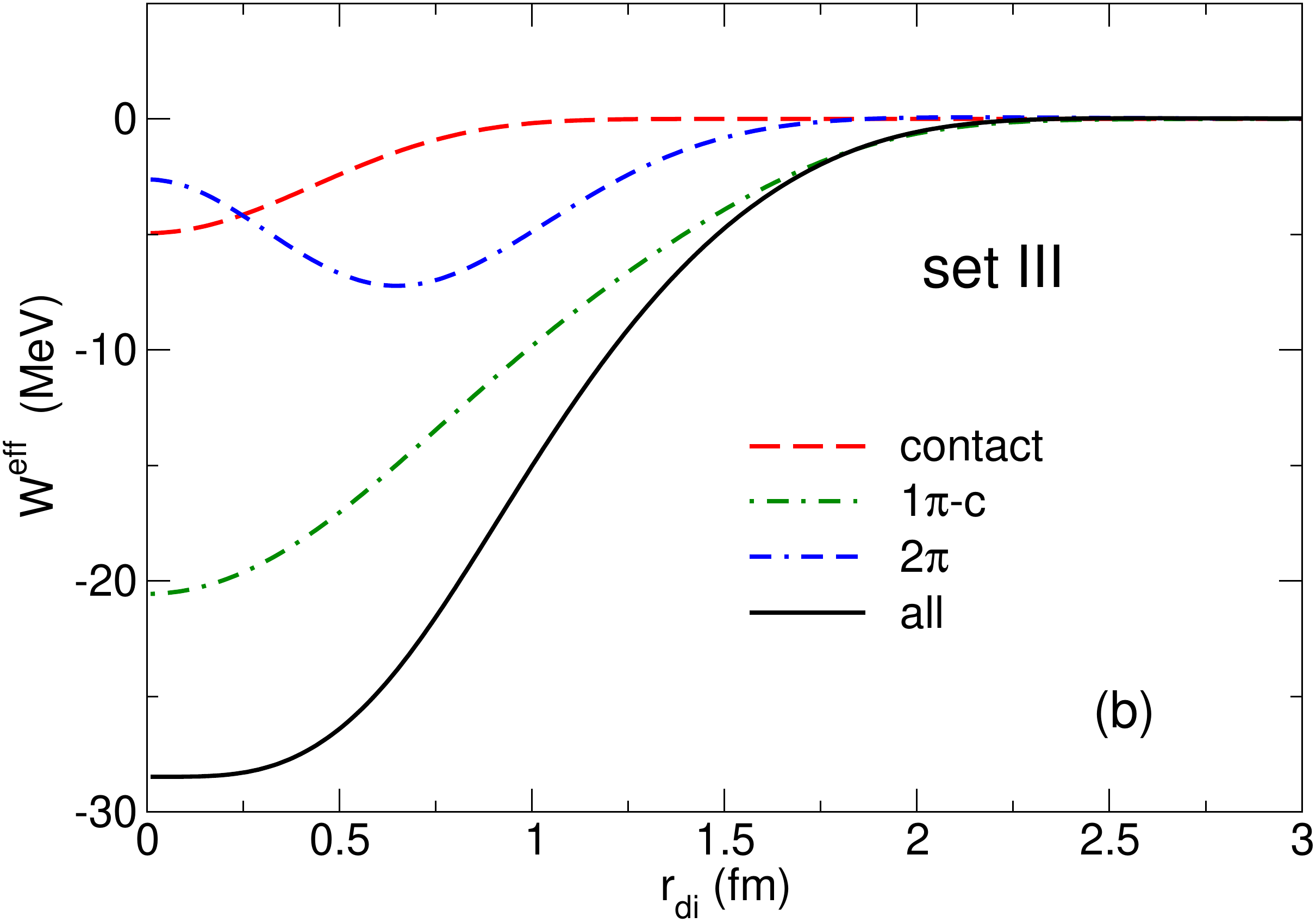}
\caption{
The individual contributions from the contact, 1$\pi$-c and 2$\pi$ parts of the 3N potential to the monopole  part of the adiabatic effective $d$-$i$ potential $W^{\rm eff}_{di}$ obtained for sets I ($a$) and III ($b$). The sums of these contributions are shown by solid lines.  
}
\label{fig:Weff-di}
\end{figure}

 The sum of the $d$-$i$ interactions $W^{\rm eff}_{di}$ obtained with contact, 1$\pi$-c and 2$\pi$ contributions is shown in Fig. \ref{fig:Weff-di} for two sets (I and III) with $c_D \neq 0$. 
The contribution from the contact interaction  is repulsive for set I, but attractive for set III due to the different signs of the low-energy constant $c_E$ in these two cases.  This has a profound effect on the total $d$-$i$ potential, $W^{\rm eff}_{di}$, which has a repulsive core for set I, but is purely attractive for set III. While in the first case all three 3N force components contribute to the $W^{\rm eff}_{di}$ shape, in the second case this shape is dominated by the 1$\pi$-c part.

\begin{figure*}[htb]
{\includegraphics[width=0.49\textwidth]{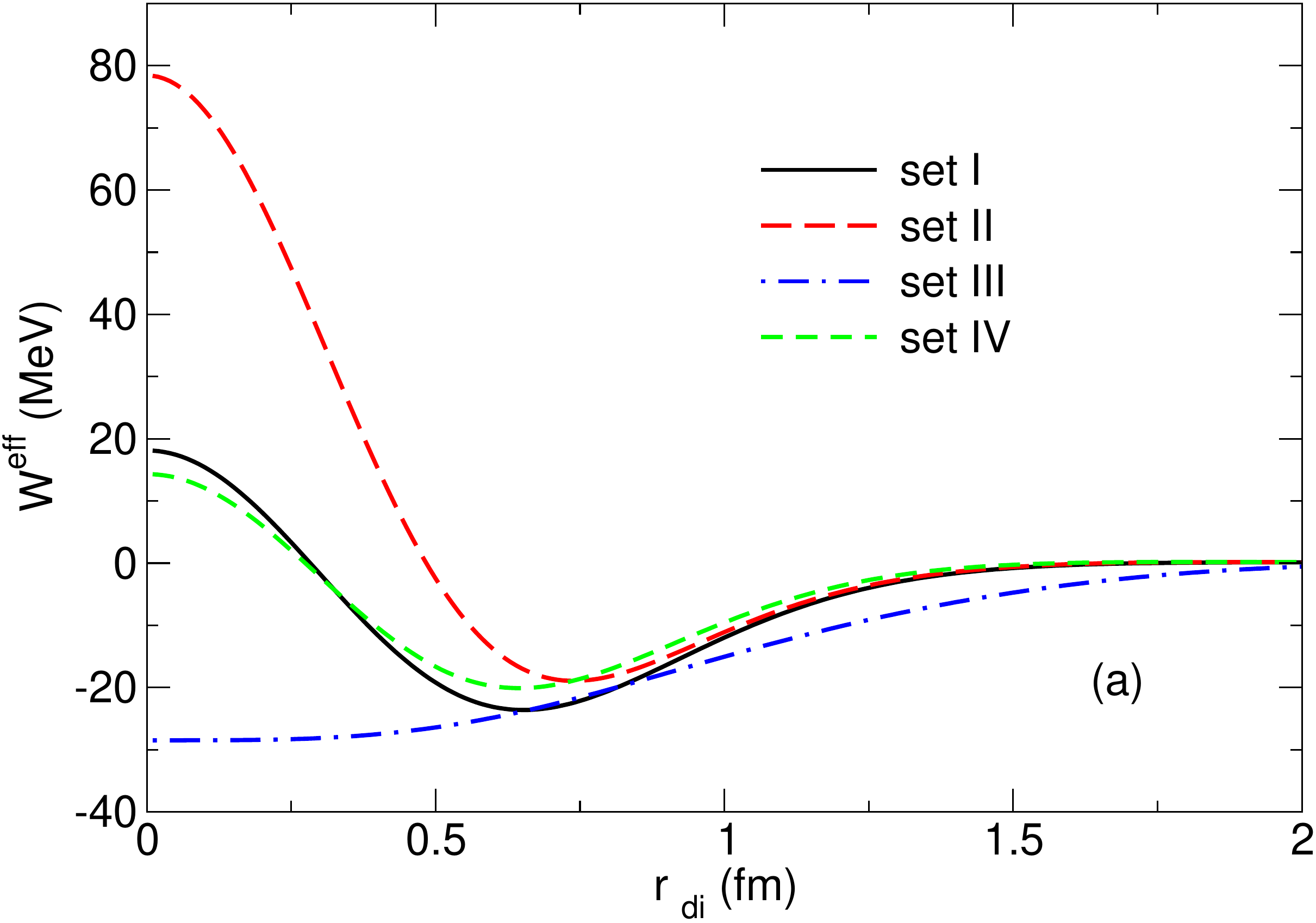}
\includegraphics[width=0.49\textwidth]{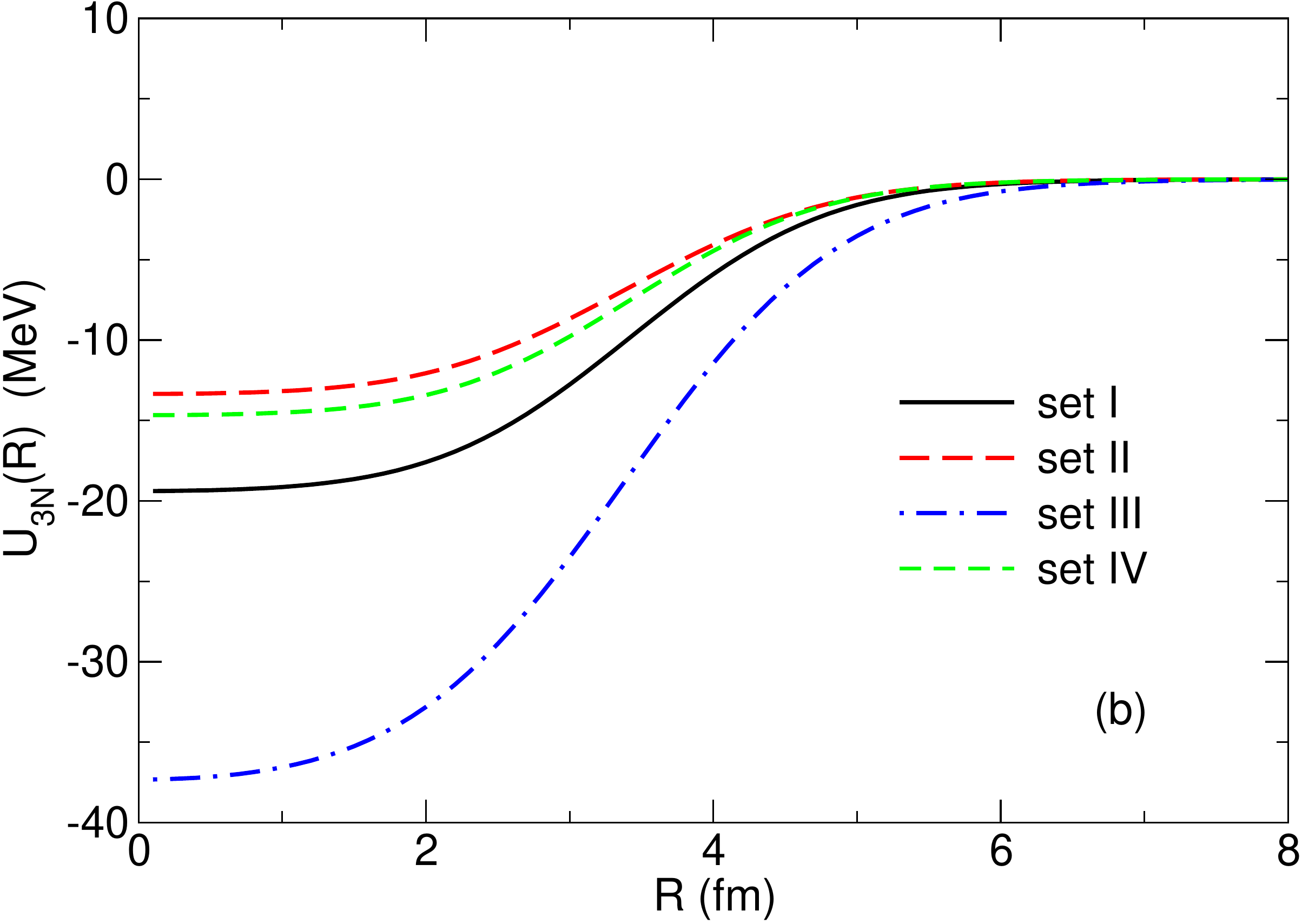}}
\caption{
The monopole part of the total effective $d$-$i$ potential $W^{\rm eff}_{di}$ ($a$) and the corresponding adiabatic potential $U_{3N}(R)$ ($b$) calculated for the $d +^{40}$Ca system using four sets of the $\chi$EFT 3N interaction.
}
\label{fig:dA-ADWA}
\end{figure*}

 The total $W^{\rm eff}_{di}$ interactions calculated with all four sets of the 3N potentials are compared to each other in Fig. \ref{fig:dA-ADWA}a. They depends strongly on the choice of 3N force. Their volume integrals and r.m.s. radii are shown in Table II. 
 Since these radii are smaller than the typical sizes of nuclei
 the shape of the $d$-$A$ potentials  $U_{3N}$ will be mainly determined by the nuclear density $\rho$. 
 Since the volume integrals of $W^{\rm eff}_{di}$ and $U_{3N}$ are related by a simple factor equal to the number of nucleons, $A$, in the target, the depths of the $U_{3N}$ potentials obtained for different 3N sets are in the same proportions as the corresponding volume integrals of $W^{\rm eff}_{di}$. This can be seen in Fig. \ref{fig:dA-ADWA}b where $U_{3N}$ are shown for the case of $d-^{40}$Ca for all four  3N sets.

\begin{table*}[htb]
\caption{Absolute values of the volume integrals (V.I., in MeV) and r.m.s. radii (in fm) of the effective $d$-$i$ interactions $W^{\rm eff}_{di}$ and the $d-^{40}$Ca folding potentials $U_{3N}$ calculated in the ADWA and Watanabe models.
}
\centering
\begin{tabular} {p{1.5 cm } p {1.5 cm} p{ 2.2 cm} p{ 1.5 cm} p{ 3.0 cm} p{1.5 cm }  p{2.2 cm } p{1.5 cm } p{1.2 cm }} 
\hline
 & \multicolumn{4}{c}{ADWA $\,\,\,\,\,\,\,\,\,\,\,\,\,\,\,\,\,\,\,\,\,\,\,\,$} & \multicolumn{4}{c}{Watanabe} \\
\hline
  &  \multicolumn{2}{c}{$W^{\rm eff}_{di} \,\,\,\,\,\,\,\,\,\,\,\,\,\,\,\,\,\,\,\,\,\,\,\,$}  & \multicolumn{2}{c}{$U_{3N}  \,\,\,\,\,\,\,\,\,\,\,\,\,\,\,\,\,\,\,\,\,\,\,\,\,\,\,\,\,\,\,\,\,\,\,\,$} & \multicolumn{2}{c}{$W^{\rm eff}_{di} \,\,\,\,\,\,\,\,\,\,\,\,\,\,\,\,\,\,\,\,\,\,\,\,\,\,\,\,\,\,\,\,$} & \multicolumn{2}{c}{$U_{3N\,\,\,\,\,\,\,\,\,\,\,\,\,}$}\\
set & V.I. &  radius & V.I. &  radius & V.I. &  radius & V.I. &  radius \\
\hline
I   & 110.8 & 0.838 & 4430 & 3.531 & 0.343 & 1.228 & 13.72 & 3. \\
II  & 76.2  & 0.875 & 3047 & 3.540 & 2.137 & 1.469 & 85.47 & 3.732 \\
III & 215.7 & 1.183 & 8626 & 3.628 & 34.46 & 1.259 & 1378. & 3.653 \\
IV  & 83.5  & 0.711 & 3341 & 3.503 & 2.957 & 0.888 & 118.3 & 3.315 \\
\hline\hline
\end{tabular}
\label{tab:vints}
\end{table*}

 \section{The ADWA calculations with the $d-A$ potentials arising from the 3N force}

\begin{figure*}[t]
{\includegraphics[width=0.378\textwidth]{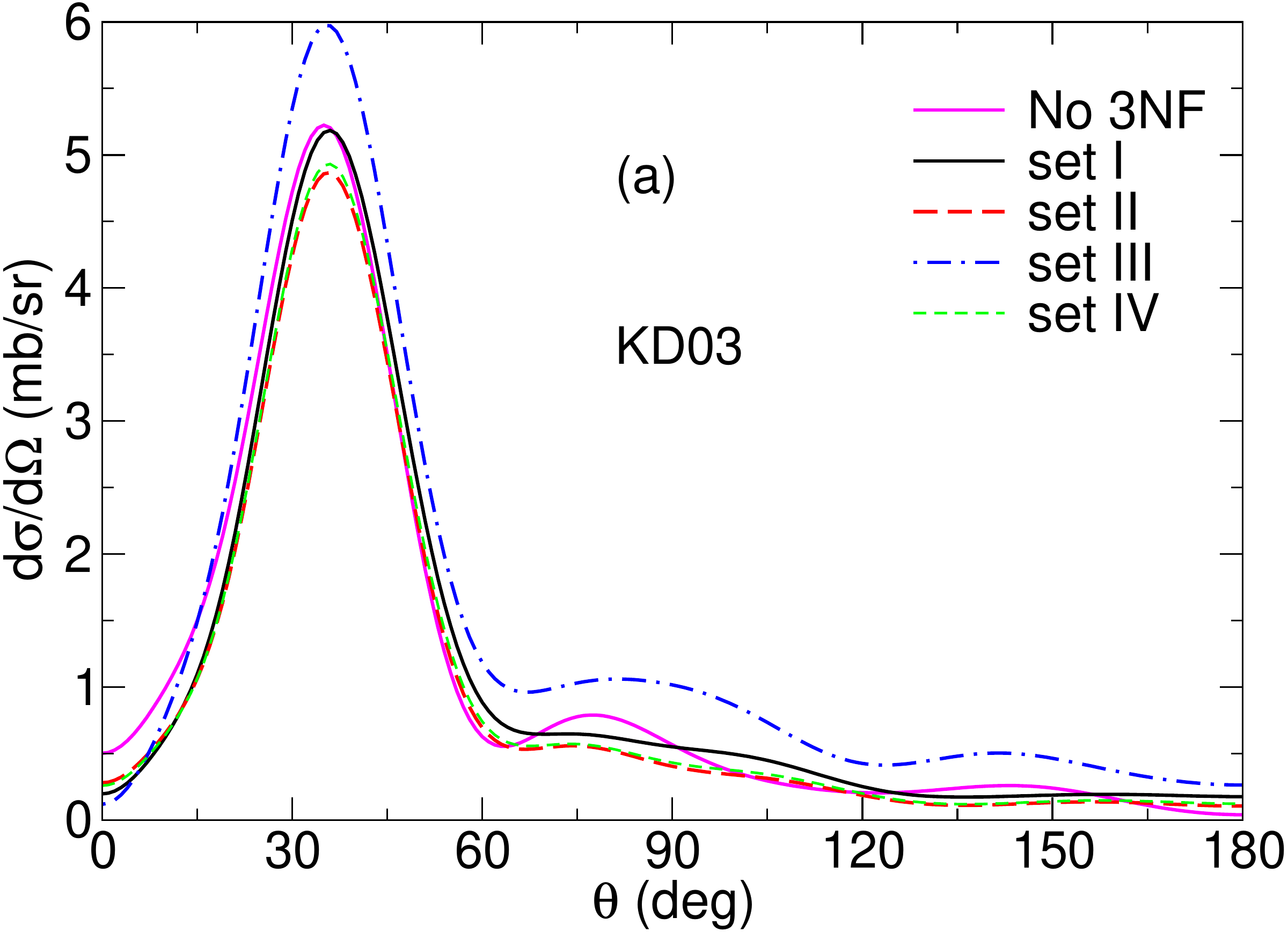}
\includegraphics[width=0.378\textwidth]{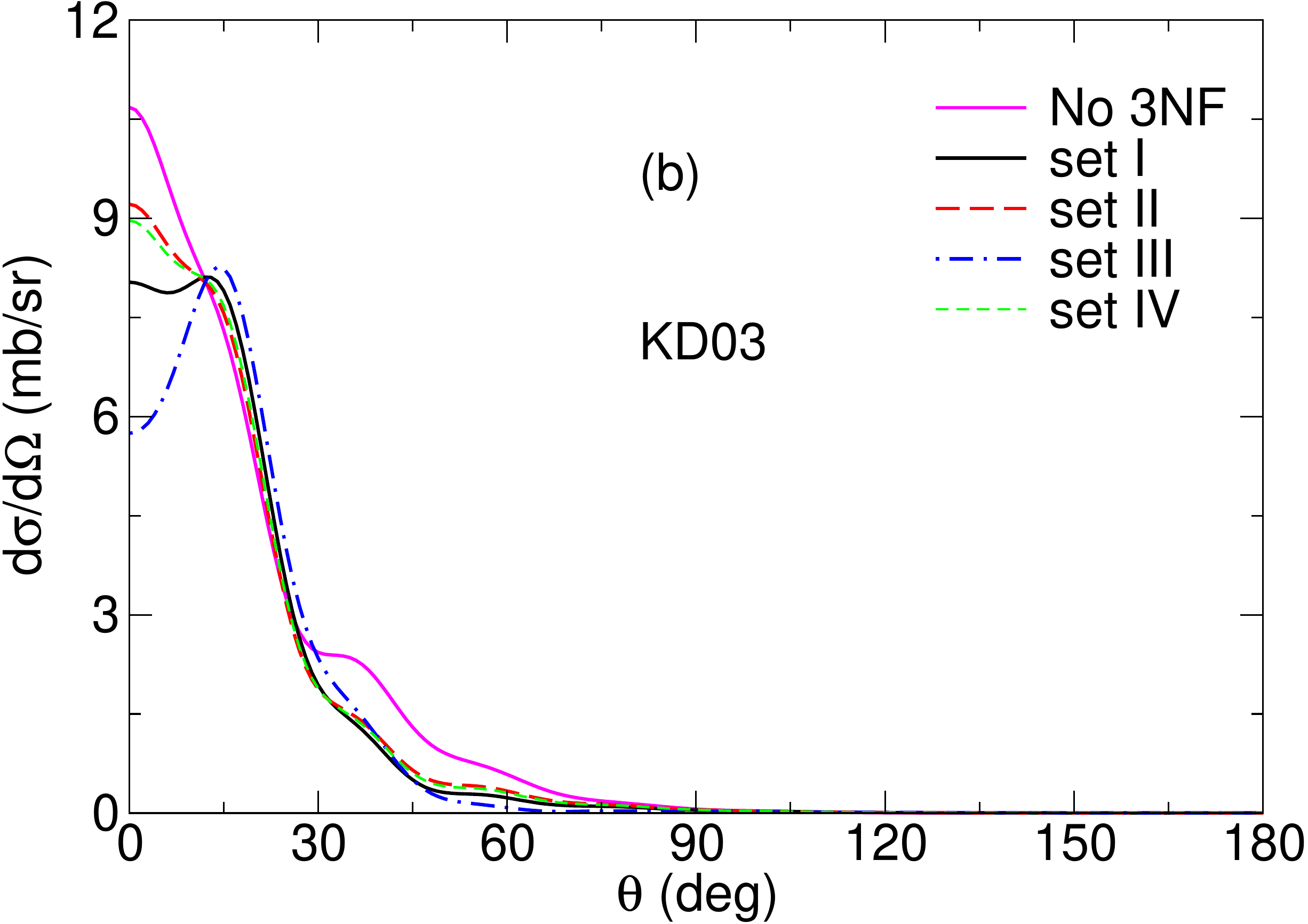}}
{\includegraphics[width=0.378\textwidth]{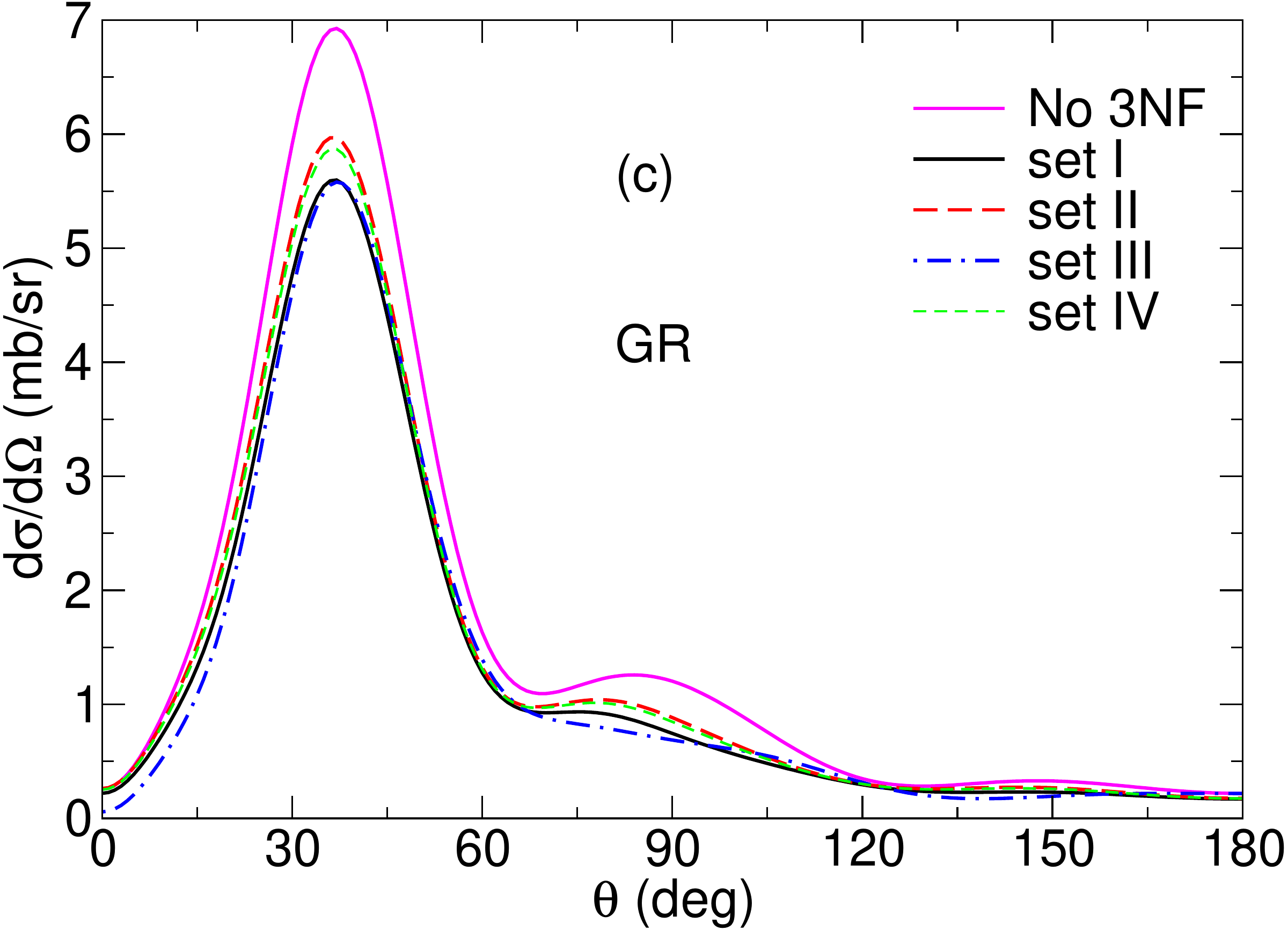}
\includegraphics[width=0.378\textwidth]{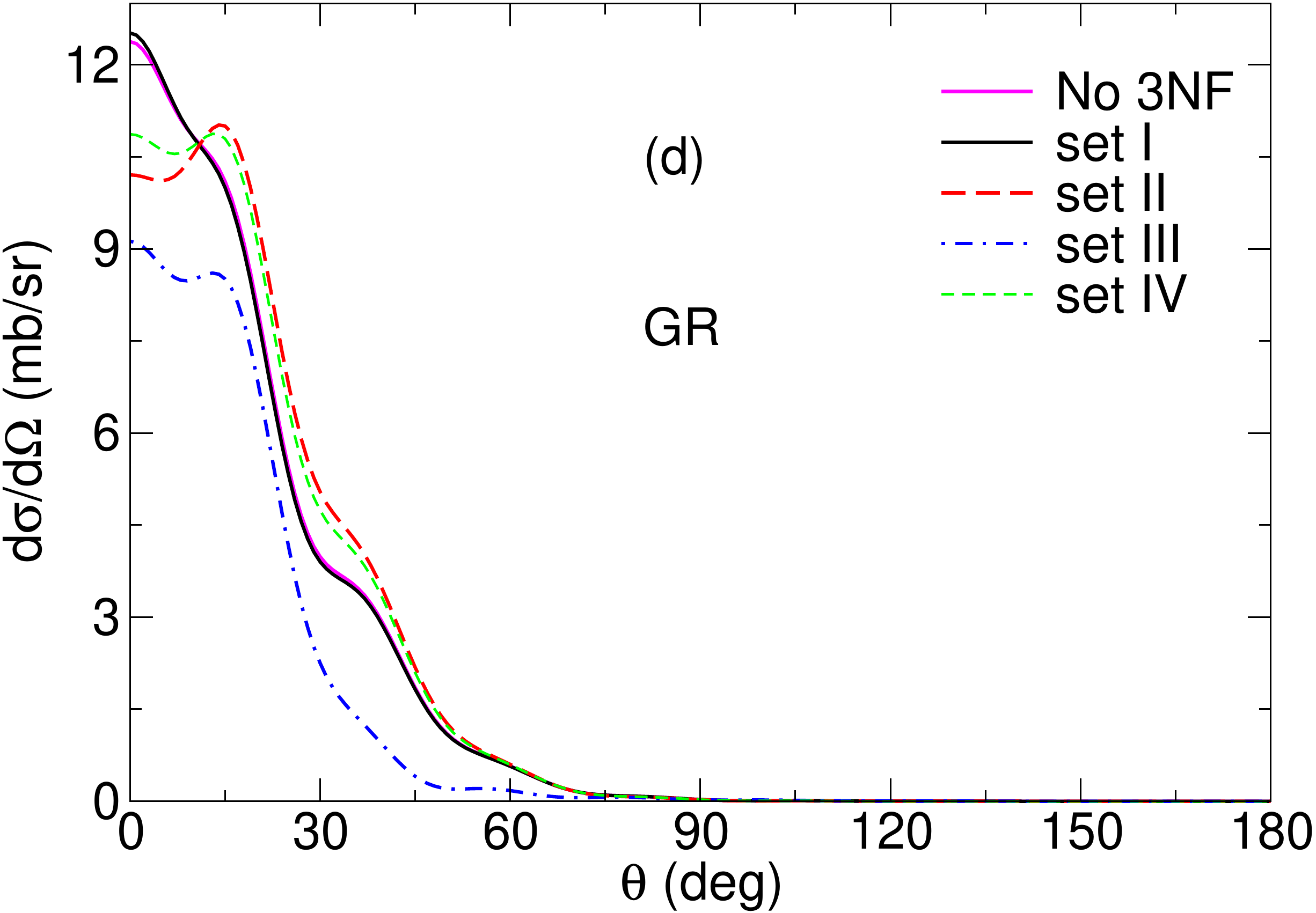}}
{\includegraphics[width=0.378\textwidth]{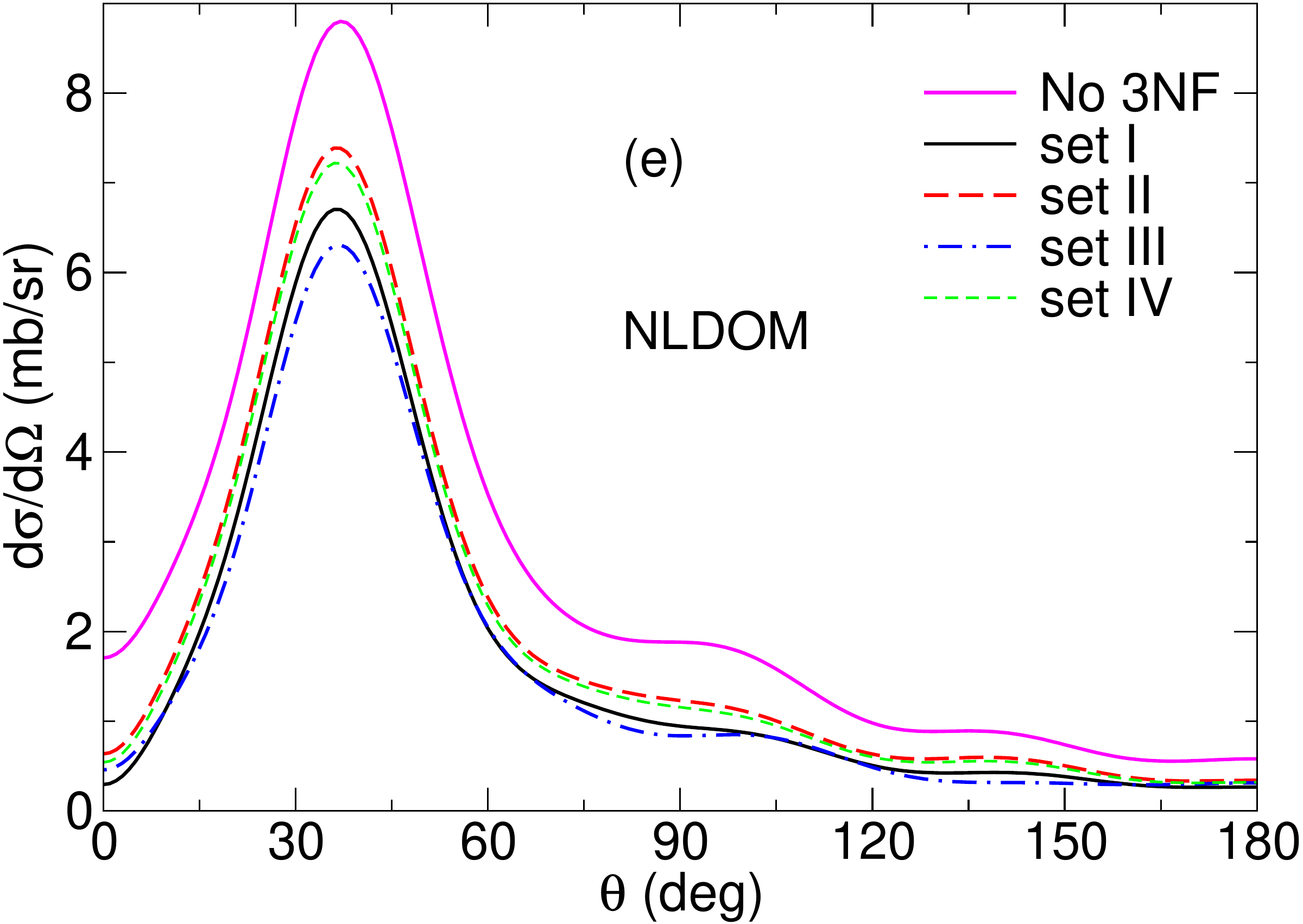}
\includegraphics[width=0.378\textwidth]{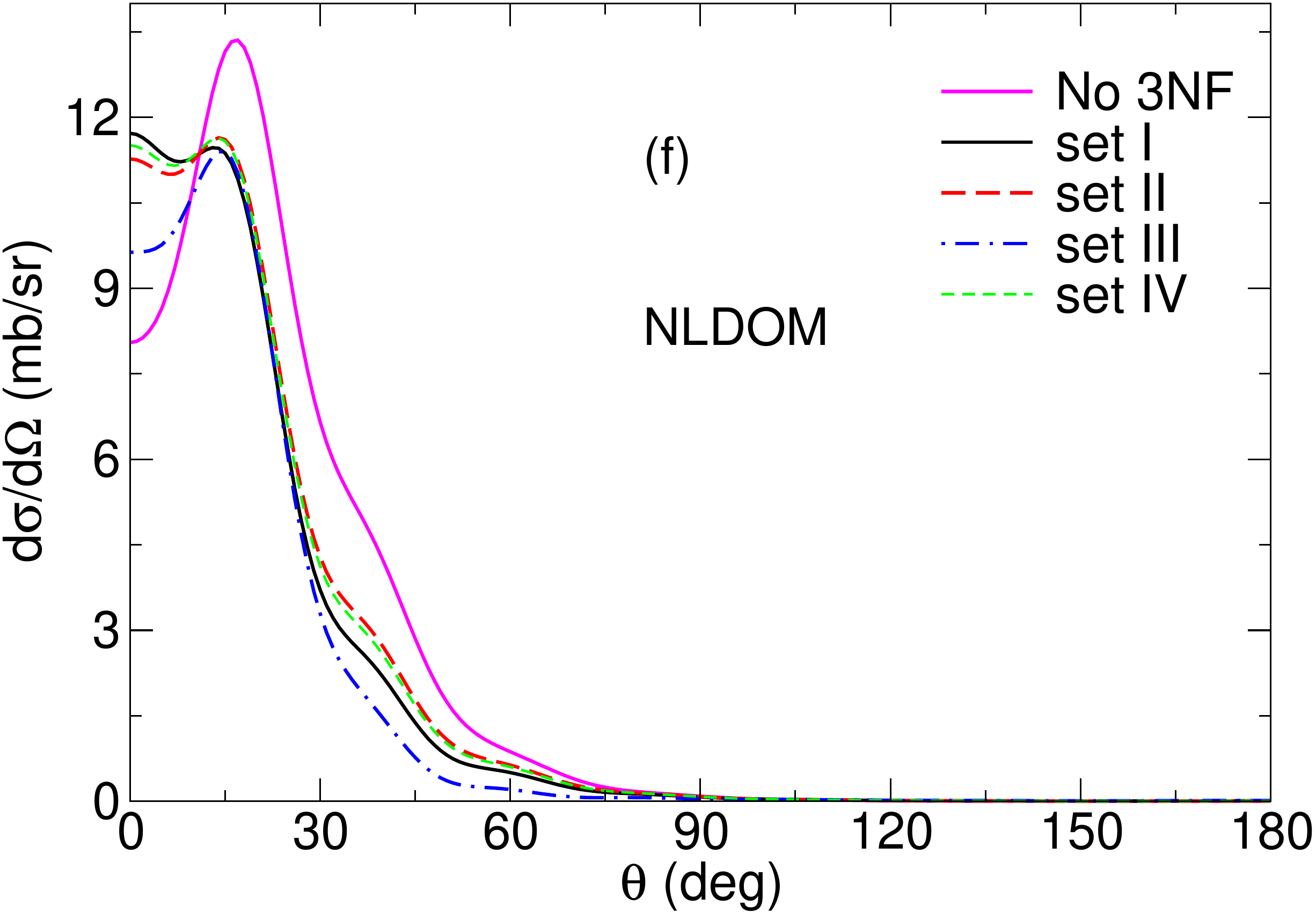}}
{\includegraphics[width=0.378\textwidth]{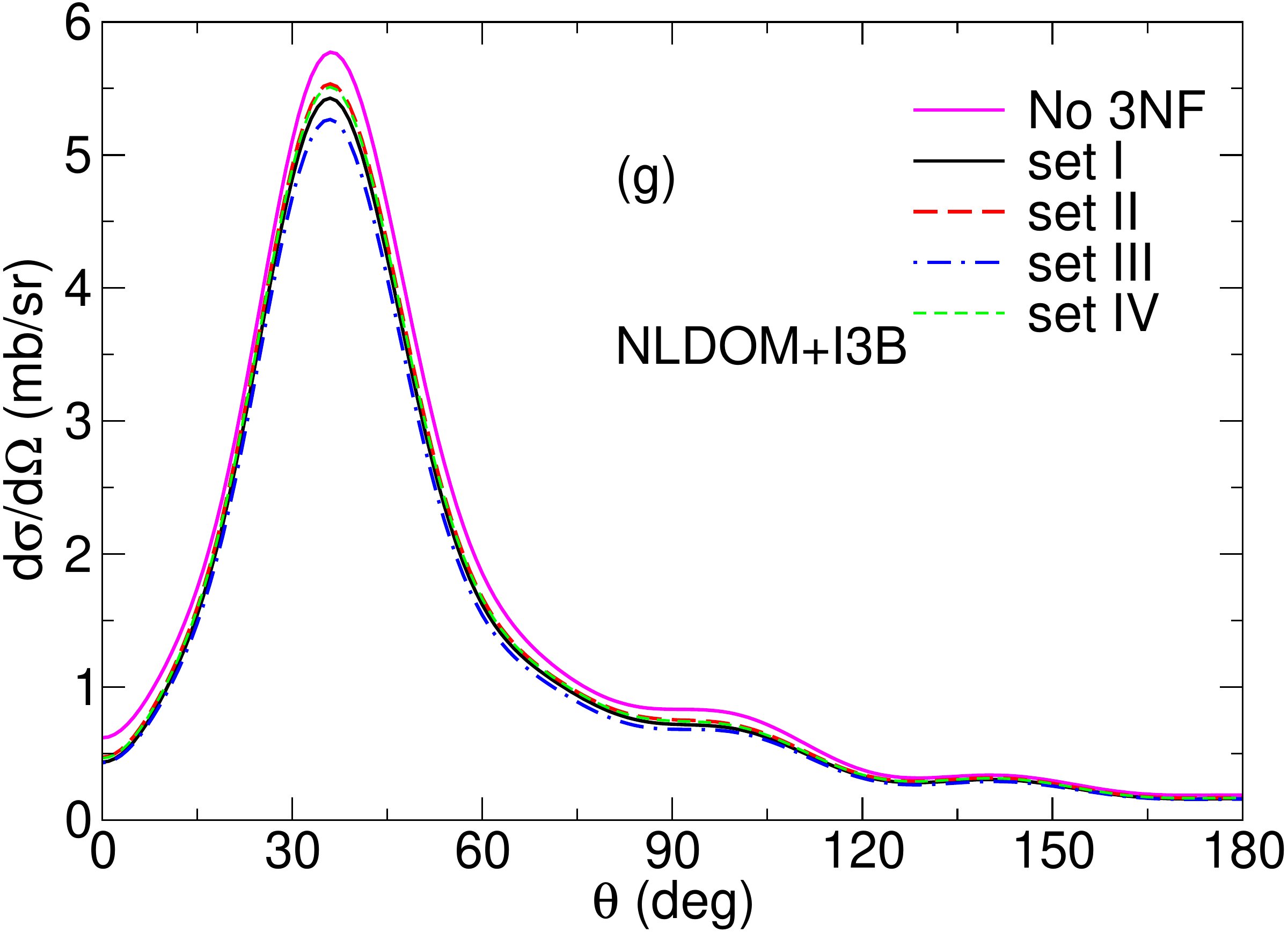}
\includegraphics[width=0.378\textwidth]{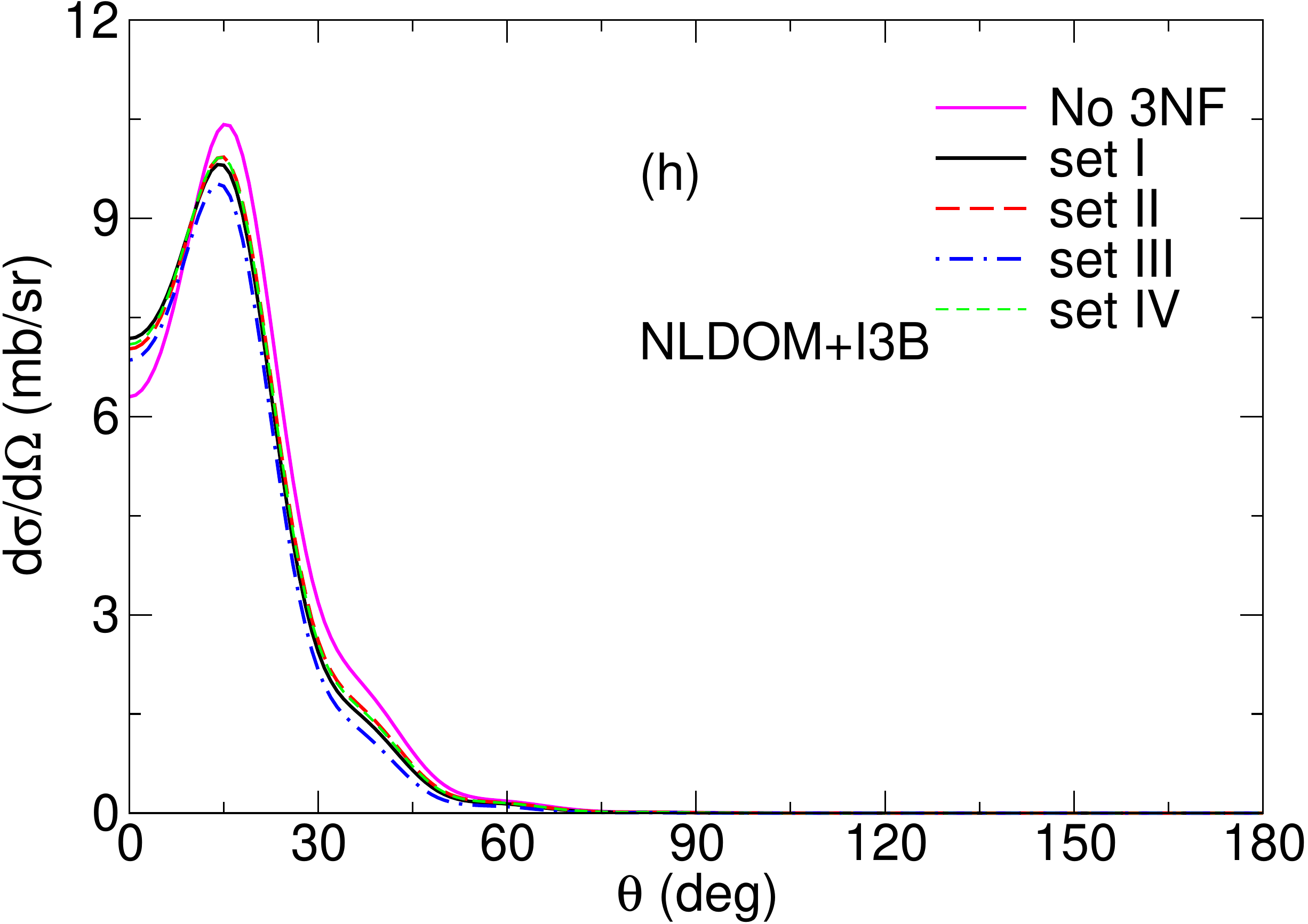}}
\caption{
The $^{40}$Ca$(d,p)^{41}$Ca reaction at $E_d = 11.8$ MeV (left column) and 56 MeV (right column) calculated with nucleon optical potentials KD03 ($a,b$), GR ($c,d$), NLDOM ($e,f$) and NLDOM with I3B effects described in the text ($g,h$). For each case, four 3N force sets are used and calculations without 3N force (3NF) are also shown.
}
\label{fig:dp-ca40}
\end{figure*}

\begin{figure*}[t]
{\includegraphics[width=0.4\textwidth]{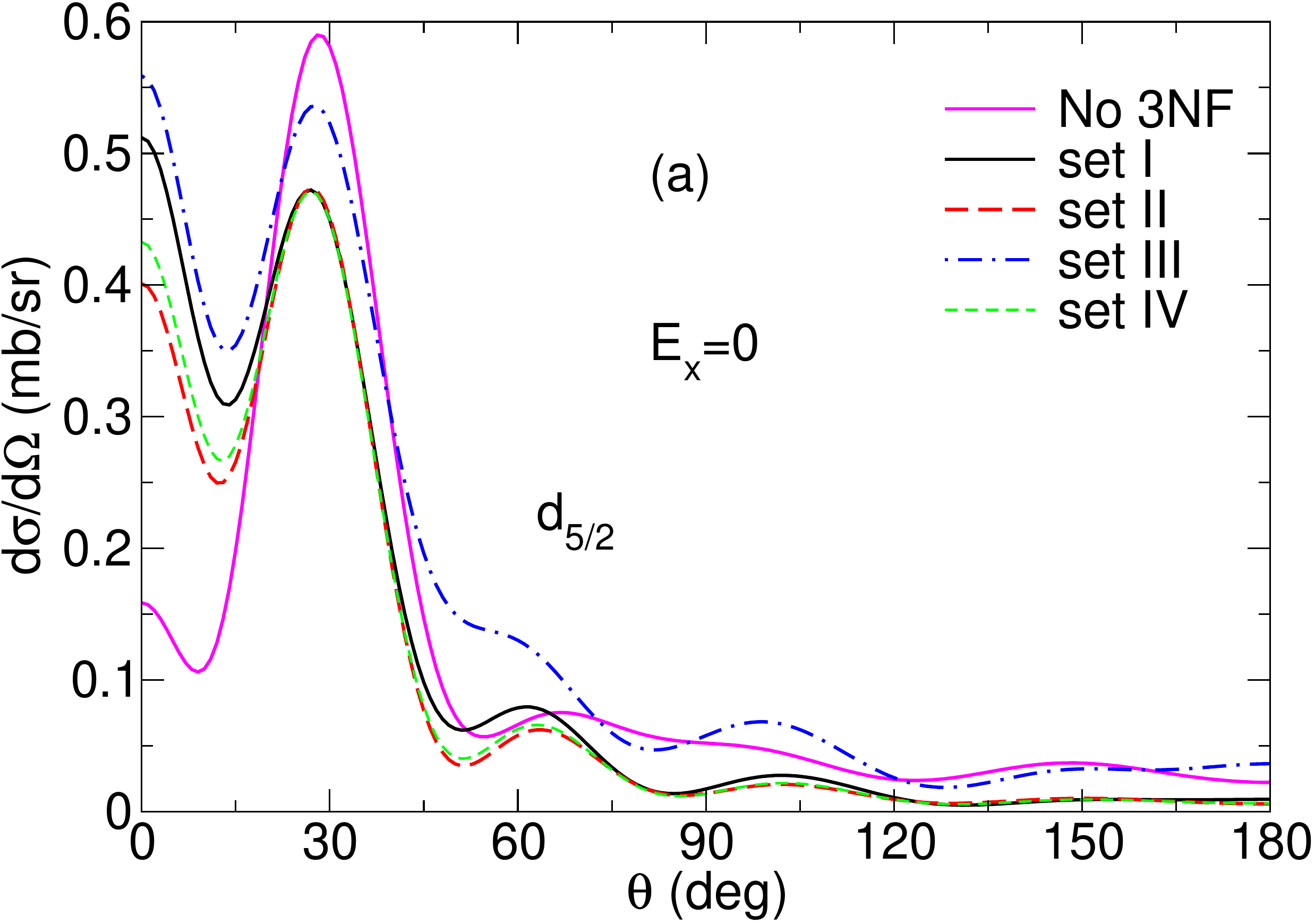}
\includegraphics[width=0.4\textwidth]{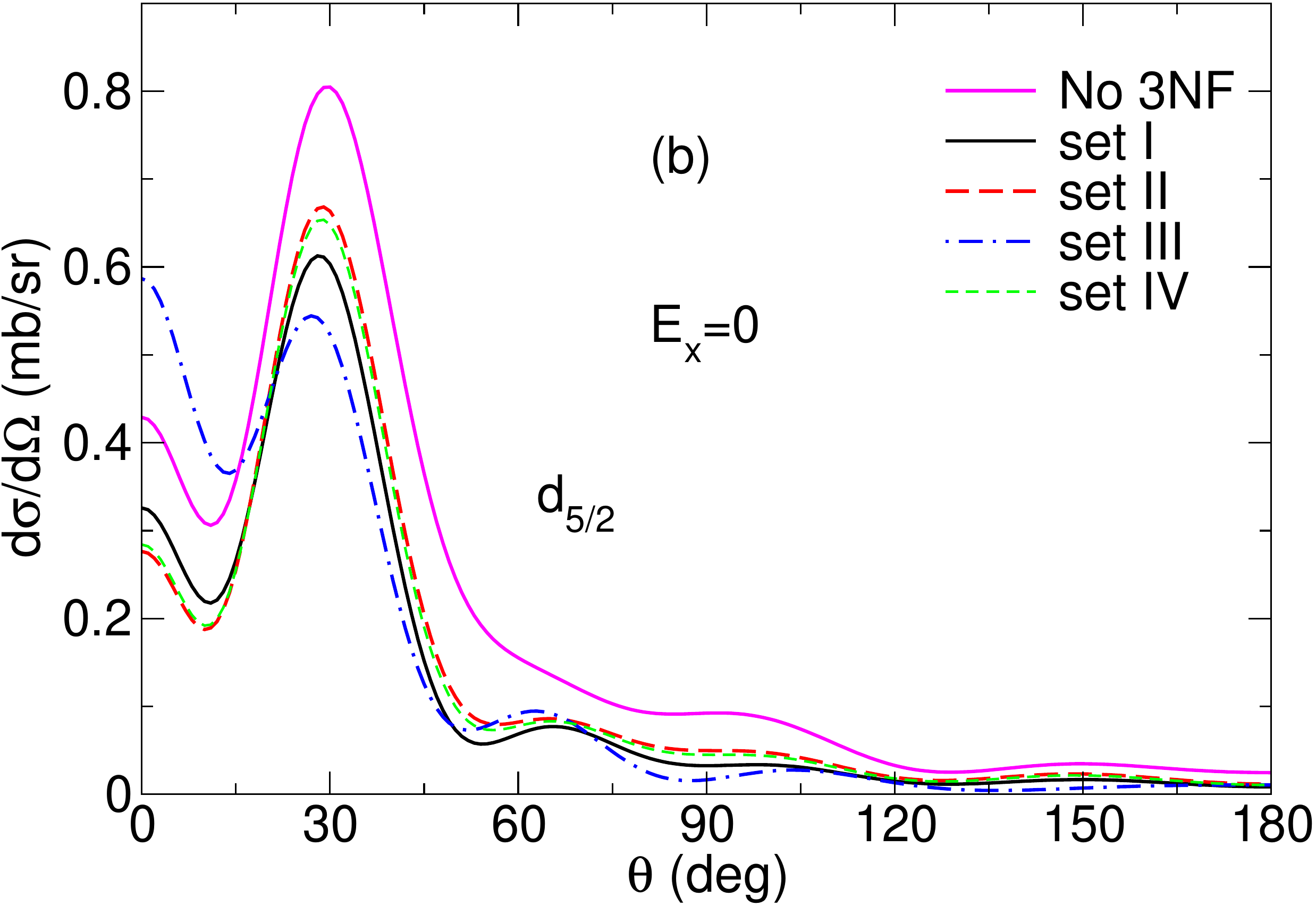}}
{\includegraphics[width=0.4\textwidth]{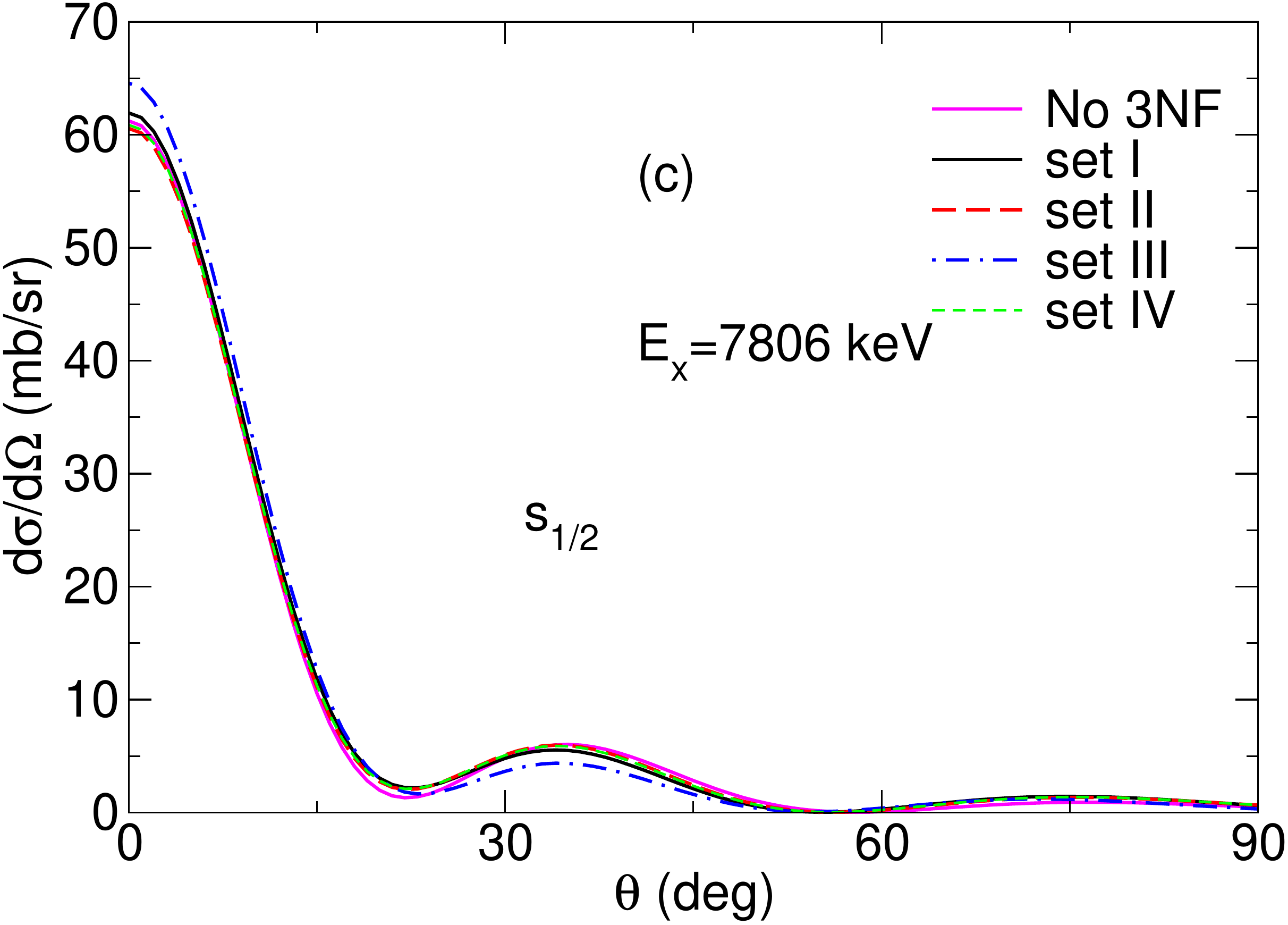}
\includegraphics[width=0.4\textwidth]{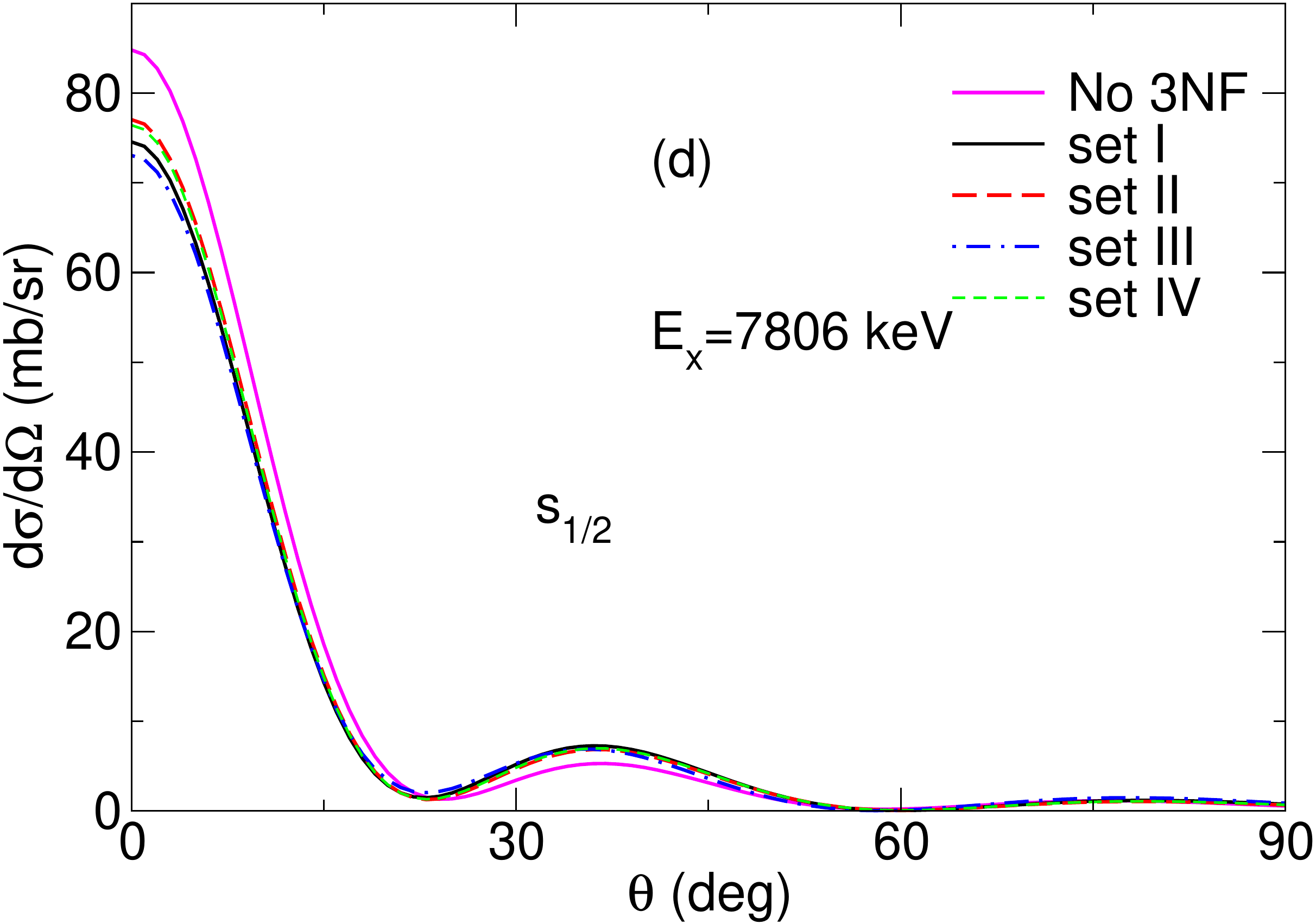}}
{\includegraphics[width=0.4\textwidth]{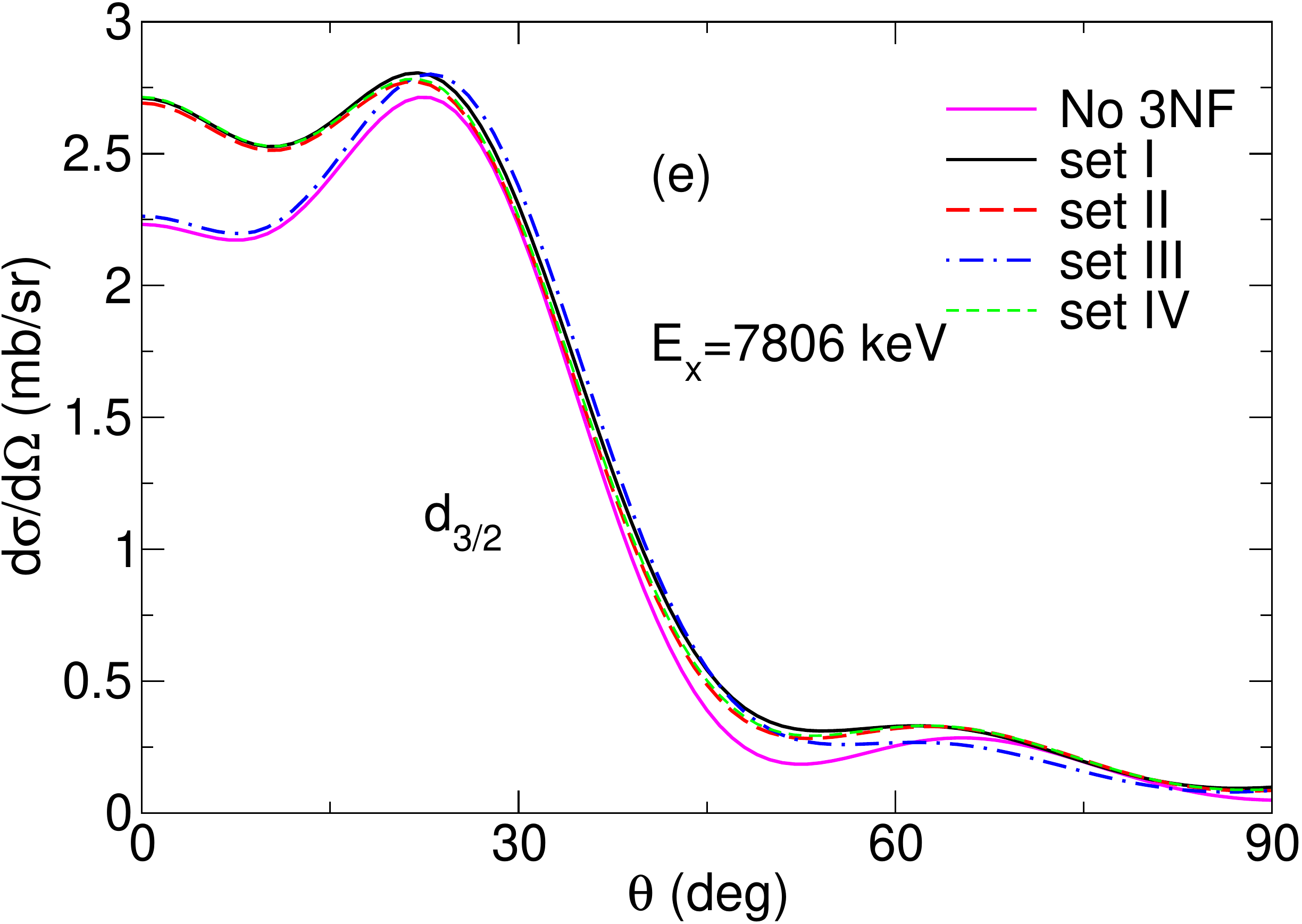}
\includegraphics[width=0.4\textwidth]{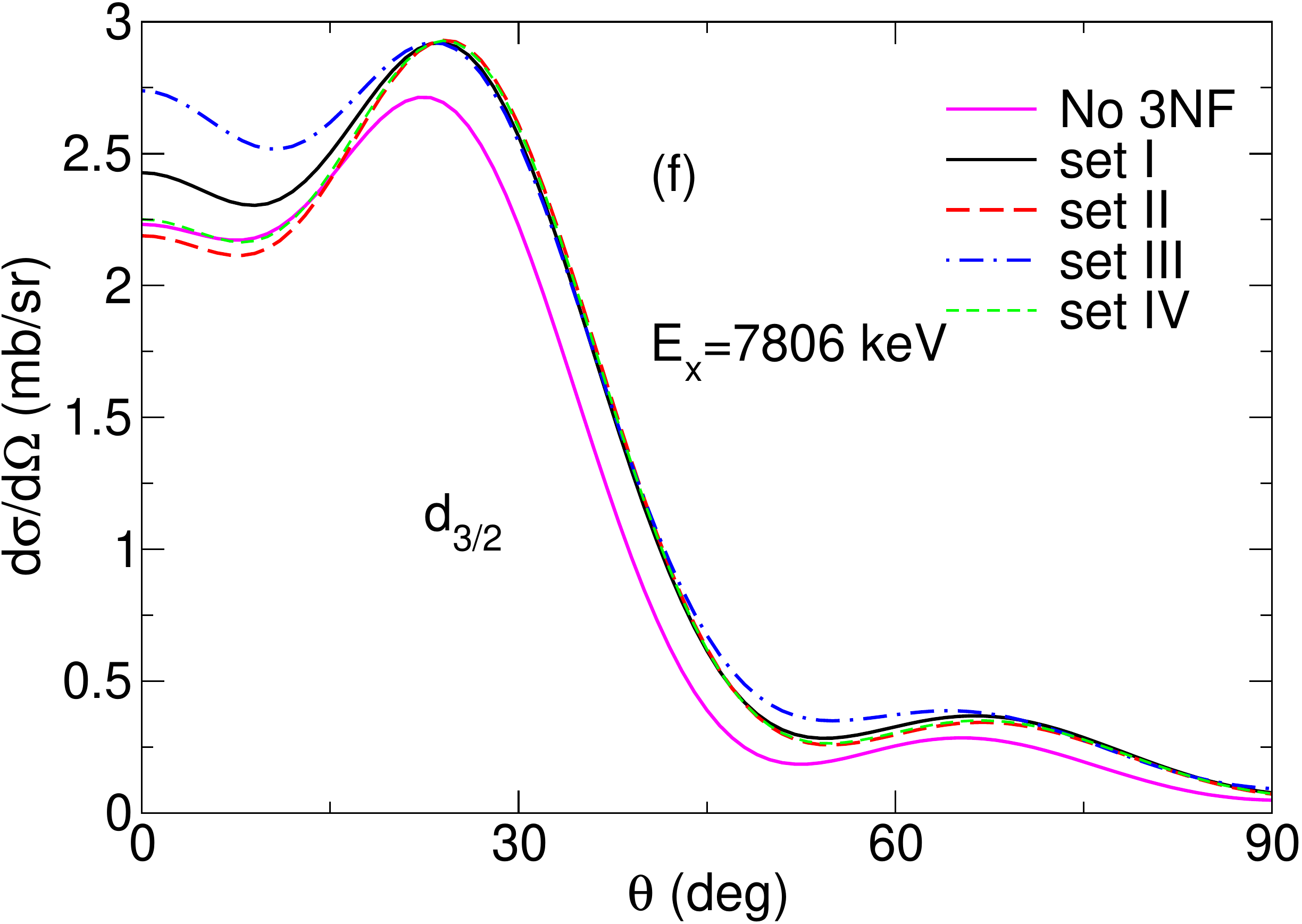}}
\caption{
The $^{26}$Al$(d,p)^{27}$Al reaction at $E_d = 12$ MeV calculated with nucleon optical potentials KD03 (left column) and GR (right column), each with four sets of 3N interactions. The final states in $^{27}$Al are the ground state, ($a$) and ($b$), and the $E_x = 7806$ KeV state for which the $s_{1/2}$ and $d_{5/2}$ transfers are presented separately in the middle row, ($c$) and ($d$), and bottom row, ($e$) and ($f$), respectively. Calculations without 3N force (3NF) are also shown.
}
\label{fig:dp-al26}
\end{figure*}

The example of the $d$-$A$ potential shown in Fig. \ref{fig:dA-ADWA}b 
 shows that additional contribution to the adiabatic potential due to the 3N force can be sufficiently strong to give a noticeable contribution to the $(d,p)$ cross section.
Intuitively, one could expect that
the largest influence on $(d,p)$ cross sections can be expected from the 3N force set III.
However,  we discovered that importance of the 3N contribution  varies with 
 the choice of the nucleon-target optical potentials used to construct the adiabatic potential $U^{\rm ADWA} = \la \phi_1 | U_{nA} + U_{pA}| \phi_0 \ra$. Adding a 3N contribution to this potential can be simulated by renormalizing the real part of $U^{\rm ADWA}$ by some real factor $N_R$. By changing $N_R$ monotonously within the range of 0.9 to 1.6 we learned that the corresponding changes in the  $(d,p)$ cross sections are not linear, moreover, they can behave differently for different  $U^{\rm ADWA}$ choices. This observation explains the cross sections results shown in Figs. \ref{fig:dp-ca40} and \ref{fig:dp-al26}.

 In Fig. \ref{fig:dp-ca40} we have plotted the ADWA  zero-range  differential cross sections for $^{40}$Ca($d,p)^{41}$Ca reaction calculated at two incident deuteron energies, 11.8  MeV and 56 MeV, 
 using the code TWOFNR \cite{TWOFNR}. Three different nucleon-target optical potentials were used both in the deuteron and proton channels: $(i)$ local potentials from the Koening-Delaroche (KD03) nucleon optical model systematics \cite{KD03}, $(ii)$ nonlocal energy-independent optical potentials from the  $N=Z$ Gianinni-Ricco (GR) systematics  \cite{GR} and $(iii)$ energy-dependent nonlocal dispersive optical model (NLDOM) potential \cite{NLDOM} with modified parameters from \cite{Wal16} that  reproduce the neutron separation from $^{41}$Ca. When employing nonlocal potentials we used their leading order local-equivalent representations both in the entrance deuteron and exit proton channels as explained in \cite{PB} and \cite{Tim13,Wal16}, respectively.
 The overlap between $^{40}$Ca and $^{41}$Ca was taken from \cite{Wal16}.

 The KD03 Johnson-Tandy potential for $E_d$ = 11.8 MeV has a depth of about 107 MeV and one could expect that adding  attractive 3N contributions with depths  between 13 and 37 MeV will introduce proportionate changes to the cross sections. However,  adding set I with the depth of 19 MeV does not change the cross sections much, while slightly shallower sets II and  IV affect the cross sections  decreasing it by $\sim 8\%$ (see Fig. \ref{fig:dp-ca40}a). At the same time adding deep attractive Set III  ($\sim 37$ MeV) increases the cross sections by 15$\%$,
  which is the result of non-linear nature of the optical potential behaviour with the renormalization factor $N_R$. The situation  changes with the deuteron incident energy, when for $E_d = 56$ MeV  the $U^{\rm ADWA}$ depth is about 10$\%$ smaller than that at 11.8 MeV. Figure \ref{fig:dp-ca40}b shows that while the slope of all four $^{40}$Ca$(d,p)^{41}$Ca cross sections in the angular range of $20^{\circ } < \theta < 30^{\circ }$ is similar their behaviour at $\theta < 20^{\circ}$ is very different and it does seem to scale with the depth of the added 3N contribution.

 For the case of nonlocal energy-independent optical $p$-$A$ and $n$-$A$ potentials the local-equivalent $d$-$A$ potential is shallower than the Johnson-Tandy potential constructed from local-equivalents of nonlocal $p$-$A$ and $n$-$A$ potentials \cite{Tim13}. Thus, for the case of $d-^{40}$Ca at $E_d = 11.8$ MeV considered above the depth of the GR local-equivalent potential is 86 MeV as compared to 107 MeV associated with the KD03 potential. For  $E_d = 56$ MeV these depths are 73 and 91 MeV, respectively.
 Unlike in the case of KD03, the spread between the 11.8 MeV cross sections calculated with four different 3N sets is smaller and they all  differ more noticeably from those calculated without the 3N force (see Fig. \ref{fig:dp-ca40}c).   Stronger 3N effects are seen in Fig.  \ref{fig:dp-ca40}d at 56 MeV for sets II, III and IV while set I does not affect the cross section despite the $U_{3N}$ depth is about 26$\%$ of $U^{\rm ADWA}$.

 \begin{figure*}[bth]
 {\includegraphics[width=0.45\textwidth]{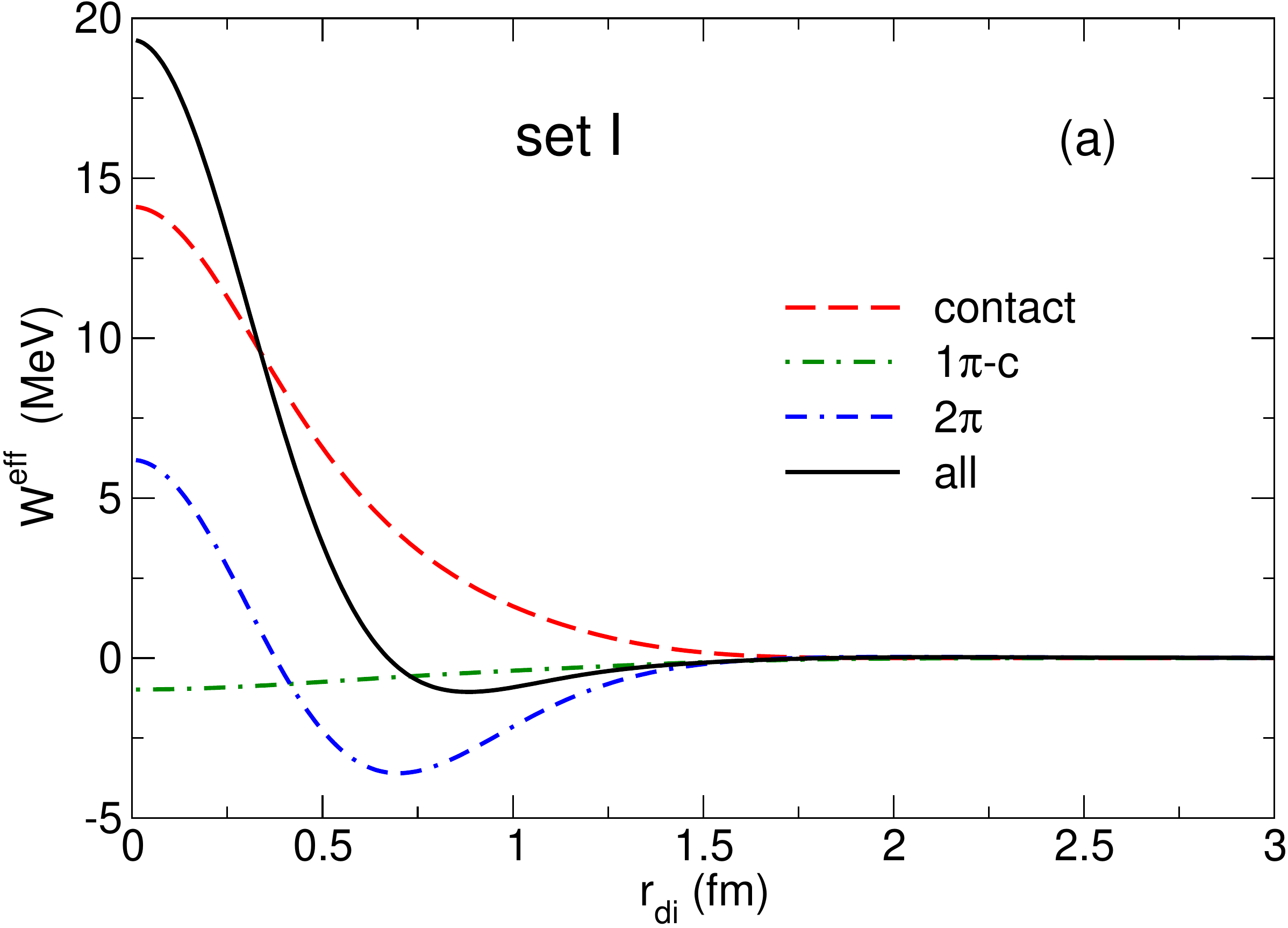}
\includegraphics[width=0.45\textwidth]{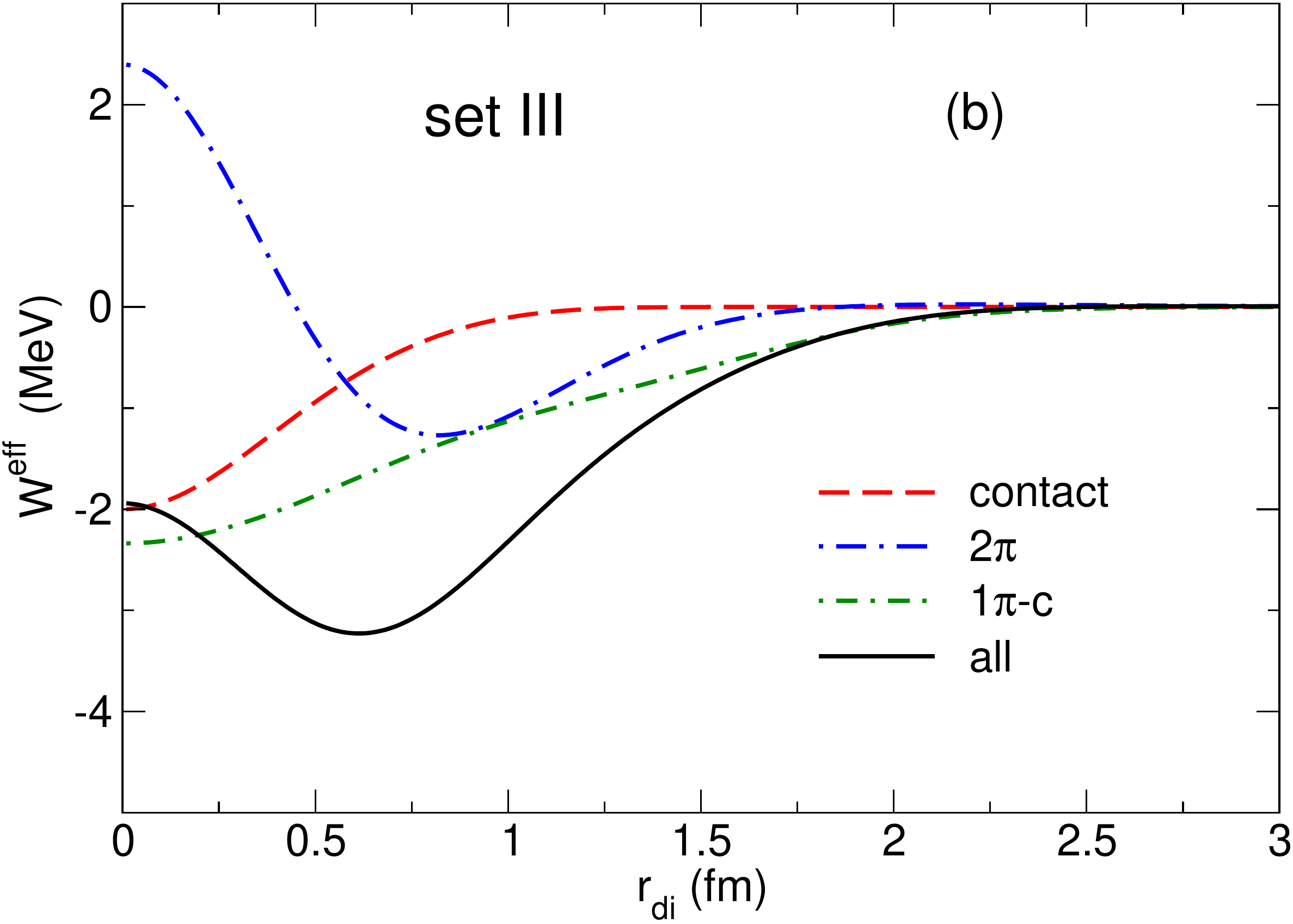}}
{\includegraphics[width=0.45\textwidth]{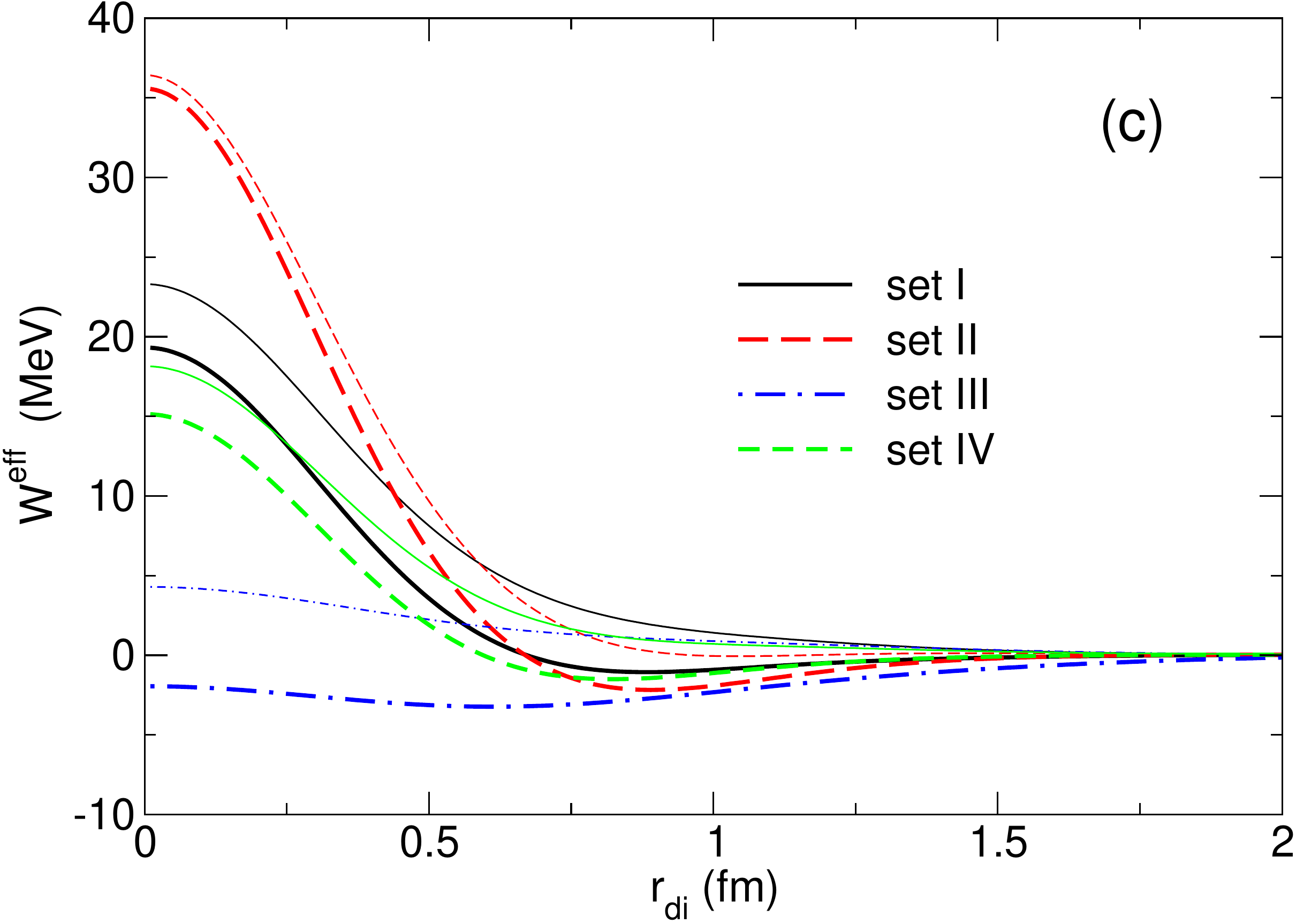}
\includegraphics[width=0.45\textwidth]{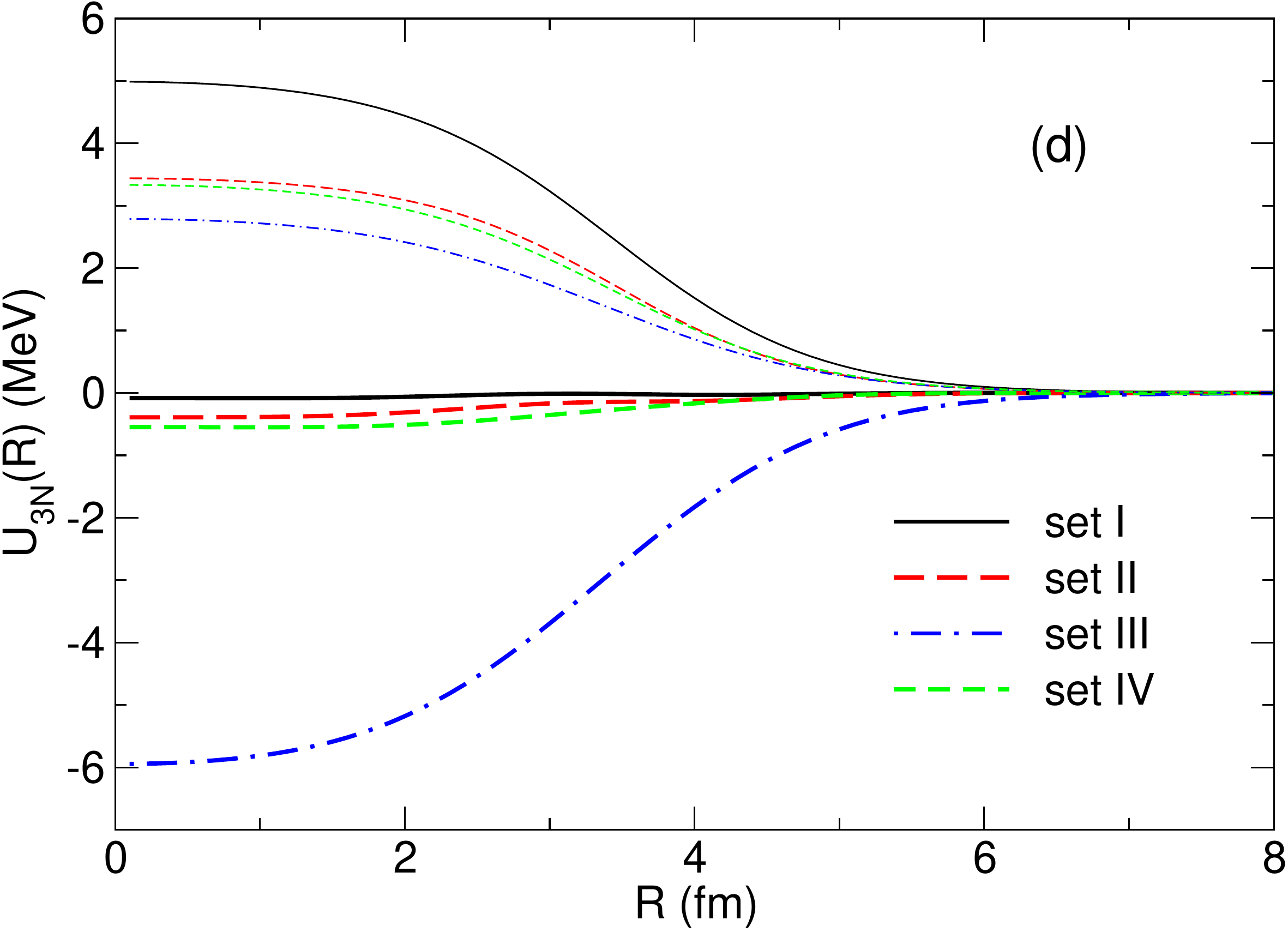}}
\caption{
The effective $d$-$i$ potentials $W^{\rm eff}_{di}$ calculated in the Watanabe model. Decomposition into contact,  1$\pi$-c and $2\pi$ exchange is shown in ($a$) and ($b$) for set I and set III, respectively. Comparison between $W^{\rm eff}_{di}$ obtained with  all four 3N sets is shown in ($c$). Thin lines represent calculations with the deuteron $s$-state only. Figure ($d$)  shows
 $U_{3N}(R)$ calculated for the $d +^{40}$Ca system using the same four sets and with $s$-state only shown by thin lines.
}
\label{fig:dA-Wat}
\end{figure*}
 
 It has been shown in \cite{Joh14} that in the adiabatic limit the energy-dependence of nonlocal optical potentials requires using the nucleon optical potentials evaluated at an energy shifted with respect to the traditionally used $E_d/2$ value by a large number given by the $n$-$p$ kinetic energy averaged over the short range of their interaction. This number is 57 MeV for a deuteron model without high $n$-$p$ momenta, but it can  be much larger \cite{Bai16}. The depth of an optical potential usually decreases with the nucleon energy, and for $d-^{40}$Ca ADWA potential with NLDOM it is 82 and 69 MeV for $E_d = 11.8$ and 56 MeV, respectively. 
 Adding 3N force at $E_d = 11.8$ Mev is more noticeable than in the case of the GR nonlocal potential and the spread between $(d,p)$ cross sections obtained with  four 3N sets is larger (see Fig. \ref{fig:dp-ca40}e). Even bigger 3N effect is seen in Fig. \ref{fig:dp-ca40}f at $E_d = 56$ MeV but the spread between the four cross sections is smaller.

 The $(d,p)$ cross sections calculated with nonlocal optical potentials evaluated at shifted energies  are overestimated \cite{Joh14,Wal16}. They lack the contribution from the induced $n+p+A$ three-body force (I3B) arising due to interaction between $n$ and $p$ via excited  states of the target $A$. It was shown in \cite{Din19} that accounting for this force in the ADWA  increases the imaginary part of the adiabatic deuteron potential by a factor of two, thus reducing the $(d,p)$ cross sections. It was also shown in \cite{Din19} that with increased absorption the $(d,p)$ cross sections are less sensitive to small changes in the real potential. Here, we increase the dynamical  part of NLDOM by a factor of two to account for I3B and observe that in this case adding $U_{3N}$ produces a smaller effect on  the $(d,p)$ cross sections for both deuteron incident energies (see Fig. \ref{fig:dp-ca40}~g and h).

 The calculations with the $^{40}$Ca target involve  neutron transfer to the $f_{7/2}$ orbit in $^{41}$Ca. To get an idea of how the 3N force manifests itself in $(d,p)$ reactions for other angular momenta we have considered the $d_{5/2}$ and $s_{1/2}$ transfers in the $^{26}$Al$(d,p)^{27}$Al reaction at $E_d = 12 $ MeV for two $^{27}$Al final states: the ground state and the astrophysically important state at $E_x = 7806$ keV. The calculations are performed with the local KD03 and nonlocal GR optical potentials only. The standard geometry of the potential well for the transferred neutron (radius  $r_0 = 1.25$ fm and diffuseness $a = 0.65$ fm) was used with the spectroscopic factor set to one.  The results, shown in Fig. \ref{fig:dp-al26}, suggest that the $(d,p)$  sensitivity to the 3N force is smaller for the $s_{1/2}$ transfers, while $d_{5/2}$ transfer can be more strongly influenced by it, with the ground state of $^{27}$Al being particularly affected. Again, as in the $^{40}$Ca$(d,p)^{41}$Ca case,  quantative conclusions about importance of the 3N contribution depend on the choice of the optical potential.

\section{The Watanabe model with 3N force}

The adiabatic calculations of the $d$-$A$ folding potential, $U_{3N}$, revealed its strong sensitivity to the choice of the 3N model and that for all models a large contribution to  $U_{3N}$  comes from the deuteron $d$-state. This sensitivity arises because the ADWA uses the short-range function $v_l(r)$ representing the $n$-$p$ interaction, which is large when $l=2$.

   \begin{figure*}[t]
{\includegraphics[width=0.4\textwidth]{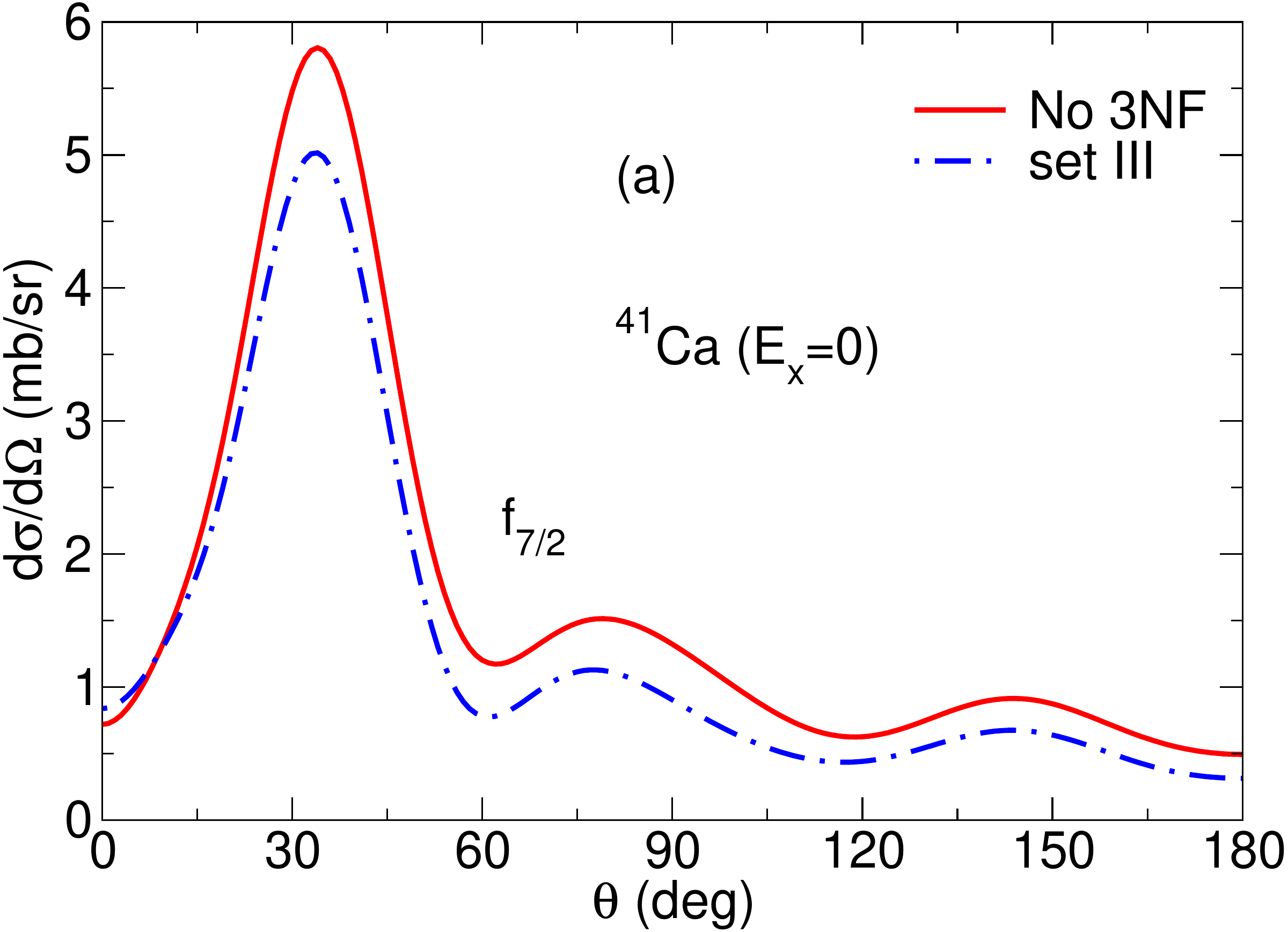}
\includegraphics[width=0.4\textwidth]{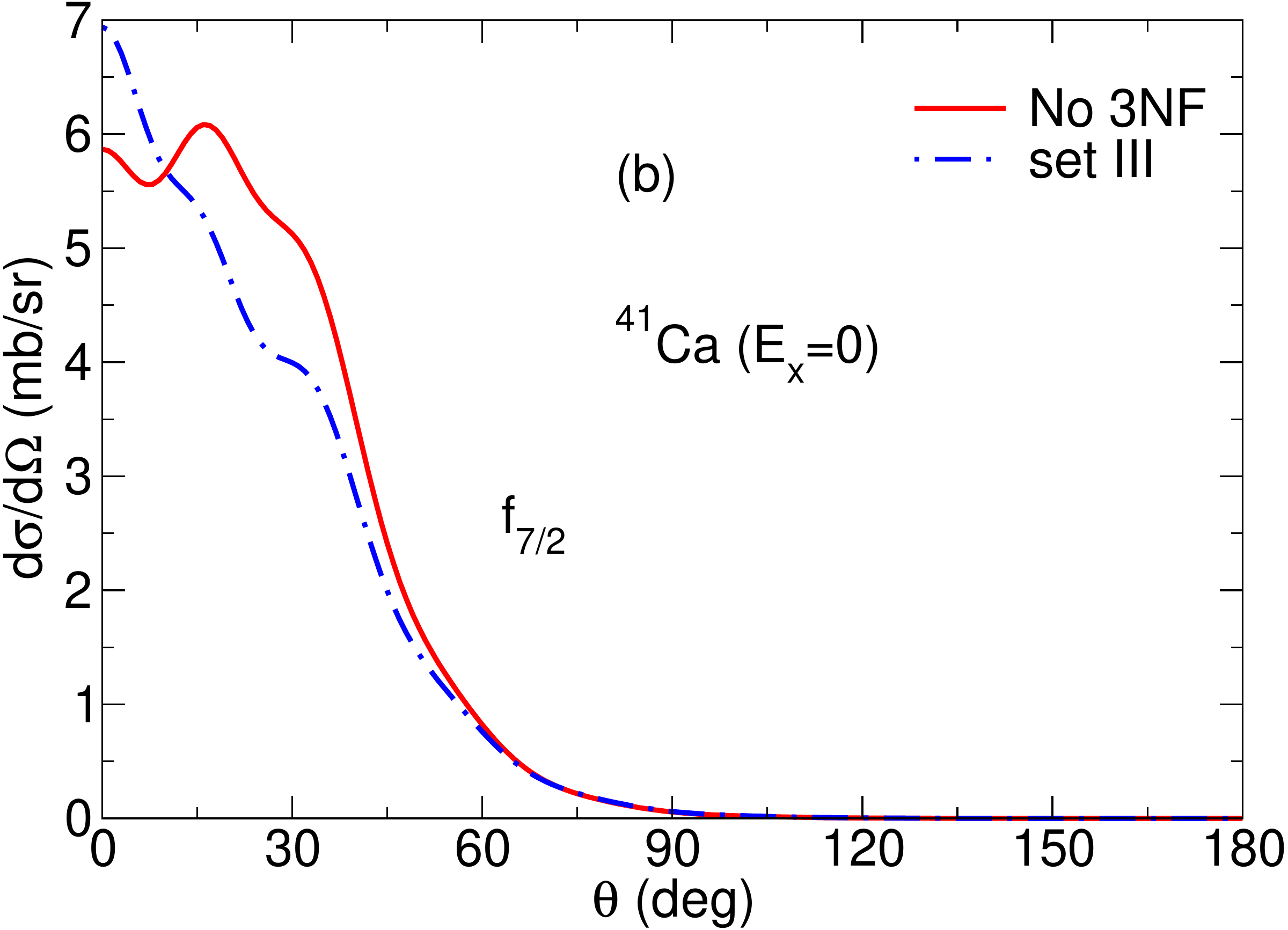}}
\includegraphics[width=0.4\textwidth]{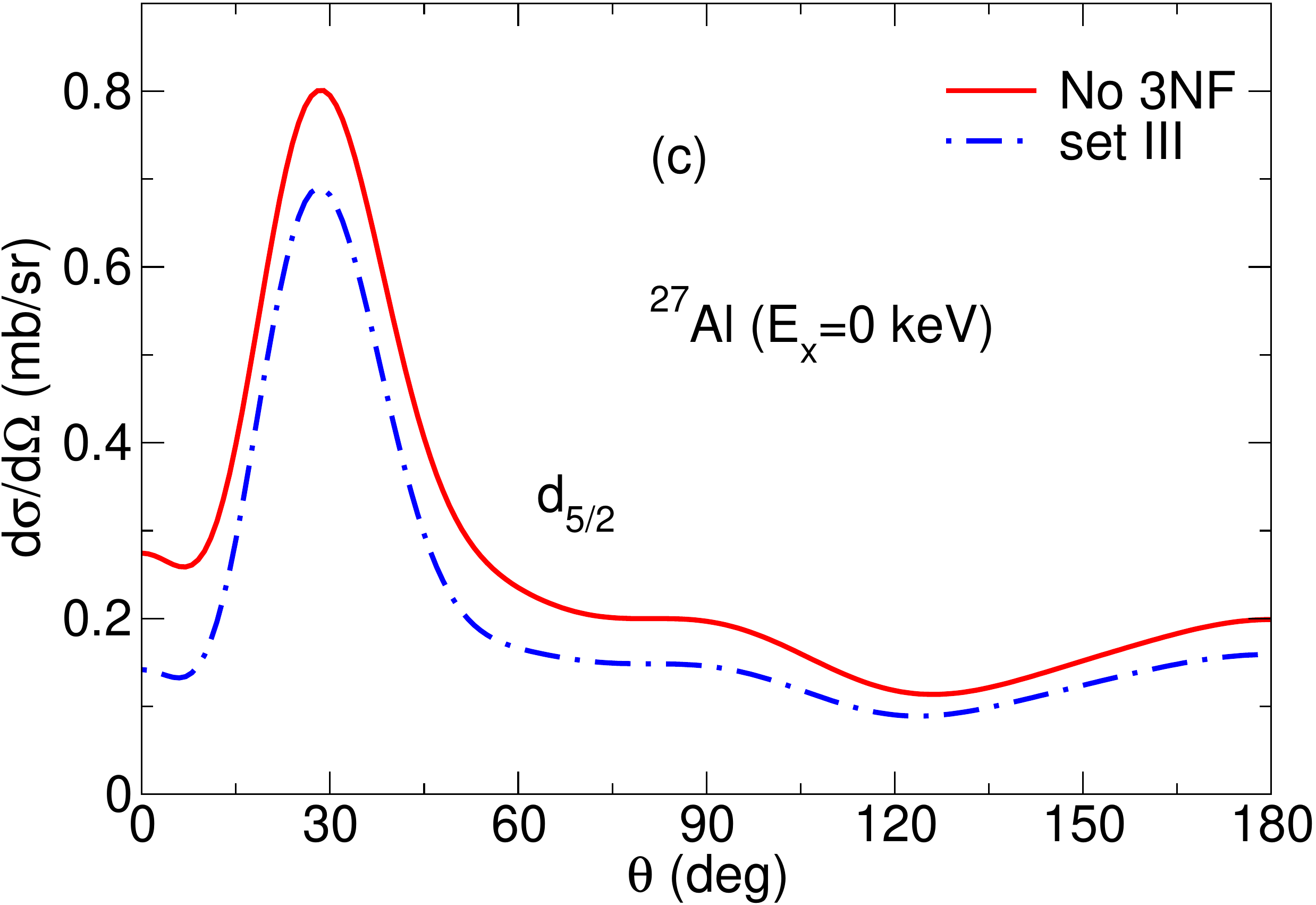} \\
{\includegraphics[width=0.4\textwidth]{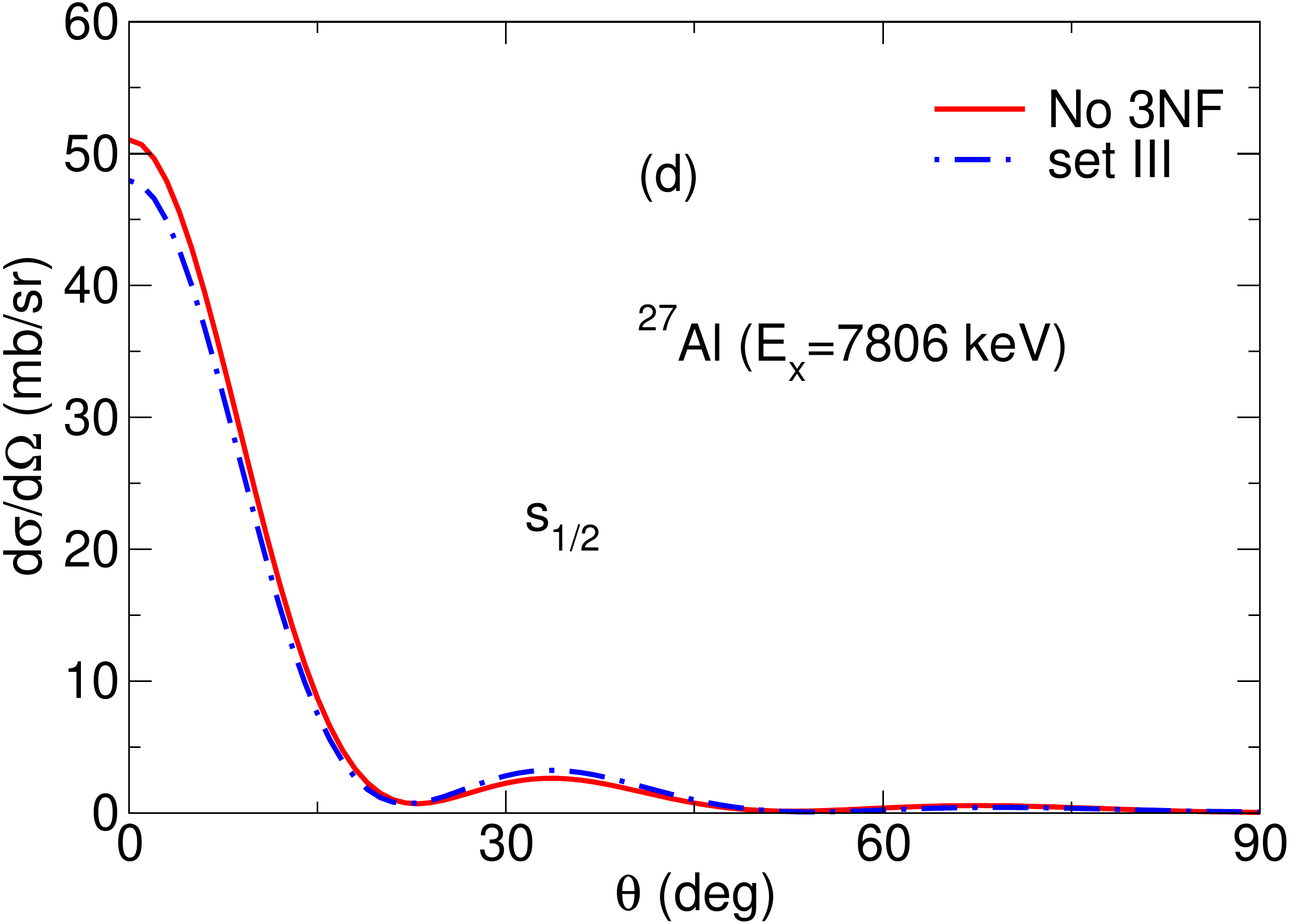}
\includegraphics[width=0.4\textwidth]{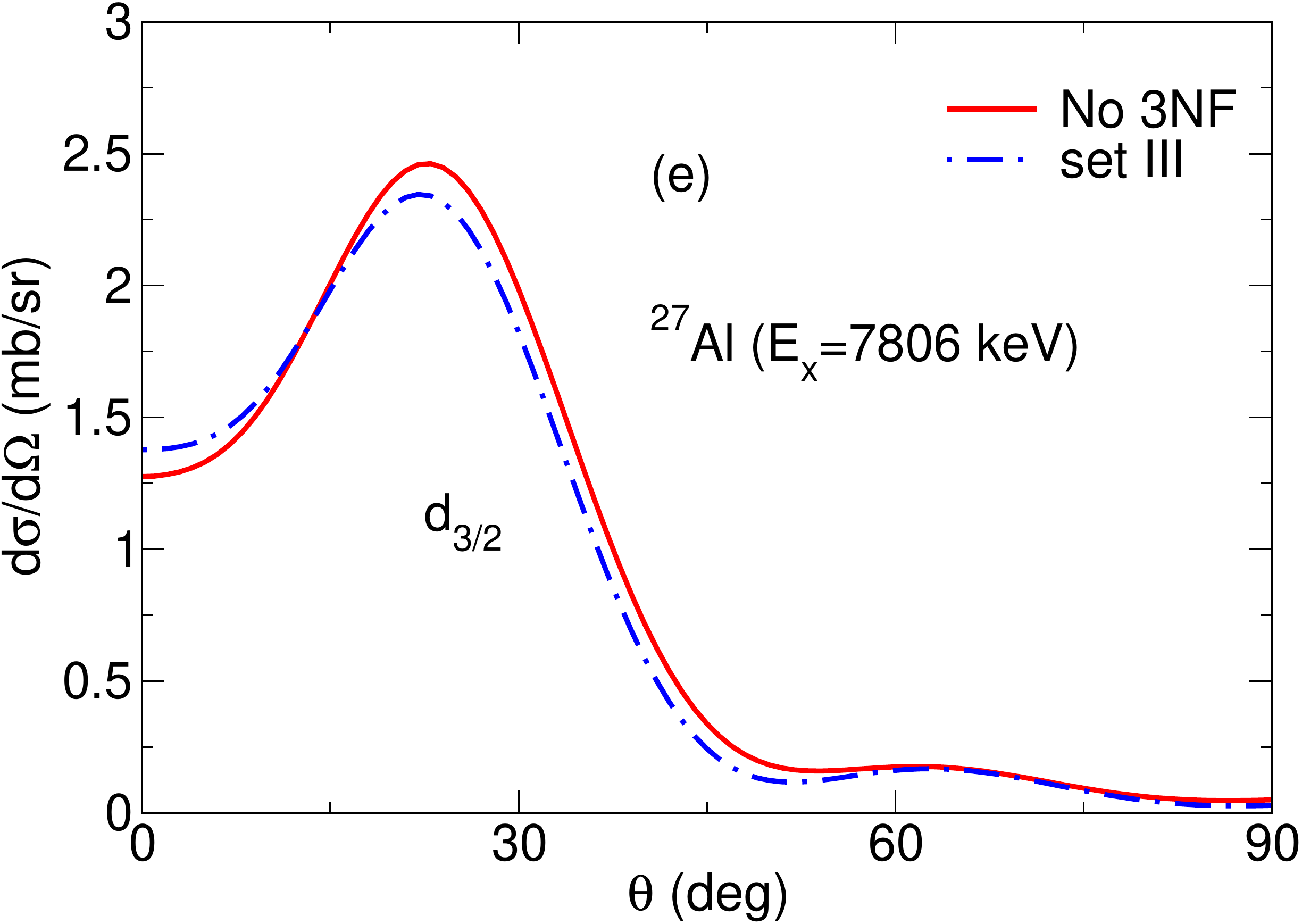}}
\caption{
The $^{40}$Ca$(d,p)^{41}$Ca cross sections at $E_d = 11.8$ MeV ($a$) and 56 MeV ($b$) and the $^{26}$Al$(d,p)^{27}$Al  cross sections at $E_d = 12$ MeV ($c, d, e$) calculated in the Watanabe model  calculated without 3N force (3NF) and with  set III.  The final states in $^{27}$Al are the ground state ($c$) and the $E_x = 7806$ KeV state for which the $s_{1/2}$ and $d_{5/2}$ transfers are presented separately in ($d$) and ($e$), respectively.
}
\label{fig:dp-watanabe}
\end{figure*}

The ADWA accounts for deuteron breakup only in an approximate way. An exact three-body treatment of deuteron breakup in Faddeev  and continuum-discretized coupled channel (CDCC) approaches using local nucleon optical potentials suggests that non-adiabatic corrections can be important (see, for example, \cite{Upa12,Cha17}). Non-adiabatic effects are even stronger when $N$-$A$ optical potentials, employed in $(d,p)$ calculations, are nonlocal \cite{Gom18}.
To understand if 3N-model  sensitivity persists  beyond the ADWA we have performed a few Watanabe folding calculations of $W^{\rm eff}_{di}$ and $U_{3N}$. Although the Watanabe folding model does not connect the deuteron ground state wave function with its breakup states, it represents the leading order term in the CDCC expansion. One might expect that dependence on the 3N force would be similar for all CDCC matrix elements involving the lowest continuum bins since for a fixed NN interaction  the $n$-$p$ continuum wave functions are similar at low $n$-$p$ energies and at small $n$-$p$ separations that determine $W^{\rm eff}_{di}$.

The expressions for $W^{\rm eff}_{di}$ in the Watanabe folding model are obtained by replacing all $v_l$ by $u_l$ in the corresponding  equations of Sections IV, V and VI.  The $W^{\rm eff}_{di}$ are shown in Fig. \ref{fig:dA-Wat}  for all four sets of the 3N force and in the case of sets I and III, where $c_D \neq 0$, separate contributions from the contact, 1$\pi$-c and $2\pi$-exchange are shown. For the total $W^{\rm eff}_{di}$ force we also show the calculations with deuteron $s$-state only.

The Watanabe  potentials $W^{\rm eff}_{di}$ 
  are wider than those obtained in ADWA
because they contain a product of two long-range functions $u_l(r)$ and $ u_{l'}(r)$ (see Table II for r.m.s. radii of $W^{\rm eff}_{di}$). The overlap of such a product  with the short-range 3N force is weaker than the one for $u_l(r)v_{l'}(r)$  so that the  resulting Watanabe  $W^{\rm eff}_{di}$ potentials are weaker as well (see Table II for their volume integrals).
The probability of the $d$-state in $u_l$ is smaller than in $v_l$ but the $d$-state still gives noticeable contribution to $W^{\rm eff}_{di}$ mainly through the tensor operator $S_{np}$ in the 1$\pi$-c and 2$\pi$ cases.

The $d$-state influence propagates to the $d$-$A$ potentials $U_{3N}$, shown in Fig.  \ref{fig:dA-Wat}d. Without it $U_{3N}$ is repulsive with the height showing a small spread between 3 and 5 MeV. With $d$-state included, the contribution from the 2$\pi$-exchange almost cancels out the contribution from the contact interaction for sets I, II and IV, leading to  negligible contribution from the 3N force. A different story occurs for set III where all three contributions, the contact, the 1$\pi$-c and 2$\pi$, are attractive, resulting in a noticeable total 3N force. It is still significantly smaller than any $U_{3N}$ of the adiabatic case but does affect the $(d,p)$ cross sections, which is shown in  Fig. \ref{fig:dp-watanabe}
for the same targets, $^{40}$Ca and $^{26}$Al, and at the same deuteron incident energies that are considered in the previous section. 
In these calculations
we  used the Watanabe model  to calculate the folding potentials
$\la \phi_0 | U_{nA} + U_{pA} | \phi_0\ra$, with KD03  chosen for $U_{nA}$ and $U_{pA}$. 
For $^{40}$Ca target the effect of including the 3N force (set III) has an opposite effect to the one seen in the ADWA for both deuteron energies. For $^{26}$Al, this force
has a weaker effect on   populating the $^{27}$Al ground state than the ADWA does but it  shows a slightly stronger influence  for the  final $^{27}$Al state at $E_x = 7806$ keV. In the latter case the 3N effect is still small, no more than 10$\%$ in the maximum.

 \section{Summary, conclusions and future challenges}
 
 We have presented the first calculations of a three-body optical potential due to bare 3N interactions of the neutron and proton in the incoming deuteron with target nucleons. With the aim of   using this potential in modelling $(d,p)$ reactions, we have evaluated its expectation value in  ADWA  motivated by the widespread use of the latter in the analysis of deuteron stripping  experiments. Unlike standard $N$-$A$ folding potentials, which are mainly sensitive to the volume integral of the NN force along with the target density, the three-body optical potential arising due to bare 3N interactions probes the short-range part of the deuteron wave function and both the strength and the range of the 3N interactions. A simple hypercentral model considered in Sec. III shows the necessity of a consistent approach to the choice of both the deuteron model and the 3N force. 
 
In the present work the deuteron wave function has been taken from the $\chi$EFT at N2LO where 3N force arises naturally. We used consistently four N2LO sets of local 3N interactions from  \cite{Lyn16} that describe equally well some light nuclei and nuclear matter.
  These sets involve three different formats of the contact interactions, with two sets having a $1\pi$-c component as well, and 2$\pi$-exchange. The  last two are heavily dominated by the contribution from the deuteron $d$-wave. Within ADWA, the deuteron $d$-wave state also gives an important contribution to the contact interaction.

 We have used all  four 3N force sets in the ADWA  calculations of $(d,p)$ cross sections and have shown that, potentially, they could distinguish between the  choices made for the 3N force. However, the relative importance of these choices also depends on the nucleon-target optical potential employed in calculations. Future development of {\it ab-initio} theories of optical potentials, such as those initiated in \cite{Rot17,Rot18,Idi19}, may on the one hand fix this optical potential dependence, but on the other they might introduce an additional  new 3N-dependence in the optical potentials.

 The ADWA overemphasizes the role of short $n$-$p$ distances, which may lead to overestimation of 3N  effects. We have compared the ADWA predictions to those of the Watanabe model, which also treats $d$-$A$ as a three-body system but does not allows for deuteron excitation to the $n$-$p$ continuum. The 3N Watanabe $d$-$A$ potential is close to zero for three sets of the 3N force (I, II and IV), where near-cancellation occurs between the contributions from the contact and 2$\pi$ forces, but for set III  it is sufficiently large  to affect the $(d,p)$ cross sections. This particular set has a large contribution from the 1$\pi$-c force and attractive contact  interaction so that amplification rather than cancellation takes place when adding all three 3N contributions. A noticeable effect on $(d,p)$ cross sections in the Watanabe model suggests that it can also be manifest beyond the ADWA.

 This work is the first step in the investigation of the bare 3N force effects in the deuteron channel of $(d,p)$ reactions, but it does not yet give the full picture of this contribution. Other effects, such as non-adiabatic deuteron breakup into  $n$-$p$  states, including those with different  orbital momenta and spins (for example breakup into singlet channels), and additional 3N term in the $T$-matrix, as introduced in \cite{Tim18}, could modify the conclusions of this work. Both of these effects require an extension of the folding formalism considered in this paper.

 We should also point out that a fixed phenomenological density of the target $A$ was used in our calculations. Ideally, the target density should be calculated in a microscopic model that uses the same NN and 3N interactions. This could induce an additional 3N-dependence sensitivity of the $U_{3N}$ potential and the corresponding $(d,p)$ cross sections.
 The target density for non-zero spin nuclei can also have significant non-spherical components, which may give rise to stronger quadrupole contributions to the incident channel deuteron scattering. These are intriguing problems awaiting further exploration.
 
 Finally, we should point out that choosing a different NN model and the 3N interactions associated with it  can result in a different $(d,p)$ sensitivity to the 3N force. Changing the NN model, for example by going beyond the N2LO within the $\chi$EFT, requires calculation of new matrix elements involving 3N forces. They contain high-momentum cut-off regulators and uncertainty in the  choice of the latter can significantly affect 
the short-range form of the deuteron wave function  important for $W^{\rm eff}_{di}$ and $U_{3N}$ calculations. It would be fascinating to see if $(d,p)$ reactions could offer any help in restricting the model parameters of the NN and 3N force.

 \section*{Acknowledgements.}
 The $\chi$EFT N2LO deuteron wave function has been obtained from J. Lynn, which is gratefully appreciated. We are also  thankful to I. Tews for valuable comments about 3N interaction. 
 This work was supported by the United Kingdom Science and Technology Facilities Council (STFC) under Grant No. ST/L005743/1.

 \section*{Appendix}

 Here expressions for $H^{(i)}(x_-,x_+)$ are given for $x_{\pm} = \sqrt{r_{di}^2  \pm  r r_{di} + \frac{r^2}{4}}$. The notation $F_{\pm}$ everywhere stands for $F(x_{\pm})$ and $F(x)$ can be any of the functions $U(x)$, $T(x)$, $Y(x)$ or $\delta_{R_{3N}}(x)$. So,
 \beq
 H^{(0)}(x_-,x_+) = -
 \frac{3c_1m_{\pi}^4 g_A^2}{2f^4_{\pi}(4\pi)^2}
 \frac{U_-Y_- U_+Y_+}{x_-
x_+},
 \eeqn{H0}
\begin{widetext}
\beq
 H^{(1)}(x_-,x_+) =  -
\frac{3c_3 g_A^2}{36f^4_{\pi}}
\left\{
\frac{m_{\pi}^4 }{16\pi^2} \left[ 9\left(r_{di}^2 - \frac{r^2}{4}\right) \frac{Y_-Y_+T_-T_+}{x^2_-x^2_+}+ 3Y_-Y_+\left( \frac{T_-(1-T_+)}{x^2_-} +\frac{T_+(1-T_-)}{x^2_+}\right)  \right]
\right. \eol
-
\left.
\frac{3m_{\pi}^2 }{4\pi} \left( \frac{Y_-T_-\delta_+}{x^2_-} +\frac{Y_+T_+\delta_-}{x^2_+}\right) \right\},
 \eeqn{H1}
   \beq
 H^{(2)}(x_-,x_+) =  -
\frac{3c_3 g_A^2}{36f^4_{\pi}}
\left\{
\frac{m_{\pi}^4 }{16\pi^2} \left[ -\frac{9}{4}\left(r_{di}^2 - \frac{r^2}{4}\right) \frac{Y_-Y_+T_-T_+}{x^2_-x^2_+}+ \frac{3}{4}Y_-Y_+\left( \frac{T_-(1-T_+)}{x^2_-} +\frac{T_+(1-T_-)}{x^2_+}\right)  \right]
\right. \eol
-
\left.
\frac{3m_{\pi}^2 }{4\pi} \left( \frac{Y_-T_-\delta_+}{x^2_-} +\frac{Y_+T_+\delta_-}{x^2_+}\right) \right\},
 \eeqn{H2}
 \beq
H^{(3)}(x_-,x_+) =  -
 \frac{3c_3 g_A^2}{36f^4_{\pi}}
 \left[
- \frac{3m_{\pi}^4 }{16\pi^2} Y_-Y_+ \left( \frac{T_-(1-T_+)}{x^2_-} -\frac{T_+(1-T_-)}{x^2_+}\right)  +\frac{3m_{\pi}^2 }{4\pi} \left( \frac{Y_-T_-\delta_+}{x^2_-} -\frac{Y_+T_+\delta_-}{x^2_+}\right)
\right],
 \eeqn{H3}
\beq
H^{(4)}(x_-,x_+) = - \frac{3c_3 g_A^2}{36f^4_{\pi}}
\left[
\frac{m_{\pi}^4 }{16\pi^2} Y_-Y_+ (1-T_-)(1-T_+) -\frac{m_{\pi}^2 }{4\pi} \left[ Y_-(1-T_-)\delta_+ +Y_+(1-T_+)\delta_-\right]- +\delta_-\delta_+
\right].
\eeqn{H4}
\end{widetext}

\bibliographystyle{apsrev4-1}

\begin{thebibliography}{50}
\bibitem{Bar13} B.R. Barrett, P. Navratil and J.P.  Vary, Prog. Part.  Nucl. Phys. {\bf 69}, 131 (2013).

\bibitem{Lyn16} J.E. Lynn, I. Tews, J. Carlson, S. Gandolfi, A. Gezerlis, K.E. Schmidt and A. Schwenk,  Phys. Rev. Lett. {\bf 116},  062501 (2016).
 \bibitem{Tew16} I. Tews, S. Gandolfi, A. Gezerlis and A. Schwenk, Phys. Rev. C {\bf 93}, 024305 (2016)
\bibitem{Pia20} M. Piarulli and I. Tews,
Frontiers in Physics, {\bf 7}, 245 (2020). 

\bibitem{Del15} A. Deltuva and P.U. Sauer, 
 Phys. Rev. C {\bf 91}, 034002 (2015).
\bibitem{Viv18} M. Viviani, L. Girlanda, A. Kievsky, L.E. Marcucci, J. Dohet-Eraly, Few-Body Syst. {\bf 59}, 73 (2018)
\bibitem{Bar16} D. W. Bardayan,
J. Phys. G: Nucl. Part. Phys. {\bf 43}, 043001 (2016)

\bibitem{Austern} N. Austern, { \em Direct nuclear reaction theories},
(Wiley, New York, 1970)
\bibitem{JT} R.C. Johnson and P.C. Tandy, Nucl. Phys. A \textbf{235}, 56 (1974).
\bibitem{Tim18} N.K. Timofeyuk, Phys. Rev. C {\bf 97}, {054601},  (2018)
\bibitem{Din19} M.J.  Dinmore, N.K. Timofeyuk, J.S. Al-Khalili and R.C. Johnson,  Phys. Rev.  C {\bf 99}, 064612 (2019) 
\bibitem{Tim20} N.K. Timofeyuk and R.C. Johnson, Prog. Part. Nucl. Phys. {\bf 111}, 103738 (2020)

\bibitem{Pan13} D.Y. Pang, N.K. Timofeyuk, R.C. Johnson, J.A. Tostevin, Phys.Rev. C {\bf 87}, 064613 (2013)


 \bibitem{deJ74} C. D. Jager, H. D. Vries, and C. D. Vries, At. Data and Nucl. Data Tables, {\bf 14},  479 (1974)
 \bibitem{Epe05} E. Epelbaum, W. Gl\"ockle and U.-G. Mei\ss ner, Nucl. Phys. A {\bf 747}, 362 (2005)

  \bibitem{TWOFNR} J.A. Tostevin, University of Surrey version of the code {\sc twofnr} (of M. Toyama, M. Igarashi and N. Kishida) and code {\sc front}, 
  http://nucleartheory.eps.surrey.ac.uk/NPG/code.htm
 \bibitem{KD03} A.J. Koning and J.P. Delaroche,  Nucl. Phys. A {\bf 713}, {23} (2003)
 \bibitem{GR} M. M. Giannini, G. Ricco, Ann. Phys. \textbf{102}, 458 (1976).
 \bibitem{NLDOM} M. H. Mahzoon, R. J. Charity, W. H. Dickhoff, H. Dussan, S. J. Waldecker, Phys. Rev. Lett. \textbf{112}, 162503 (2014).

 \bibitem{Wal16} S.J. Waldecker and N.K. Timofeyuk, Phys. Rev. C \textbf{94}, 034609 (2016).
\bibitem{PB} F. Perey and B. Buck,  Nucl. Phys. {\bf 32}, 353 (1962).
 \bibitem{Tim13} N.K. Timofeyuk and R.C. Johnson, Phys. Rev. Lett. {\bf 110}, {112501} (2013)
 
\bibitem{Joh14} R.C. Johnson and N.K. Timofeyuk, Phys. Rev. C \textbf{89}, 024605 (2014).
\bibitem{Bai16}  G. W. Bailey, N. K. Timofeyuk, J. A. Tostevin, Phys. Rev. Lett. \textbf{117}, 162502 (2016).


\bibitem{Upa12} N.J. Upadhyay, A. Deltuva, F.M. Nunes, Phys. Rev. C {\bf 85}, 054621 (2012) 
\bibitem{Cha17} Y. Chazono, K. Yoshida, K. Ogata, Phys. Rev. C {\bf 95},  064608 (2017).
\bibitem{Gom18} M. G\'omez-Ramos and N.K. Timofeyuk, Phys. Rev. C \textbf{98}, 011601(R) (2018).
\bibitem{Rot17} J. Rotureau, P. Danielewicz, G. Hagen, F.M. Nunes, T. Papenbrock, Phys. Rev. C {\bf 95}, 024315 (2017) .
\bibitem{Rot18} J. Rotureau, P. Danielewicz, G. Hagen, G.R. Jansen, F.M. Nunes, Phys. Rev. C {\bf 98},  044625 (2018).
\bibitem{Idi19} A. Idini, C. Barbieri, P. Navr\'atil, Phys. Rev. Lett. {\bf 123} 092501 (2019).
\end{thebibliography}

 \end{document}